\documentclass[sigconf]{acmart}

\copyrightyear{2021}
\acmYear{2021}
\setcopyright{acmcopyright}
\acmConference[CCS '21]{Proceedings of the 2021 ACM SIGSAC Conference on Computer and Communications Security}{November 15--19, 2021}{Virtual Event, Republic of Korea}
\acmBooktitle{Proceedings of the 2021 ACM SIGSAC Conference on Computer and Communications Security (CCS '21), November 15--19, 2021, Virtual Event, Republic of Korea}
\acmPrice{15.00}
\acmDOI{10.1145/3460120.3484577}
\acmISBN{978-1-4503-8454-4/21/11}

\usepackage{tikz}
\usepackage{amsmath}
\usepackage{todonotes}
\usepackage[normalem]{ulem}
\usepackage{tabularx}
\usepackage{tikz}
\usepackage{amsmath}

\usepackage[utf8]{inputenc}

\usepackage{booktabs} % For formal tables
\usepackage{graphicx}
\usepackage{amsfonts}
\usepackage{xspace}
\usepackage{color,soul}
\usepackage[ruled, vlined, linesnumbered]{algorithm2e}

\usepackage{mdsymbol}
\usepackage{amsmath}
\usepackage{mathtools}
\usepackage{listings}

\usepackage{threeparttable}
\usepackage{tabularx}
\usepackage{pifont}
\usepackage{multirow}

\usepackage{enumitem}
\usepackage{cancel}
\usepackage[normalem]{ulem}
\usepackage{xcolor,colortbl}
\usepackage{arydshln}

\usepackage[normalem]{ulem}

\usepackage{hyperref,xcolor}% http://ctan.org/pkg/{hyperref,xcolor}
\definecolor{winered}{rgb}{0.7,0,0}

\definecolor{gray}{gray}{0.7}

\hypersetup
{
colorlinks=true,
linkcolor=winered,
urlcolor={winered},
filecolor={winered},
citecolor={winered},
allcolors={winered}
}

\newcommand{\gcell}{\cellcolor{gray!35}}

\usepackage{tikz}
\usepackage{comment}
\usepackage[framemethod=tikz]{mdframed}

\usepackage{flushend}

\definecolor{darkpastelgreen}{rgb}{0.01, 0.75, 0.24}
\definecolor{cadmiumgreen}{rgb}{0.0, 0.42, 0.24}
\definecolor{brickred}{rgb}{0.8, 0.25, 0.33}
\definecolor{cornellred}{rgb}{0.7, 0.11, 0.11}
\definecolor{burgundy}{rgb}{0.5, 0.0, 0.13}
\definecolor{frenchblue}{rgb}{0.0, 0.45, 0.73}
\definecolor{light-gray}{gray}{0.92}
\definecolor{lightlight-gray}{gray}{0.97}
\definecolor{codegray}{gray}{0.90}
\definecolor{inputgray}{gray}{0.90}

\global\mdfdefinestyle{rtboxstyle}{%
linecolor=black,%
leftmargin=0cm,rightmargin=0cm,linewidth=0.5pt,
roundcorner=3,
skipbelow=0pt,backgroundcolor=lightlight-gray
}

\usepackage[most]{tcolorbox}
\newcommand{\hlbox}[1]{
\begin{tcolorbox}[enhanced,
  breakable,
  colback=white,
  boxrule=0pt,
  frame hidden,
  top=0mm,left=1.5mm,right=0mm,bottom=0mm,arc=0mm,
  borderline west={0.75mm}{0mm}{gray}]
#1
\end{tcolorbox}
}

\newcommand{\code}[1]{\texttt{#1}}

\mdfdefinestyle{MyFrame}{%
     roundcorner=1pt,
     innertopmargin=6pt,
     innerbottommargin=6pt,
     innerrightmargin=6pt,
     innerleftmargin=6pt,
    }

\newcommand*\Let[2]{#1 $\gets$ #2}
\newcommand*\blkcc[1]{\tikz[baseline=(char.base)]{
            \node[shape=circle,fill,inner sep=1pt] (char) {\textcolor{white}{#1}};}}

\newcommand{\sysname}{{{\textsc{Spinner}}}\xspace}

\newcommand{\cmt}[1]{}

\newcommand{\homedir}{\raise.17ex\hbox{$\scriptstyle\sim$}}

\newcommand{\optmap}[2]{`\code{#1}' $\mapsto$ `\code{\color{red} #2}'}

\definecolor{darkgreen}{RGB}{40,125,40}

\newcommand{\updated}[2]{{{#2}}}
\newcommand{\revised}[1]{#1}
\newcommand{\avgtotal}[2]{#1 & (#2)}

\begin{document}
\fancyhead{}

%
% paper title
% can use linebreaks \\ within to get better formatting as desired
%\input{revision/revision.tex}
\setcounter{figure}{0}
\setcounter{table}{0}
\setcounter{section}{0}
\setcounter{page}{1}
\title{\sysname: Automated Dynamic Command Subsystem Perturbation}%\vspace{-2em}}

\author{Meng Wang, Chijung Jung, Ali Ahad, and Yonghwi Kwon}
\affiliation{%
	\institution{University of Virginia}
	\city{Charlottesville} 
	\state{Virginia}
	\postcode{22904}
	\country{USA}
}
\email{{mw6td, cj5kd, aa5rn, yongkwon}@virginia.edu}

\begin{abstract}
\noindent
Injection attacks have been a major threat to web applications. Despite the significant effort in \updated{fundamentally}{} thwarting injection attacks, protection against injection attacks remains challenging due to the sophisticated attacks that exploit the existing protection techniques' design and implementation flaws. 
In this paper, we develop \sysname, a system that provides general protection against \updated{command}{input} injection attacks, including OS/shell command, SQL, and XXE injection.
Instead of focusing on \updated{}{detecting} malicious inputs\updated{ and injected commands}{}, \sysname constantly randomizes underlying subsystems so that injected \updated{commands}{inputs (e.g., commands or SQL queries)} that are not properly randomized will not be executed, hence prevented. 
\updated{Further,}{} We revisit the design and implementation choices of previous randomization-based techniques and develop a more robust and practical \updated{randomization scheme to prevent}{protection against} various sophisticated \updated{command}{input} injection attacks.
\updated{}{To handle complex real-world applications,} we develop \updated{effective}{a bidirectional analysis that combines forward and backward} static analysis techniques \updated{that}{to} identify intended commands \updated{}{or SQL queries} \updated{in a target program}{} to ensure the \updated{intended commands are executed correctly}{correct execution of the randomized target program}. 
We implement \sysname for the shell command processor and two different database engines (MySQL and SQLite)\updated{}{ and in diverse programming languages including C/C++, PHP, JavaScript and Lua}. \updated{\sysname also supports diverse programming languages, including C/C++, PHP, JavaScript and Lua.}{}
Our evaluation results on \updated{32}{42} real-world applications including 27 vulnerable ones show that it effectively prevents a variety of \updated{command}{input} injection attacks with low \updated{}{runtime} overhead (\updated{3.74}{around 5}\%).

\end{abstract}

%\end{abstract}

%%%%%%%%%%%%%%
% TODO: replace this section with code generated by the tool at https://dl.acm.org/ccs.cfm

% I used this: https://dlb4.acm.org/ccs/ccs_flat.cfm#10002978


%\ccsdesc{Security and privacy~Use https://dl.acm.org/ccs.cfm to generate actual concepts section for your paper}
% -- end of section to replace with generated code

 % TODO: replace with your keywords
%%%%%%%%%%%%%%
%\pagenumbering{gobble} % To remove page numbers
% make the title area
\maketitle
\vspace{-1em}
\section{Introduction}
\label{sec:intro}
Injection attacks have been a long-standing security problem, listed as the first security risk in the OWASP Top 10 security risks~\cite{owasp_top_ten}. 
Among them, \updated{command}{input} injection \updated{}{(e.g., shell command/SQL injection)} is one of the most prevalent injection attacks. 
It happens when malicious inputs \updated{commands}{} (\updated{OS/}{}shell commands or SQL queries) are injected and executed on the victim system. 
Despite the effort in thwarting injection attacks~\cite{nguyen-tuong-sql,  Haldar_2005, chin_2009, csse, sqlcheck, sqlrand, halfond06fse, halfond_wasp, sekar_ndss, AMNESIA, CANDID, webssari, xie_aiken, pixy, wassermann_2007,wassermann_2008,minamide_2005, noxes, saner_2008}, \updated{command}{} injection vulnerabilities \updated{can be}{are} still \updated{exploited}{pervasive} in practice because, in part, the ever-evolving attacks \updated{are exploiting}{exploit} the limitations of the prevention measures.

\noindent
{\bf Existing Prevention Techniques.}
Input sanitization/validation is a recommended practice to prevent \updated{command}{input} injection attacks~\cite{Alkhalaf2014AutomaticDA, webssari, saner_2008}. However, implementing a sanitizer that can filter out all malicious inputs is extremely challenging due to the large and complex input space (e.g., grammars for OS/shell commands and SQL are expressive\updated{}{, allowing various inputs}). 
\revised{
\updated{}{Another straightforward approach is first identifying all allowed inputs on each call-site of APIs and only allowing them. However, this cannot prevent attacks that inject the allowed inputs twice. For instance, attackers can inject new ``\code{rm}'' commands to a vulnerable code snippet ``\code{system("rm logfile \$opt")}'' (Details can be found in Appendix~\ref{appendix:whitelist_approach}).}  
}
% and the complex input requirements and specifications (e.g., prohibited inputs in one context might be desirable in another context). 
%
%For example, filtering out special characters (e.g., `\code{'}' or `\code{;}') is a typical way of preventing shell command injection. However, in practice, completely restricting the use of special characters is difficult as some contents may require them (e.g., passwords are often encouraged to have special characters).
%Moreover, as input specifications (e.g., SQL grammars) are often evolving, introducing new keywords that can be abused in injection attacks, prevention techniques should be updated accordingly.
%
There are more advanced prevention techniques, such as those leveraging dynamic taint analysis~\cite{nguyen-tuong-sql,  Haldar_2005, chin_2009, csse, sqlcheck, halfond06fse}. However, they suffer from over/under tainting issues and runtime overhead.
Techniques that build models of benign commands/SQL queries to detect anomalies~\cite{sekar_ndss,  AMNESIA, CANDID} require accurate modeling of ever evolving attackers and target applications. %, which have been evolving over the years.
% that do not follow the pre-generated models/patterns of benign commands.

\noindent
{\bf Randomization-based Prevention.}
There are techniques~\cite{sqlrand,autorand} that randomize SQL keywords (in SQL engine and benign SQL queries) to prevent the execution of injected SQL queries that are not randomized. 
While the idea is effective, \updated{it has a critical limitation}{they have a critical limitation in their design choices}. 
To deploy the techniques, they rely on a proxy to translate a randomized query to a standard query using a \emph{parser}. %, instead of randomizing SQL engines directly.
\revised{
If the proxy's translator fails because of sophisticated SQL queries and grammar differences between SQLs (e.g., SQL dialects~\cite{sql_dialect} as discussed in Section~\ref{subsubsec:advanced_sql_injection}), malicious queries can be injected or benign queries may not be properly executed.
Diglossia~\cite{diglossia} is an injection attack prevention technique that proposes the dual-parsing approach. Unfortunately, it also relies on the accuracy of the parser used in the dual-parser (Details on how Diglossia will fail are presented in Section~\ref{subsec:comparison_existing}).
%
%{A straightforward approach to eliminate the proxy is to change the source code of SQL engines directly.}
%Unfortunately, we find that it is practically challenging because some SQL engines are closed source software (e.g., MSSQL), and \updated{the complexity of}{} parsers in SQL engines \updated{are difficult to handle}{contain various ad-hoc SQL query parsing code snippets, making it difficult to randomize them all}. 
%We have tried to randomize keywords in MySQL source code \updated{}{by replacing constant string keywords in the base parser created by Lex/Yacc}\updated{ and recompiled it. However,}{, but} \updated{the MySQL version}{it} did not function correctly.
%
\updated{Moreover,}{Other} randomization techniques~\cite{sqlrand,autorand} are susceptible to attacks that leak randomization key because their randomization scheme is not dynamically changing. Attackers can then prepare and inject a randomized command. 
}

%Moreover, we find that the limitation is caused by the design choice of using parsers and cannot be mitigated by engineering effort.
%avoiding detection. Worse, building a perfect parser is also an extremely challenging task, particularly with attackers exploiting bugs in parsers and bypassing detection techniques~\cite{xxx}.

\noindent
{\bf Our Approach.}
We propose a robust and practical randomization-based technique called \sysname to prevent \updated{command}{input} injection attacks.
The technique works by randomizing words in inputs (e.g., commands and SQL queries) and the subsystems  (e.g., shell process and SQL engine) that \updated{execute}{parse and run} the inputs.
%====
\updated{Once we randomize the command names and keywords in the subsystem}{The randomized subsystems does not allow} commands that are not properly randomized \updated{will not work}{to be executed}. For instance, if `\code{rm}' is randomized to `\code{\color{red}xc}' (\code{rm} $\mapsto$ \code{\color{red}xc}), the original command `\code{rm}' will result in an error (i.e., the command not found error) while `\code{\color{red}xc}' command will work as same as the original `\code{rm}.'
To ensure the \emph{intended benign commands} from applications work correctly with the randomization, we analyze target programs to identify and instrument the intended commands to be randomized. 
%Specifically, given a command found in a target program, if the definition of the command is originated from a trusted source (e.g., hardcoded commands or originated from trusted sources), we randomize it at runtime (Details in Section~\ref{subsec:instrumentation_comp}). 
To this end, legitimate commands are correctly randomized at runtime, while injected commands are not randomized and prevented from being executed.

\revised{
{\it 1) Revisiting Design Choices:}
To mitigate sophisticated attacks evading existing randomization based preventions~\cite{sqlrand, autorand}, we revisit the design choices made by existing techniques. 
%Specifically, we reduce the attack surface caused by the proxy and parser for randomization.
%
First, we eliminate the proxy and parser requirement for the shell process randomization by hooking APIs called before and after the shell process's original parser. 
%directly integrate the system into the shell process without using a proxy and parser for the shell process randomization.
%Specifically, we hook APIs that are called before and after the shell process's original parser. 
% of randomizing the command execution system from outside (e.g., via proxies),
%Second, for SQL engines that there are difficult to blend our technique in, we implement a proxy by hooking APIs that execute SQL queries (e.g., \code{mysql\_query()}), similar to the existing techniques~\cite{sqlrand-llvm,autorand}. 
%However, we overcome the critical weakness of the existing technique, leveraging parsers. Instead, we develop dual randomization scheme (Details in Section~\ref{subsubsec:database_random}) to prevent sophisticated attacks that exploit bugs or flaws of parsers to sneak malicious SQL queries. 
Second, for SQL engines that are difficult to blend our technique in, we develop a \emph{bidirectional randomization} scheme \updated{}{based on a scanner} (Details in Section~\ref{subsubsec:database_random}) to prevent sophisticated attacks (e.g., those exploiting bugs/flaws of parsers). % to sneak malicious SQL queries. 
Third, \updated{our technique}{\sysname} changes the randomization scheme at runtime so that even if an attacker learns a previously used randomization key and \updated{uses it to inject}{injects} a randomized command, the attack will fail.
}

%execute commands (e.g., \code{system()} and \code{mysql\_query()}) to randomize command names and SQL keywords without relying on parsers. 
%Specifically, 
%For applications (i.e., database engines) that we have to randomize from the outside, we leverage a scanner-based dual randomization technique, instead of using parser-based techniques that are vulnerable (Details in Section~\ref{subsubsec:database_random}). 
%Our technique is grammar and keyword agnostic. %preventing sophisticated attacks that existing techniques fail to handle.

%\sysname automatically randomizes the specification of commands by instrumenting the intended commands to be randomized in a target program. %, ensuring the correct execution of the intended commands. 
{\it 2) \updated{Practical and Generic}{} Program Analysis Approaches for \updated{Command}{Input} Randomization:}
We propose practical program analysis techniques that can effectively analyze large and complex programs for \updated{command}{input} randomization. 
In particular, we show that our approach, bidirectional data flow analysis (Section~\ref{subsubsec:composition}), is scalable to real-world applications including WordPress~\cite{wordpress}.
\updated{Our algorithm is also generic. We implement \sysname in multiple languages (including C/C++, PHP, JavaScript, and Lua).}{}
\updated{Our experiments show that \sysname effectively prevents all the attacks we tested with low overhead, 3.74\% for 32 real-world programs.}{}
%\vspace{-1em}
Our contributions are summarized as follows:

\vspace{-0.5em}
\begin{itemize}[leftmargin=*]
    \item We propose an approach that can prevent various types of \updated{command}{input} injection attacks by randomizing \updated{how applications access the command targets (e.g., via command names, binary file paths, and SQL statements)}{the subsystems that run or process the inputs}\updated{ (e.g., shell commands or SQL queries)}{}.
    
    \item We design and develop \updated{a}{an effective} static data flow analysis technique \updated{}{called bidirectional analysis} that can identify intended commands in complex real-world applications. 
    \updated{We combine forward and backward data flow analyses to identify command compositions and prevent encoded commands from being used in non-command execution APIs.}{} 
    
    \item  We implement a prototype of \sysname in diverse programming languages, including C/C++, PHP, JavaScript and Lua. % to prevent three types of \updated{command}{input} injection attacks. 
    
    \item Our evaluation results show that it prevents 27 \updated{command}{input} injection attacks with low overhead (\updated{3.64}{$\approx$5}\%).
    
    \item  We release our implementation and data-sets publicly~\cite{csr-tool}.
    
\end{itemize}

\section{Definitions and Backgrounds}
\label{sec:background}

\begin{figure*}[htp]
    \centering
    \includegraphics[width=1.0\textwidth]{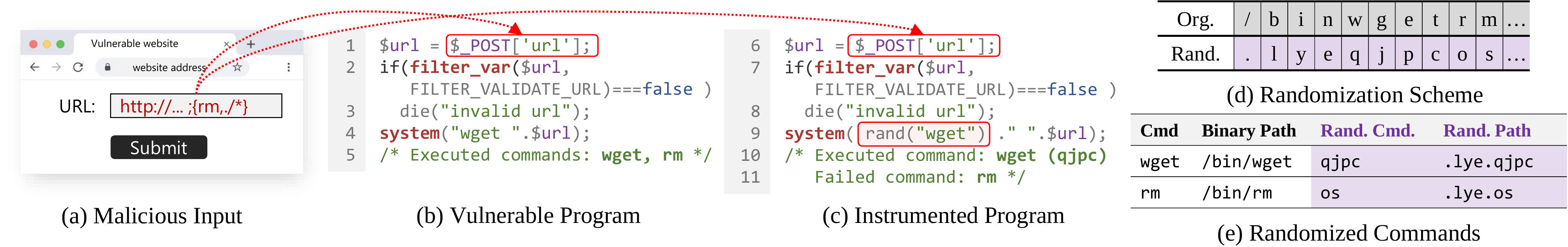}
    \vspace{-2em}
     \caption{Example of \sysname Preventing a Command Injection Attack}
     \vspace{-1em}
     \label{fig:moti_example}
\end{figure*}

%\autoref{fig:softwarestack} shows a typical workflow of an application that receives inputs and executes commands (e.g., shell commands or SQL queries). 
%Specifically, a program (\blkcc{2}) receives inputs from untrusted sources (\blkcc{1}) such as a remote network.

\noindent
{\bf Scope of \updated{Command}{Inputs for Randomization}.}
We consider three types of \updated{commands}{inputs to randomize}: OS/shell commands, SQL queries, and XML queries.
This is because they are commonly exploited in web server applications that \sysname aims to protect, according to the OWASP Top 10 document~\cite{owasp_top_ten}. % (i.e., OS/shell commands and SQL queries are mentioned in A1. XXE attacks is mentioned in A4).
Those inputs are used by a program to leverage external programs' functionalities. For example, a program can compress files by executing a shell command that executes `\code{gzip}'. 
SQL queries are \updated{commands to use the}{for} SQL engines to store and retrieve values to/from the database. An XML query is an interface for interacting with XML entities (e.g., reading and writing values in the entities).

\revised{
\noindent
\textbf{Choice of Term `Command.'}
\sysname focuses on preventing three different input injection attacks: shell injection, SQL injection, and XXE injection attacks.
In this paper, we use the term \emph{command} to include the three input types to facilitate the discussion. We consider SQL queries and XXE entities \emph{commands} as they eventually make the subsystem run or execute particular code.

\noindent
{\bf Command Execution APIs.}
We define a term \emph{Command Execution API} to describe APIs that execute a command or a query.
%Typically, it takes a string argument that holds a command \updated{}{or SQL query} that is composed by concatenating \updated{command names and options}{strings originated from various sources including remote users, constants, and configuration files}\updated{ in various ways}{}.
%It then calls APIs that execute commands, which we call \emph{Command Execution APIs}, with the composed commands as arguments.
%\code{system()} and \code{mysql\_query()} are examples and % for shell commands and SQL queries respectively. 
A list of command execution APIs is shown in Table~\ref{table:sinkfunctions}.

\noindent
{\bf Command Specification.}
A command passed to a command execution API should follow a certain specification. Specifically, shell commands should use correct command names or external executable binary file names. SQL queries should follow the predefined SQL keywords and grammar. 
If a command does not follow the specification (e.g., a wrong file name), its execution will fail.

\noindent
{\bf \updated{Command}{Input} Injection Vulnerability.}
\updated{Command}{Input} injection happens when an attacker injects malicious \updated{commands}{inputs} to the composed command string \updated{}{or SQL query string} passed to a command execution API as an argument.
In practice, programs may try to validate and sanitize suspicious inputs that might contain malicious \updated{commands}{inputs}.  % find some command injection attacks that bypasses.
% https://support.portswigger.net/customer/portal/articles/2590739-sql-injection-bypassing-common-filters-
%https://medium.com/@hninja049/command-injection-bypass-cheatsheet-4414e1c22c99
Typically, when a program composes a command, the command name (e.g., `\code{gzip}') is defined as a constant string or loaded from configuration files that are not accessible to attackers (hence can be trusted).
%and arguments are defined as constant or loaded from trusted files (e.g., configuration files (\blkcc{3})).
However, some programs allow users to define arguments of the command. % (e.g., file or folder names of the \code{gzip} command). 
As a result, attackers aim to inject malicious commands through the arguments.
After a command is composed, the program calls command execution APIs  (e.g., {\tt exec()}, {\tt system()}, or {\tt mysql\_query()}) to fulfill the command execution.
}

\noindent
{\bf Limitations of Existing Randomization Techniques.}
%The idea of randomizing input specification to prevent injection attacks is not new. 
There are existing techniques~\cite{sqlrand, autorand, isr2} that randomize the keywords and grammars.
% using \textit{parsers}. 
%
% this is wrong suggestion. Not 'ly': previous\hl{ly}
While we share the similar idea to them, our work differs from them as we aim to solve the following three limitations.

First, existing techniques leverage parsers to randomize/derandomize commands. Unfortunately, attackers often exploit bugs or design flaws in parsers to evade the prevention techniques that rely on them.
%Without a perfect parser, the existing techniques may fail to prevent advanced command injection attacks.
In particular, the parsers may not handle complicated benign inputs (e.g., because of the use of SQL dialects~\cite{sql_dialect}), breaking benign functionalities or allowing injection attacks. We elaborate details of such weaknesses of existing techniques in Section~\ref{subsubsec:advanced_sql_injection}. 
\vspace{-0.3em}
\hlbox{\sysname handles this by integrating our randomization scheme to the internal parser in the shell process and leveraging our bidirectional randomization scheme for SQL engines. The bidirectional randomization is grammar and keywords agnostic, meaning that attacks exploiting flaws of parsers will be prevented.
}%\sysname handles this limitation by cooperating existing parsers in OS/shell process

Second, existing automated approaches~\cite{autorand,sqlrand-llvm} leverage static analysis techniques to identify intended (i.e., benign) commands in the source code. We find that their static analysis techniques are not scalable to complex real-world applications. %The existing techniques may break the benign functionalities as they were not able to identify and randomize benign commands.
\vspace{-0.3em}
\hlbox{We propose a practical and scalable bi-directional data flow analysis (Details in Section~\ref{subsubsec:composition}) that can effectively identify benign commands in complex real-world applications.
}
Third, existing techniques randomize the command specification statically, meaning that the commands are randomized only once. If an attacker learns the randomization scheme (e.g., via information leak vulnerabilities), the attacker can inject randomized commands which will not be prevented.
%Attackers may leverage information leak vulnerabilities to obtain the randomized commands or observe output (i.e., error messages) that contains parts of randomized commands to infer the randomization key.
\vspace{-0.3em}
\hlbox{\sysname dynamically randomizes the command specification whenever a command execution API is called. To this end, even if an attacker learns a previously used randomization key, it will not help subsequent attacks.
}
% may learn the randomization scheme by sending  
%with a separate proxy for randomization/derandomization can be evaded by attacks that inject malicious commands 
%\sysname uses a SQL grammar agnostic scanner that is less fragile than a parser. %Moreover, \sysname's SQL randomization does not require to know a list of SQL keywords.

%\sysname prevents sophisticated command injection attacks (in Appendix~\ref{appendix:sqlrand-llvm}) by (1) directly hooking and randomizing the shell process and (2) leveraging the SQL agnostic dual randomization scheme (Section~\ref{subsubsec:database_random}). 
%As a result, \sysname is more robust in preventing sophisticated attacks.

%To this end, existing techniques implement proxies to implement the randomization scheme. More importantly, they leverage \emph{parsers} to pick keywords to randomize and derandomize. 
%Unfortunately, we find that those parsers are the weakest link that can be exploited by attackers. 
%Specifically, if an attacker provides a malicious input that cannot be properly handled by proxies (and their parsers), the attack can be successful. 

%\vspace{-0.5em}
\noindent
{\bf Threat Model.}
\sysname aims to prevent remote \updated{command}{input} injection attacks (including SQL/XXE injections) on server-side applications. 
We expect server-admins and web-developers as typical users of \sysname.
Client-side attacks such as XSS (Cross Site Scripting) and XSRF (Cross Site Request Forgery) are out of the scope.
We assume the subject program and inputs from trusted sources (defined by the user) are benign, but inputs from untrusted sources can include malicious commands. Typical trusted sources are local configuration files. 
%\sysname requires a target program that has intended commands originated from trusted sources. If a program allows an untrusted input to be executed directly, applying \sysname will simply prevent the functionality completely because there are no commands that can be trusted. For example, \sysname cannot be applied to a webshell (e.g., \code{system(\$\_GET[...])}) as it will prevent the \code{system()} completely.
We trust local software stacks, including OS kernel, applications, and libraries. If they are compromised, attackers can disable \sysname.
%
%This paper focuses on preventing remote command injection attacks. 
\sysname does not focus on preventing attacks that compromise non-command parts such as arguments of commands (e.g., a directory traversal attack).
\sysname does not aim to prevent \updated{}{binary} code injection\updated{attacks }{} (e.g., shellcode injection). \updated{There are existing randomization based techniques preventing code injection attacks~\cite{conf_ndss17_a2c,isr} and \sysname is specialized in preventing command injection attacks}{}

%\vspace{-1em}
\section{Motivating Example}
\label{sec:motivation}

\autoref{fig:moti_example} shows an example of a \updated{}{shell} command injection attack to a vulnerable server-side program written in PHP. 
%The attack begins from the website that sends a \code{POST} request to the server-side program. %
%
From a website shown in \autoref{fig:moti_example}-(a), an attacker sends a malicious input (with a malicious command) through the textbox on the webpage. The server-side program, shown in \autoref{fig:moti_example}-(b), is vulnerable because it directly passes the input to {\tt system()}, executing the injected command (line 4). It tries to sanitize inputs via {\tt filter\_var()} at line 2 (commonly recommended~\cite{php_sanitize1, php_sanitize2, php_sanitize3}), but it fails. % to filter the malicious input. %: \redcode{http://...;\{rm,./*\}}. % successfully makes its way to the exploitation. %, showing that input sanitizations in practice often fail to work properly.

\noindent
{\bf Command Specification Randomization.} 
%We randomize the command targets (e.g., OS shell in this example) to prevent injected commands that are not aware of our randomization from being executed. 
In this example, \sysname randomizes commands in the shell process. % to prevent an injected OS (or shell) command from being executed.
%\autoref{fig:moti_example}-(d) and (e) show an example of how \sysname works. 
%\noindent
%{\it -- Shell Command Types:}
There are two types of shell commands~\cite{linux_commands_external_internal}: (1) internal commands that are implemented inside of OS/shells such as `\code{cd}' and (2) external commands that are implemented by separate binaries such as `\code{grep}'. 
% linux_commands_external_internal: https://linuxconfig.org/internal-vs-external-linux-shell-commands

For internal commands, we randomize the command names by hooking and overriding APIs in the shell process (Details in Section~\ref{subsubsec:shellcommand_rand}).
For external commands, since the shell process will look up a binary file for the external command to execute (i.e., check whether a binary for the command exists), we randomize the binary file names and paths (e.g., `\code{rm}' $\mapsto$ `\code{\color{red}os}' as shown in \autoref{fig:moti_example}-(e)) in the file I/O APIs. 
%For both types, we randomize the name of commands (e.g., name of internal commands and binary file paths of the external commands) in the OS/shell subsystem.
This will prevent injected malicious (\updated{}{and} {\it not randomized}) commands from being executed. 
%
%Details of how we mitigate the concerns of sophisticated attackers injecting randomized commands are discussed in Section~\ref{xxx}.
%To execute an external command (i.e., the second type), a program will look up a binary file for the external command first (i.e., check whether a binary for the command exists) and then execute the program. 
%\sysname also randomizes the file path of the supported commands, as shown in the fourth column of \autoref{fig:moti_example}-(e).
For the randomization, we use a one-time substitution cipher. % (i.e., 1:1 mapping between the characters that can be used to specify commands). 
%Specifically, 
%We conservatively assume that any printable ASCII code can be used. 
%Hence, 
Specifically, as shown in \autoref{fig:moti_example}-(d), we create a mapping between the original input and its randomized character. %: \optmap{/}{.}, \optmap{b}{l}, \optmap{i}{y}, \optmap{n}{e}, and so on.
%
%Specifically, characters `{\tt a}', `{\tt b}', and `{\tt c}' will be mapped to  `{\tt G}', `{\tt s}', and `{\tt ;}' respectively.  
To execute a command ``\code{wget}'' under this randomization scheme, one should execute ``\code{qjpc}'' as shown in  \autoref{fig:moti_example}-(e). 
To prevent brute-force attacks against the randomized commands, \sysname provides two mitigations. 
First, \sysname creates a new randomization scheme on \emph{every new command} to mitigate attacks leveraging previously used randomized commands.
Second, to further make the brute-force attacks difficult, \sysname supports one to multiple bytes translation, enlarging the searching space.
Details can be found in Appendix~\ref{appendix:bruteforce_attack}.
% shows how commands and binary paths are randomized according to \autoref{fig:moti_example}-(d).

\noindent
{\bf Instrumentation by \sysname.}
Once the commands are randomized in the shell process, the system cannot understand commands that are not randomized. In other words, it affects every command in the program including intended and benign commands, breaking benign functionalities.
To ensure the correct execution of intended commands, we statically analyze the program to identify intended (hence benign and trusted) commands that are originated from trusted sources (e.g., defined as a constant string or loaded from a trusted configuration file). We describe our bidirectional command composition analysis for identifying intended commands in Section~\ref{subsubsec:composition}. 
Then, we instrument the target program to randomize intended commands.
%MA: are you really instrumenting the rand() function? Don't you instrument the binary to detect "wget" as a benign function to randomize it later?
\autoref{fig:moti_example}-(c) shows the instrumented program. At line 9, as ``\code{wget}'' is the intended command (because it is a constant string), it is  instrumented with ``\code{rand()}''. Note that \code{\$url} that includes an injected command ``\code{\{rm,./*\}}'' is not instrumented because it is originated from an untrusted source (\code{\$\_POST[`url']}). %, hence it will not be executed.

\begin{figure}[h]
    \centering
    \vspace{1em}
    \includegraphics[width=0.85\columnwidth]{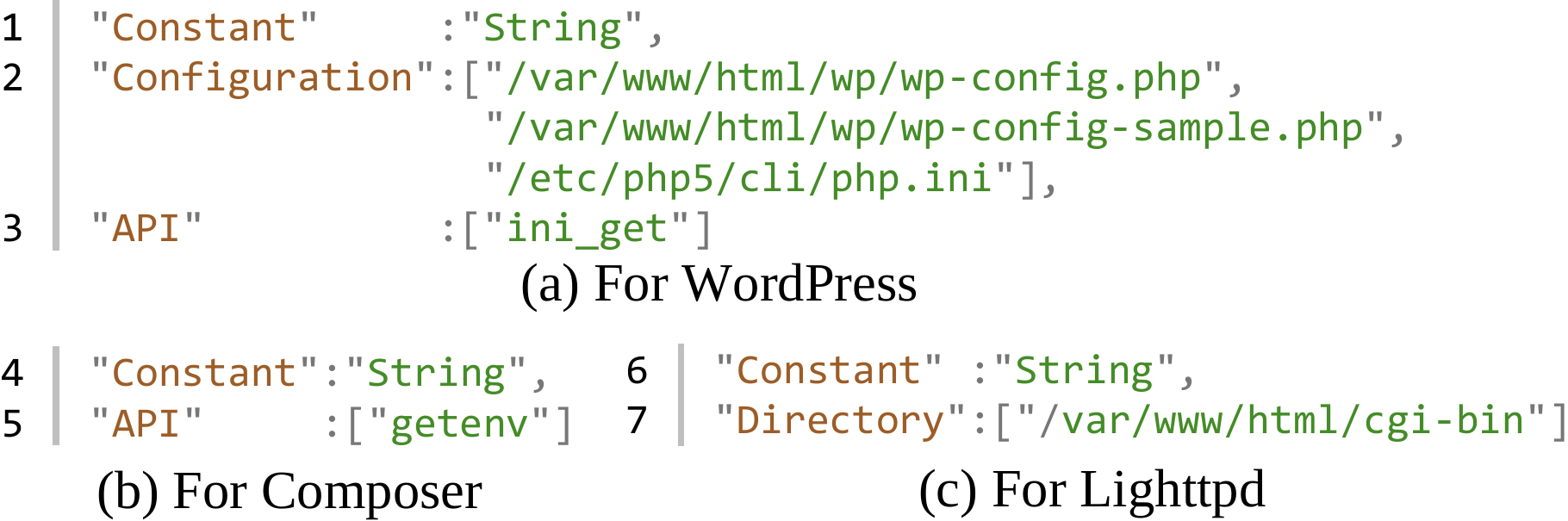}
    \vspace{-1em}
    \caption{\revised{Trusted Command Specification Examples}}
    \vspace{-1em}
    \label{fig:spec_example}
\end{figure}

\begin{figure*}[ht]
    \centering
    \includegraphics[width=0.9\textwidth]{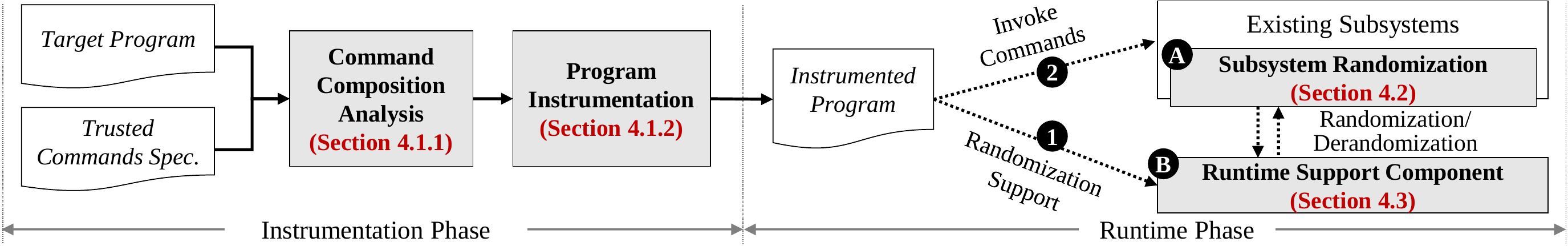}
    \vspace{-0.5em}
     \caption{Overview and Workflow of \sysname (Design details are presented in the annotated sections)}
     \vspace{-1em}
     \label{fig:overview}
\end{figure*}

%\vspace{1em}
\section{Design}
\label{sec:design}
\autoref{fig:overview} shows the workflow of \sysname with two phases. %: Instrumentation (Section~\ref{subsec:instrumentation_phase}) and Runtime (Section~\ref{subsec:runtime_phase}).
%There are two phases: Instrumentation and Runtime phases. 

%\vspace{-1em}
\subsection{Instrumentation Phase}
\label{subsec:instrumentation_phase}
\sysname takes a target program to protect and specification of trusted commands as input. It \updated{then}{} analyzes the target program to identify intended commands to instrument randomization primitives. % to \updated{the identified commands}{them}. %(e.g., shell/SQL commands that can be trusted and should be executed according to the provided specification). 
%Details of our static analysis based instrumentation are described in Section~\ref{subsubsec:composition}. 
%The outcome of this phase is a program with all trusted (or intended) commands instrumented.

%
%To this end, the runtime support returns the correctly randomized command to the instrumented program, ensuring the trusted command's execution. Untrusted commands are not instrumented and will not be randomized.
%, preventing injected commands from being executed.
%
%We use different methods to implement a runtime randomization module for different command interfaces. 
%Specifically, for the OS/shell command interface, we leverage the library hooking technique~\cite{ld_preload_github} to implement OS/shell command randomization. For database engines, we modify source code of a SQL engine, specifically SQLite~\cite{sqlite}, to create a shim database front-end 
%to randomize keywords used in SQL statements (e.g., {\tt select} and {\tt update}).

% \subsubsection{Instrumentation Component}
% \label{subsec:instrumentation_comp}
% We analyze a target program to identify variables used to compose commands. We check their origins and determine whether they can be trusted (Section~\ref{subsubsec:composition}), then we instrument the trusted variables (Section~\ref{subsubsec:instrumenation}). 

\noindent
{\bf Target Program.}
\sysname analyzes and instruments target programs' source code. Hence, it requires the target program's source code. 
Note that we do not require source code of the subsystems (e.g., shell process and database engines). 

\revised{
\noindent
{\bf Trusted Command Specification.}
Another input that \sysname takes is the trusted command specification which is a list of trusted source \emph{definitions} as shown in \autoref{fig:spec_example}. 
We provide a semi-automated tool to derive the trusted command specification as well (Details can be found in Appendix~\ref{appendix:tcstool}).
There are four types of trusted source definitions: (1) a constant string containing known command names such as hard-coded command (Lines 1, 4, and 6), (2) a path of a configuration file that contains definitions of trusted commands (Line 2), (3) a path of a folder where all the files in that folder are trusted (Line 7), and (4) APIs that read and return values from trusted sources (Lines 3 and 5).

The first type represents a command in a constant string. We consider a hard-coded command is \emph{an intended command by the developer}. 
%Note that constant strings that do not contain command names do not belong to this type. We obtain the known command names using a list of known internal Linux commands~\cite{manpages_ubuntu_com} and enumerating executable binaries on the target system (e.g., executables in `\code{/usr/bin/}'). Since available commands are not frequently updated, obtaining the list is mostly a one-time effort. 
The second type is to handle a command defined in the program's configuration file. 
For example, a PHP interpreter can execute other applications (e.g., \code{sendmail} for \code{mail()}) which is defined in \code{php.ini}. We include the \code{php.ini} file in our analysis as shown in \autoref{fig:spec_example} at line 2.
%a PHP program using a PHP file path defined in its configuration file. 
%We consider the PHP file path as trusted as we trust the content in the configuration file.
The third type is a folder. It is to define all files under the folder to be trusted. For instance, a web-server may trust all the CGI (Common Gateway Interface) programs found at the time of offline analysis with a configuration shown in \autoref{fig:spec_example} at line 7.
The fourth type describes APIs that read trusted sources. For instance, \code{getenv()} returns values of local environment variables. If a user assumes that the local environment variables cannot be modified at runtime, it can add the API to the trusted sources.

For most applications, specifying the first type (i.e., hard-coded commands) as a trusted source is sufficient.
For some applications, configuration files may define trusted commands. In such a case, the command specification should include the trusted configuration files' file paths so that, at runtime, commands originated from the configuration file are trusted.
\updated{}{Note that \sysname checks whether a trusted source (e.g., configuration file) can be modified by remote users. If there is any data flow between untrusted sources and trusted sources, \sysname notifies the users to redefine the trusted command specification. 
Similar to the values from configuration files, values from databases are trusted, only if there is no data-flow from untrusted sources to the database.
We define \emph{untrusted sources} as any sources \emph{controlled by remote users (i.e., potential attackers)}.
Further, to prevent modifications of trusted sources, we hook APIs that can change the trusted sources (e.g., \code{setenv()}) to make them read-only and detect attempts to modify during our evaluation. %We did not observe any read attempts to the trusted sources.
}
%\YK{@cj, please check where are SUNDR and additional discussion mentioned in meng's comment.}
%
%For the third type, if a directory is specified to be trusted, we enumerate all the files in the folder and specify them to be trusted.
%The third type is mostly used for XXE injection prevention, where inclusion of an untrusted file needs to be prevented.
%
%\updated{}{In Appendix~\ref{appendix:tcs_examples}, we provide examples of the trusted command specification.}
%%%%
%\updated{}{\mw{Note that  there might be data flow from the remote user to the configuration file so there is the risk that the configuration file can be corrupted. We will warn the user when we detect such flow and do not take the configuration file as a trusted source in this case. To prevent the trusted sources outside the applications from being corrupted by other programs or remote users, we hook the \code{setenv()} function to deny operations on the environment variables recognized as trusted sources in the static analysis. We also use a secure file system like SUNDR~\cite{li2004secure} which can detect unauthorized attempts to change files to store our configuration files. In Appendix~\ref{xxx} we discuss the trusted sources of the selected applications.}%}
}

\subsubsection{Command Composition Analysis}
\label{subsubsec:composition}
%In this subsection, we describe how \sysname analyzes a target program to identify intended commands passing to command execution APIs. 
Analyzing data-flows from the trusted command definitions (i.e., sources) to command execution APIs (i.e., sinks) is a challenging task, particularly for complex real-world applications.
A naive approach that uses forward analysis (e.g., taint analysis) from trusted sources to sinks often leads to the over-approximation (i.e., over-tainting),  resulting in instrumenting variables that are not relevant to the commands (i.e., false positives) hence breaking benign functionalities. % If we instrument them, it will break the target program's benign functionalities.
On the other hand, backward data-flow analysis from the sinks (i.e., tracing back the origins of variables from arguments of command execution APIs) to trusted sources is also difficult due to the complicated data dependencies.
After analyzing challenging cases from both analyses, we realize that combining two analyses can significantly reduce their limitations (i.e., over and under-approximations).

%we observe that complex data dependencies are mostly observed on the variables that do not hold commands, hence does not need to be instrumented. 
%We use backward slicing and data-flow analysis to identify program statements that are relevant to composing commands.  

%Note that this can be automatically obtained with little manual effort. 
%For most cases, we can specify that all shell commands originating from constants can be trusted. 
%Note that \sysname needs a list of commands that can be obtained by statically analyzing the shell process's source code~\cite{bashsource} (for internal commands) and scanning files in the folders containing executable binaries (for external commands), e.g., `/usr/bin'.
%For programs using databases, \sysname does not require any known SQL keywords. Typically, configuring to trust all constant terms (i.e., tokens) is sufficient.
%For XML files, the specification is automatically derived by analyzing the target program to identify trusted XML files. 

%to identify how a program composes commands and where to instrument the program to ensure their intended execution. Specifically, we use backward slicing and data-flow analysis to identify program statements that are relevant to composing commands. 

\begin{figure}[!tp]
    \centering
    \vspace{0.5em}
    \includegraphics[width=0.95\columnwidth]{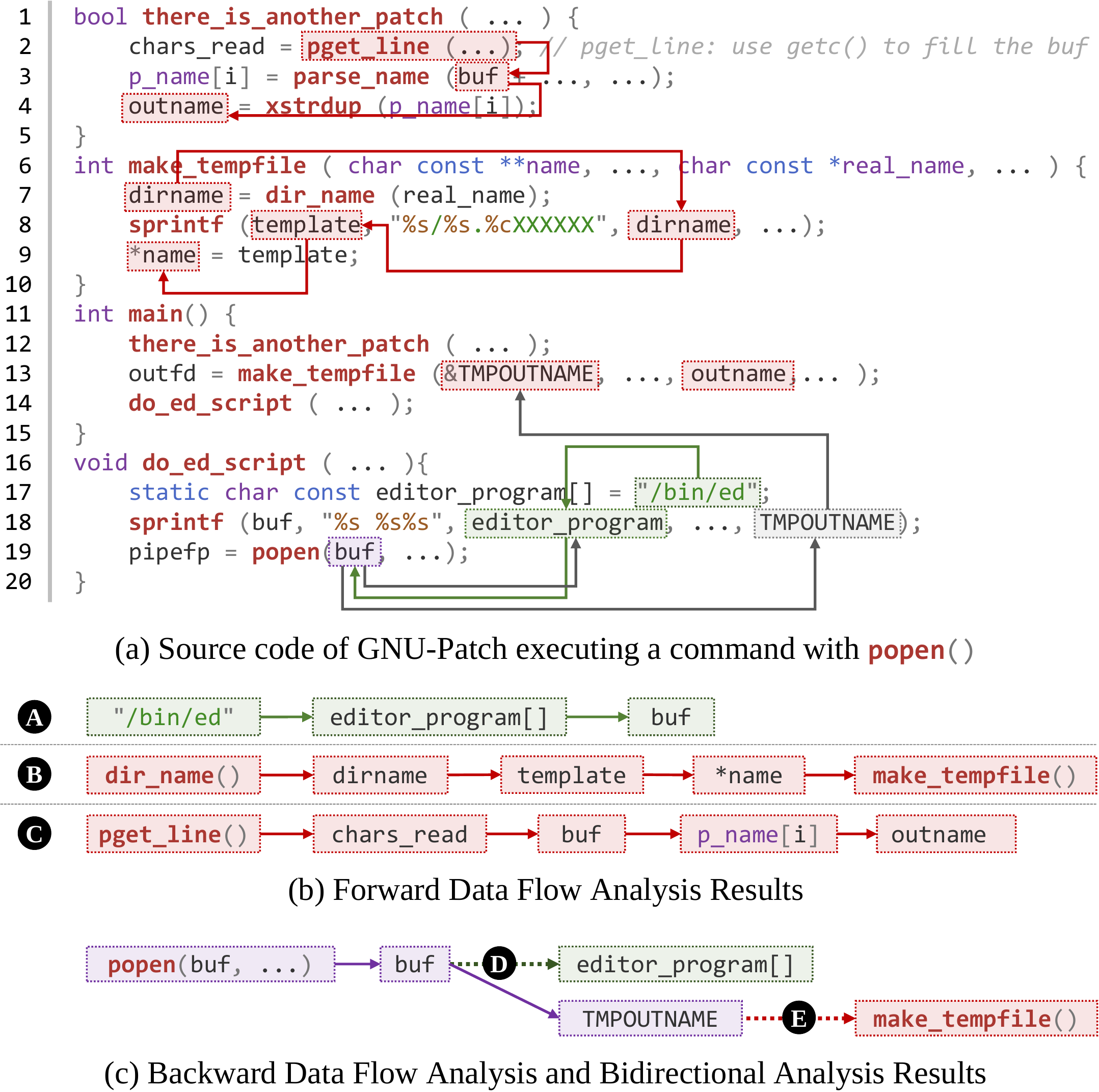}
    \vspace{-1em}
    \caption{Bidrectional Analysis on GNU-Patch~\cite{Patch}}
    \vspace{-1.5em}
    \label{fig:Code_Patch}
\end{figure}

\noindent
{\bf Bidirectional Command Composition Analysis.}
We propose a bidirectional analysis, which is the key enabling technique that makes \sysname effective in analyzing complex data-flow in real-world applications.
Specifically, we conduct forward data flow analyses (1) from trusted sources to identify variables holding trusted commands and \updated{also}{} (2) from untrusted sources to identify variables that are not relevant to commands. \sysname \emph{automatically} derives the definitions of untrusted sources: (a) return values of APIs that are not included in the trusted command specifications (e.g., \updated{return value of}{}\code{gets()}) and (b) constant strings that do not contain command names. 
From the forward data flow analysis, we obtain two sets of variables: a set of trusted variables and another set of untrusted variables.

\updated{With the results}{Next}, we conduct a backward data flow analysis from the arguments passed to command execution APIs (e.g., \code{system()}). While we analyze the program to trace back the origin of the arguments, if we encounter a node \updated{that is}{}originated from a variable in the trusted variable set, we conclude the argument is an intended command hence instrumented. If it meets a node \updated{that is}{}originated from a variable from the other set, untrusted variable set, we stop the backward analysis and conclude that the argument is not relevant to the command. % hence no instrumentation is needed.

%Our analysis quickly concludes without tracing back arguments that are not relevant to commands. 
%In the next few paragraphs, we will demonstrate the analysis with an example.

\noindent
\textbf{Running Example.}
%We use a real-world program GNU-Patch~\cite{Patch}'s code snippet to show how the command composition analysis works.
\autoref{fig:Code_Patch}-(a) shows a real-world program GNU-Patch~\cite{Patch}'s code snippet consisting of 4 functions. 
At line 19, it calls \code{popen()} which executes a shell command composed at line 18. The \code{buf} variable contains the composed command from \code{editor\_program} and \code{TMPOUTNAME} via \code{sprintf()}. 
First, \code{editor\_} \code{program} is defined as a constant string containing a known binary program path in line 17, meaning that it is an intended command and hence instrumented. 
Second, \code{TMPOUTNAME} is defined through multiple functions. It is used as an argument of \code{make\_tempfile()} function (line 13). Inside the function, it is defined by \code{sprintf()} where its value is originated from the \code{dir\_name()} function. 
As described, there are multiple functions involved to define the value \code{TMPOUTNAME}, making it difficult to trace back the origin. 

{\it 1) Forward Analysis:}
\autoref{fig:Code_Patch}-(b) shows the results of our forward data flow analysis from the trusted/untrusted sources. \blkcc{A} shows the data flow graph from the trusted source. It shows the \code{/bin/ed} command is propagated to \code{buf}.
\blkcc{B} and \blkcc{C} show two graphs from untrusted sources. Specifically, \blkcc{B} shows that the return value of \code{dir\_name()} (line 7) is propagated to \code{*name} (line 9), affecting the first argument of the \code{make\_tempfile()}. As a result, the graph include the \code{make\_tempfile()} as a node, meaning that the function's return values are untrusted and not intended commands.
\blkcc{C} also shows that the values from untrusted source \code{pget\_line()} (reading inputs from the standard input) are propagated to \code{outname}.

{\it 2) Backward Analysis:}
\autoref{fig:Code_Patch}-(c) shows our backward analysis from the sink function: \code{popen()}. \autoref{table:sinkfunctions} shows sink functions (i.e., Command Execution APIs) for each command subsystem.
From the argument of \code{popen()}, \code{buf}, we analyze how the argument is composed. 
First, it identifies \code{editor\_program[]} is concatenated via \code{sprintf()} at line 18. 
Second, it finds out that \code{TMPOUTNAME} is a part of the command and it is defined by \code{make\_tempfile()}, which can be found in the forward data flow analysis result \blkcc{B}.

{\it 3) Connecting Forward and Backward Analysis Results:}
The bidirectional analysis merges results from forward and backward analysis together as shown in \blkcc{D} and \blkcc{E}. %It essentially connect the backward analysis results from sink functions to the forward analysis results from the trusted/untrusted sources. 
Note that our backward analysis will terminate when it reaches any nodes in the forward data flow analysis results. 
This effectively reduces the complexity of the data flow analysis. 
Typically, the forward analysis is mostly localized and the backward analysis quickly reaches nodes in the forward analysis results.
Note that \sysname conducts inter-procedural analysis if function arguments (e.g., \code{name} at line 9) or global variables (e.g., \code{TMPOUTNAME} at line 18) are affected.

\begin{table}[!h]
	\centering
	\caption{Command Execution APIs (Sink Functions).}
	\label{table:sinkfunctions}
	\vspace{-1em}
	
\resizebox{1\columnwidth}{!}{%	

\begin{tabular}{l l l}
	\toprule
	{\bf Type} & 
	{\bf Function} &
	{\bf Lang.}
	\\ 
	\midrule
	
	\multirow{6}{*}{Shell} &
	{\tt exec()$^1$}, {\tt system()}, {\tt popen()} &
	C/C++ \\ \cmidrule{2-3}

 	& {\tt passthru()}, {\tt system()}, {\tt popen()}, {\tt shell\_exec()}, {\tt exec()}, {\tt proc\_open()} &
	 {PHP} \\ \cmidrule{2-3}
	 
	& {\tt os.execute()}, {\tt io.popen()} &
	 Lua \\ \cmidrule{2-3}
	 
	& {\tt spawn()$^2$}, {\tt exec()$^2$}, {\tt execFile()$^2$} &
	 \multirow{1}{*}{JS} \\  \midrule

	\multirow{3}{*}{XML} & {\tt xmlParseFile()}, {\tt xmlParseChunk()}&
	 C \\ \cmidrule{2-3}

	 & {\tt simplexml\_load\_file()}, {\tt simplexml\_load\_string()}, {\tt xpath()}, &
	 \multirow{2}{*}{PHP} \\ 
	 
	 &  {\tt xml\_parse()}&
	  \\ \midrule 
	    
	Database & {\tt mysqli::multi\_query()}, {\tt mysqli::prepare()} &
	 PHP \\ \cmidrule{2-3}
	 
	(MySQL) & {\tt mysql\_query()}, {\tt mysql\_real\_query()} &
	 C/C++ \\\midrule
	 
	Database & {\tt sqlite\_query()}, {\tt sqlite\_exec()} &
	 PHP \\ \cmidrule{2-3}
	 
	(SQLite) & {\tt sqlite3\_prepare()}, {\tt sqlite3\_exec()} &
	 C/C++ \\
	\bottomrule 
	\multicolumn{3}{l}{$^1$Including \code{exec()}, \code{execvpe()}, \code{execvp()}, \code{execv()}, \code{execlp()}, \code{execle()}, \code{execl()}, \code{execve()}.} \\
	\multicolumn{3}{l}{$^2$Including {\tt spawnSync()},{\tt execSync()}, {\tt execFileSync()}. \vspace{-2em}}
\end{tabular}
}
\end{table}

\begin{algorithm}[tb]
\caption{Bidirectional Analysis for Instrumentation}\label{alg:composition}
\footnotesize
\SetInd{0.15em}{0.85em}
\DontPrintSemicolon
\KwIn{$F_\textit{set}$: a set of functions in a target program.}
\KwOut{$\textit{Ins}_\textit{out}$: a set of variables to instrument.}
\SetKwBlock{Begin}{function}{end function}
\Begin($\textsc{BidirectionalAnalysis} {(}F_\mathit{set}{)}$)
{
  \Let{$\textit{Ins}_\textit{out}$}{ $\{\}$ }

  \For{ $\forall F_i \in F_\textit{set}$ }
  {
    \For{ $\forall v_i \in F_i$ }
    {
      \uIf{ $v_i$ \textup{is from a trusted source} }
      {
        {\textsc{ForwardAnalysis}(\textsc{CreateDepTree}($v_i$, $F_i$, \textsc{trusted}), $v_i$)}

      }
      \uElseIf{ $v_i$ \textup{is from an untrusted source} }
      {
        {\textsc{ForwardAnalysis}(\textsc{CreateDepTree}($v_i$, $F_i$, \textsc{untrusted}), $v_i$)}

      }
    }
  }
  
  \For{ $\forall F_i \in F_\textit{set}$ }
  {
    \For{ $\forall S_i \in F_i$ }
    {
      \uIf{ $S_i$ \textup{is a sink function} }
      {
        \Let{$V_\textit{args}$}{ \textsc{args}($S_i$) }
        
        \For{ $\forall V_i \in V_\textit{args}$ }
        {  
          {\textsc{BackwardAnalysis}($V_i$, $F_i$)}
        }
      }
    }
  }
  \Return {$\textit{Ins}_\textit{out}$}
}

\SetKwBlock{Proc}{procedure}{end procedure}
\Proc($\textsc{ForwardAnalysis} {(}T_\textit{cur}, V{)}$)
{
  \For{ $\forall V_\textit{use} \in $\ \textsc{GetUses}($V$) }
  {
    \uIf{$V$ \textup{is an argument} $x$ \textup{of a function} $F$}
    {
      {\textsc{ForwardAnalysis}(\textsc{AppendNode}($T_\textit{cur}$, $F$), $x$)}
    }
    \ElseIf{$V$ \textup{is a variable in an assignment} `$x = \textit{expression}$'}
    {
      {\textsc{ForwardAnalysis}(\textsc{AppendNode}($T_\textit{cur}$, $x$), $x$)}
    }
  }
}

\SetKwBlock{Proc}{procedure}{end procedure}
\Proc($\textsc{BackwardAnalysis} {(}V, F{)}$)
{
  \uIf{$V$ \textup{is a command from a trusted source}}
  {
    \Let{$\textit{Ins}_\textit{out}$}{ $\textit{Ins}_\textit{out} \ \cup \  V$ }
    
  }
  \uElseIf{$V$ \textup{is from an untrusted source}}
  {
    {\bf return}

  }
  \uElse
  {
    \Let{$V_\textit{defs}$}{\textsc{GetDefVars}($V$)}
    
    \For{ $\forall V_i \in V_\textit{defs}$ }
    {
      \uIf{ $V_i$ $\in$ \textsc{args}($F$) }
      {
         \Let{$F_\textit{callers}$}{\textsc{GetCallers}($F$)}

         \For{ $\forall F_i \in V_\textit{callers}$ }
         {
            {\textsc{BackwardAnalysis}(\textsc{GetCallerArg}($V_i$, $F_i$), $F_i$)}
         }
         
      }
      \uElseIf{ $V_i$ is a global variable }
      {
        \Let{$V_\textit{gdefs}$}{\textsc{GetGlobalDefVars}($V_i$)}
        
        \For{ $\forall V_j \in V_\textit{gdefs}$ }
         {
            {\textsc{BackwardAnalysis}($V_j$, \textsc{GetContainingFunc}($V_j$))}
         }
         
      }
      \uElse
      {
        {\textsc{BackwardAnalysis}($V_i$, $F$)}
      }
    }
  }
}
\end{algorithm}

\smallskip
\noindent
{\bf Algorithm.} 
Alg.~\ref{alg:composition} shows our bidirectional data flow analysis algorithm for identifying variables used to create commands. 

\noindent
{\it -- Step 1. Bidirectional Analysis:} 
\textsc{BidirectionalAnalysis} takes a set of functions of a target program $F_\mathit{set}$ as input. 
\updated{First}{Then}, it conducts the forward analysis (lines 3-8). Specifically, for each variable (lines 3-4), if a variable is from a trusted source (line 5), it creates a tree that describes dependencies between variables as shown in \autoref{fig:Code_Patch}-(b)-\blkcc{A}. The return value is the root node of the tree and it is passed to \textsc{ForwardAnalysis} (line 6).
Similarly, we also build trees for untrusted sources (lines 7-8) as shown in \autoref{fig:Code_Patch}-(b)-\blkcc{B} and \blkcc{C}.
\updated{Second}{Next}, it starts the backward analysis (lines 9-14). 
In each function and each statement (lines 9-10), it searches for invocations of sink functions (line 11). For each identified sink function, we obtain variables used as arguments of the function (line 12).
For each argument $V_i$, we call the \textsc{BackwardAnalysis} \updated{procedure}{} (line 14) that identifies the commands that need to be instrumented.

\noindent
{\it -- Step 2. Forward Analysis:}
Given a variable $V$, it enumerates all the statements that use $V$ via the \textsc{GetUses} function, which returns the results of the standard def-use analysis~\cite{defusechain, defuse2}.
For each statement that uses $V$, if it is used as an argument $x$ of a function call $F$ (line 18), we add the function as a node to the tree ($T_\textit{cur}$) via \textsc{AppendNode} which returns a subtree where the added node is the root of the subtree. It continues the analysis by recursively calling \textsc{ForwardAnalysis} with the subtree and the variable $x$ (line 19).
If $V$ is used in an assignment statement `$x = \textit{expression}$', where \textit{expression} contains $V$, it adds the node $x$ to the tree, and call \textsc{ForwardAnalysis} with the subtree and $x$ (lines 20--21).

\noindent
{\it -- Step 3. Backward Analysis:}
From a variable $V$ and a function $F$ containing $V$, \textsc{BackwardAnalysis} identifies variables that are used to compute the value of $V$ recursively (lines 23--39). For the identified variables, it checks whether the variable is a command and is from a trusted source (i.e., it is found during the trusted forward analysis results) (line 23). If so, the variable is added to $\textit{Ins}_\textit{out}$, which is a set that contains variables to be instrumented.
If the variable can be found in the untrusted results, it terminates (lines 25--26).

If $V$ is also computed from other variables (e.g., $V$=$V_x$+$V_y$), we also find origins of the contribution variables (e.g., $V_x$ and $V_y$) (line 28). Specifically, \textsc{GetDefVars}($V$) returns such contributing variables at the last definition of the variable $V$ (e.g., $V_x$ and $V_y$). The contributing variables are stored in $V_\textit{defs}$. 
%Note that our analysis is based on the SSA (Single Static Assignment) form~\cite{ssa_form} to handle different definitions under multiple predicates.
%
We check the variable's type of each variable $V_i$ in $V_\textit{defs}$. If it is an argument of the current function, we extend our analysis into the caller function. \textsc{GetCallers} returns all of them. To find out the corresponding variable passed to the function in the caller, we use \textsc{GetCallerArg}($V_i$). Then, we continue the analysis in the caller function $F_i$ (lines 32--33). 
If $V_i$ is a global variable, it searches all statements that define the variable, then get rvalues of the statements via \textsc{GetGlobalDefVars} (line 35). It recursively calls \textsc{BackwardAnalysis} to extend the analysis on the functions defining the global variable via {GetContainingFunc} (line 37). 
%\textsc{GetContainingFunc} returns a function that includes a variable.
%
Lastly, if $V_i$ is a local variable (line 38), it recursively conducts backward analysis (line 39).

%Note that the algorithm omits details of how corner cases are handled. In the next few paragraphs, we elaborate on how we handle challenging corner cases. 

\noindent
{\bf Inter-procedural Analysis.} 
We build a call graph~\cite{callgraph} of a target program for inter-procedural analysis. In Alg.~\ref{alg:composition}, \textsc{GetCallers} uses the call graph.
%For each statement that calls a function, a new node that represents a callee function is created. Then, it creates an edge between the caller and the callee. 
After our intra-procedural analysis, we leverage the call graph to identify callers and obtain backward slices from them. We repeat the analysis until there are no more callers to analyze.

\noindent
{\bf Indirect Calls.} 
Call targets of indirect function calls are determined at runtime. As \updated{those indirect calls}{they} are not included in the call graphs we generate, they may cause inaccurate results\updated{ later}{}.
To handle this problem, given an indirect call, we conservatively assume that the call target can be any functions in the program that have the same function signature (e.g., number of arguments and types).
%Note that if a function signature is too simple (e.g., very few arguments), there can be many candidates for caller functions. 
%This is particularly problematic in weakly typed languages such as PHP, because their variables are not explicitly typed, resulting in functions with the same number of arguments having the same function signature. 
%
However, as this is a conservative approximation, we may include more callers. 
%, eventually adding incorrect instrumentations. At runtime, those instrumentations may randomize strings that are not relevant to commands. %, leading to runtime errors. 
%
To mitigate this, we check the origins of the variables passed to callee functions. If the origins are not relevant to commands (i.e., they are not passed to command execution APIs), we prune out the caller. 

\subsubsection{Program Instrumentation}
\label{subsubsec:instrumenation}
We instrument the variables identified in the previous section (Section~\ref{subsubsec:composition}).

\begin{figure}[ht]
    \centering
    %\vspace{-0.5em}
    \includegraphics[width=1.0\columnwidth]{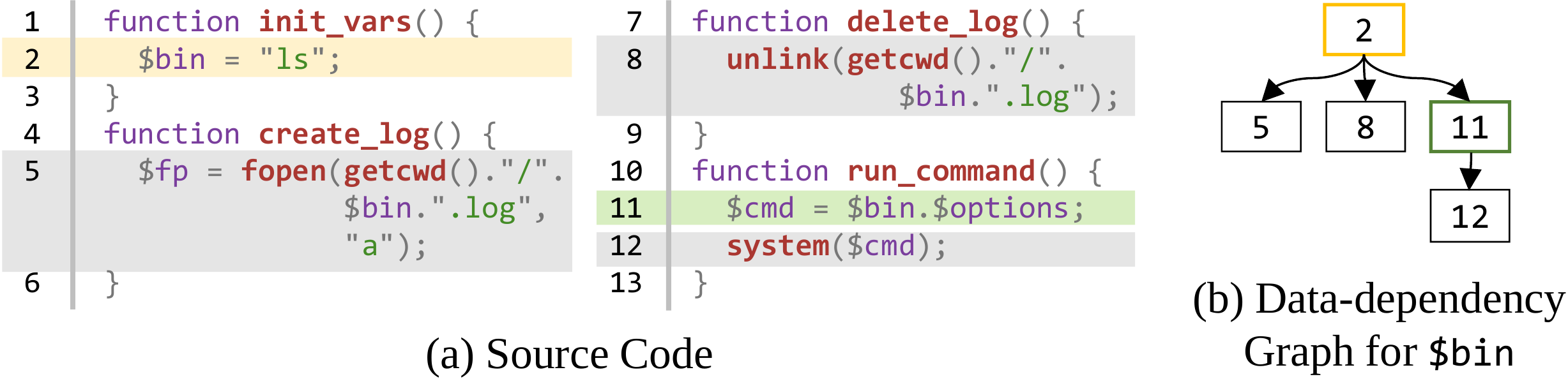}
    \vspace{-2em}
     \caption{Command used in Multiple Places}
     \vspace{-1em}
     \label{fig:instrumentation_example}
\end{figure}

\noindent
{\bf Avoiding Instrumenting Non-command Strings.}  
If an instrumented variable is used in other contexts that do not execute commands, it could break the benign execution.
\autoref{fig:instrumentation_example} shows an example. The sink function, {\tt system()}, executes {\tt \$cmdline}, which is composed by concatenating {\tt \$bin} and {\tt \$options} (line 11) where {\tt \$bin} at line 2. 
%Recall that we identify variables that are used to compose commands from the sink functions in a backward direction, then there can be cases of instrumented variables used in functions that were not properly randomized. 
%
Our analysis described in Section~\ref{subsubsec:composition} will attempt to instrument {\tt "ls"} at line 2, adding a randomization primitive to the definition of the command: ``{\tt \$bin = rand("ls")}''. 
In an original execution, ``{\tt ls.log}'' file is created at line 5 and unlinked at line 8.
However, the instrumentation at line 2 will change the file name to a randomized name. 
For instance, if the randomized name is ``{\tt mt}''  (e.g., \code{ls} $\mapsto$ \code{\color{red}mt}), the instrumented program will create and unlink ``{\tt mt.log}'', which is different from the original program. 
%to an execution at lines 5 and 8 because \sysname does not randomize non command execution APIs (e.g., \code{fopen()} and \code{unlink()}). 
%those APIs are not relevant to command execution, hence \sysname does not randomize them.

To solve this problem, we leverage dependency analysis to find a place to instrument that does not affect the other non-command execution APIs (e.g., {\tt fopen()} and {\tt unlink()} at lines 5 and 8).
%instead of instrumenting line 3, we instrument {\tt \$bin} at line 15 so that it does not affect {\tt fopen()} and {\tt unlink()} functions at lines 7 and 11. 
Specifically, we obtain a data-dependency graph, as shown in \autoref{fig:instrumentation_example}-(b). Nodes are statements in line numbers, and edges between the nodes represent the direction of data flow. From a target variable for instrumentation (\code{\$bin}), we identify statements that use the target variable.
If we instrument at the root node ({\tt \$bin}), it affects all the child nodes, including those with non-randomized functions (lines 5 and 8).
Hence, among the nodes between the root node and the node including the \code{system()} function (line 12), we pick the node line 11 to instrument.
This is because instrumenting at line 11 only affects the command execution API \code{system()}.
%. The sink function (\blkcc{5}) is a child of node \blkcc{4}. 
%Since the first node \blkcc{1} affects all of its children (i.e., the yellow shaded area) containing the non-randomized functions, we pick a child node where none of its children does not have non-randomized functions (i.e., the green shaded area), which is \blkcc{4}. 
%As a result, we choose to instrument \code{\$bin} at line 11. 
Essentially, from the root node, we pick a child node along the path to the sink function. We move toward the sink function until the picked node's children do not include any non-randomized functions.

\subsection{Runtime Phase}
\label{subsec:runtime_phase}
%An instrumented program is running with command target subsystems (e.g., OS/shell and database) with our subsystem randomization component (\blkcc{A}), which randomizes the commands inside the subsystems (e.g., commands and SQL keywords).
%To support 
%Given an instrumented program and randomized subsystems (e.g., randomized file system and database engines), 
%we run the instrumented program with our runtime support that bridges randomized subsystems and the instrumented program.
%
%For those instrumented commands, 
%\sysname runtime support randomizes commands in the subsystems\updated{ properly}{} so that the instrumented intended commands can be executed.
%
\updated{The runtime support component is synchronized with the command target randomization component, meaning that it keeps track of the current one-time pad for randomization.}{}

%\subsubsection{Command Target Randomization}
%\label{subsec:subsysrand}
%An instrumented program by \sysname calls command executing functions (e.g., {\tt system()}) with randomized commands if the commands are intended.
%
%To prevent the execution of non-randomized commands (i.e., injected commands by attackers), 
%We randomize OS/shell command processor (Section~\ref{subsubsec:shellcommand_rand}) and database engines (Section~\ref{subsubsec:database_random}). 
%In this subsection, we describe how we randomize different subsystems in detail.

\subsubsection{OS/Shell Command Processor Randomization}
\label{subsubsec:shellcommand_rand}
We randomize the OS/shell command processor by hooking two critical paths of the command execution: (1) the creation of the shell process and (2) file I/O and shell APIs that access external binary files in the shell process.
Recall that there are two types of OS/shell commands: internal and external~\cite{linux_commands_external_internal}. 
For all commands, a program spawns a shell process (e.g., `\code{/bin/sh}'). 
The shell process, which contains the implementation of internal commands, directly executes internal commands (e.g., \code{cd}).
External commands are executed by further calling APIs (e.g., \code{execve}) that run an external program.

\begin{figure}[ht]
    \centering
    %\vspace{-0.5em}
    \includegraphics[width=0.88\columnwidth]{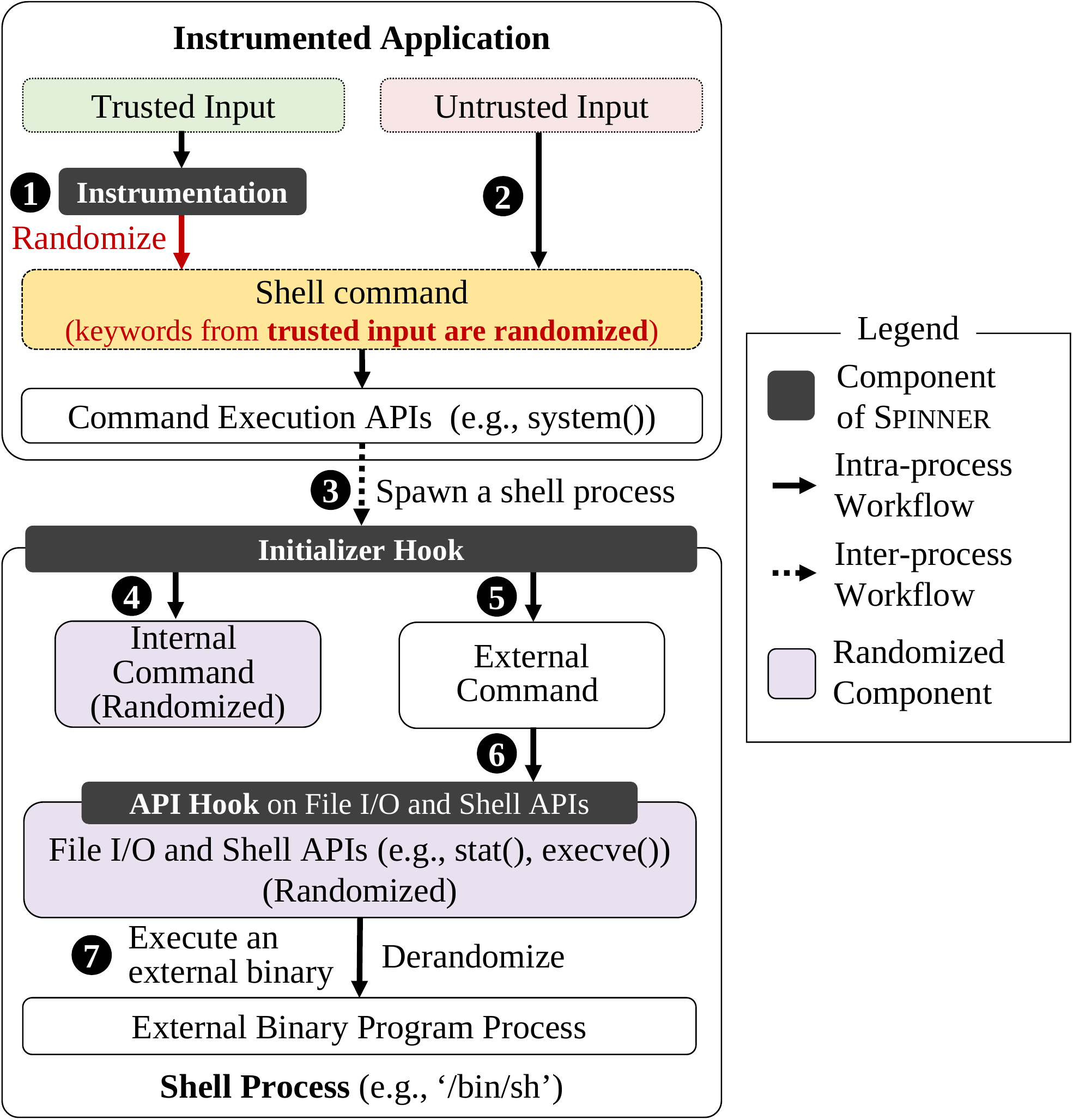}
    \vspace{-1em}
     \caption{OS/Shell Command Processor Randomizer}
     \vspace{-1em}
     \label{fig:command_randomizer}
\end{figure}

\autoref{fig:command_randomizer} shows how \sysname randomizes internal and external commands, following the typical execution flows.
To execute an OS/shell command, the program often composes a command via string operations. 
If a command is composed of trusted inputs, the command names are randomized via the instrumentation (\blkcc{1}). Commands originated from the untrusted inputs are not randomized (\blkcc{2}).
The composed command is then passed to the command execution APIs such as \code{system()}.
In the following paragraphs, we explain how \sysname works after the command execution APIs are called depending on whether the command is internal or external.
%hooking command interface functions (shown in Table~\ref{table:sinkfunctions}).
%Additional sinks are in the Appendix (Table~\ref{table:xmlsinkfunctions}).
%\autoref{fig:command_randomizer} shows the structure of our randomization module. Specifically, we hook APIs to only accept randomized commands. Note that there are two types of commands: internal and external commands~\cite{linux_commands_external_internal}. Internal commands are directly implemented by the OS/shell, while external commands are delegated to external binary programs where the OS/shell executes them.

\noindent
{\bf Internal Commands.}
To execute an internal command, an application calls a command execution API, which spawns a shell process (\blkcc{3}) and passes the command to the spawned process.
As the internal commands are implemented within the shell process, it does not make further API calls to access external binary files.
\updated{
To randomize the internal commands, \sysname intercepts the entrypoint (i.e., \code{\_start()}) of the shell process. On the entrypoint, \sysname examines the command name (\blkcc{4}) via \code{getopt()} and only allows the command if properly randomized. Note that the shell process also uses \code{getopt()} to parse commands, meaning that it does not break benign functionalities regarding command parsing.}
%Note that we infer a list of internal commands by analyzing the shell process~\cite{bashsource}. 
%The command that is not randomized will fail.
%For internal commands, we reject a command that is not randomized properly.
%The hook module a list of randomized commands and their corresponding original commands. We only allow randomized commands to be accepted. %\MA{Consider describing (here or at 4.3) how the hook module maintains that list of commands -- I haven't read much discussion about how the derandomization takes place.}

\noindent
{\bf External Commands.}
After the shell process is spawned (\blkcc{3}), if the command is an external command (\blkcc{5}), the shell process calls a few files I/O APIs such as \code{stat()} to check whether the binary file for the command exists or not (\blkcc{6}). 
If the binary exists, it will execute the binary (\blkcc{7}). 
We provide a randomized view of the underlying file system by hooking file I/O and shell APIs and only allowing access with properly randomized file paths. 
%Specifically, we hook the APIs and only allow access to the invocations with properly randomized file paths. 
If the command is not randomized, API calls such as \code{stat()} will fail, preventing the execution of the command.
A randomized command is derandomized and executed via APIs such as \code{execve()} (\blkcc{7}).
\updated{Unlike the internal commands, \sysname does not require a list of external commands for randomizing the shell process. It is only required for instrumentation (Section~\ref{subsubsec:composition}).}{}

%Specifically, if OS/shell processors recognize an external command execution request, they first look for a binary file for the external command using file API functions such as \code{stat()} (\blkcc{2}).
%Hence, we hook those file API functions (\blkcc{B}) called on behalf of the command interface functions (e.g., \code{system()}, denoted as \blkcc{3}) to randomize file paths they access (\blkcc{4}). 
%Note that we do not randomize file paths if it is not called within the command executing APIs. %We check a call stack within our API hook module to determine whether it is called within command interface functions. 

%\noindent
%{\bf Impact of the Completeness of the Parser.}
%The parser we use might be incomplete; hence, it might fail to handle tricky (often malicious) inputs. It is important to mention that, unlike techniques that use parsers to analyze malicious inputs, failing to handle tricky and malicious inputs in \sysname does not break the protection. If it fails to recognize a malicious command, the command will not be executed. As intended inputs {\it are not encoded in a complex way} (intuitively, developers do not have any motivation to make the command tricky to parse while attackers often have such motivation), we do not observe any intended commands that fail to be parsed.

\begin{figure}[ht]
    \centering
    %\vspace{-1em}
    \includegraphics[width=0.88\columnwidth]{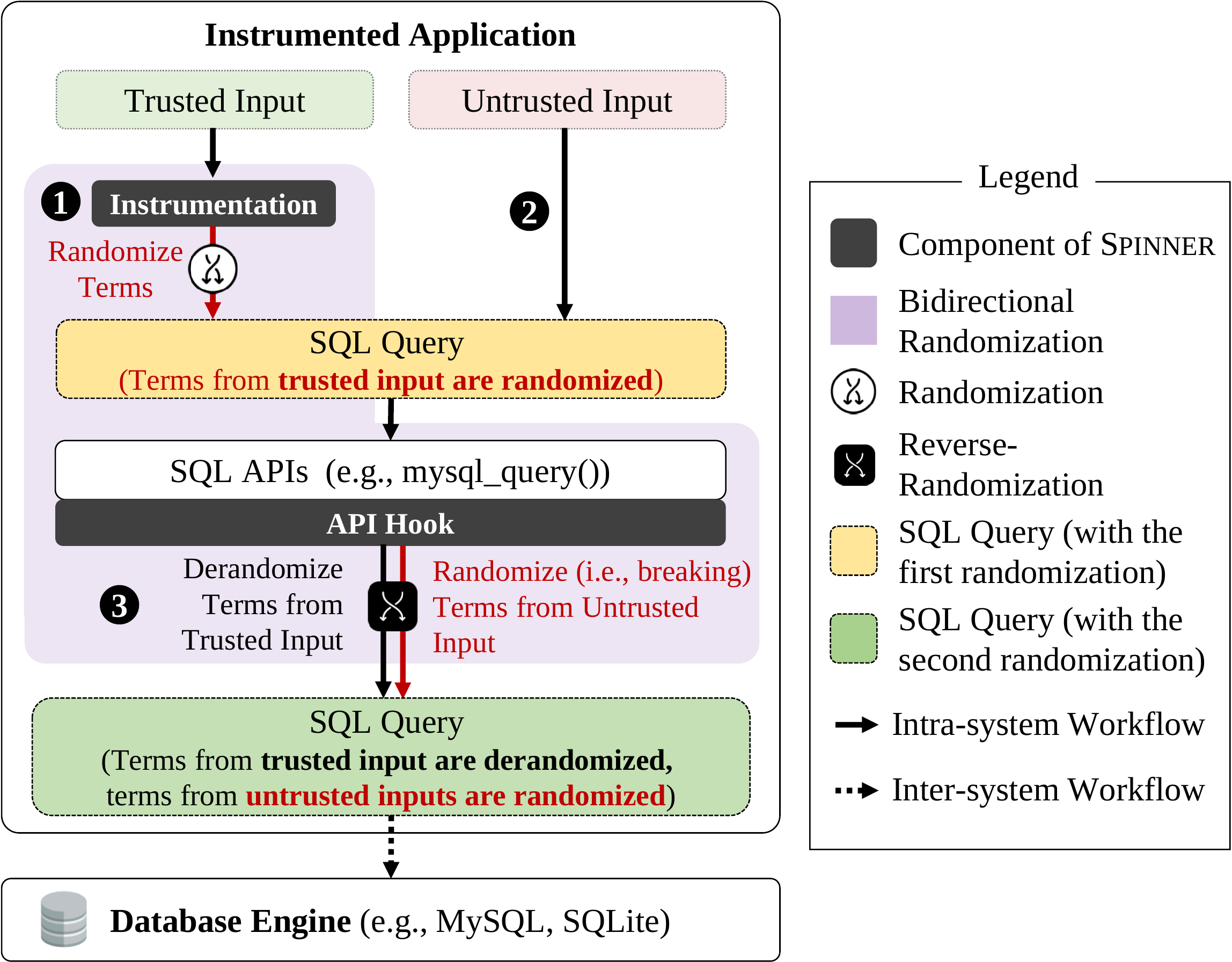}
    \vspace{-1em}
     \caption{Randomization for Database Engines}
     \vspace{-1em}
     \label{fig:database_rand}
\end{figure}

\revised{
%\vspace{-1em}
\subsubsection{Database Engine Randomization}
\label{subsubsec:database_random}
Database engines are complicated and some are \updated{closed sourced}{proprietary (i.e., closed source)}, meaning that it is difficult to randomize them in practice. % Moreover, some engines are   has a largely different design and implementation, including APIs and internal data/program structures.
%Hence, it is challenging to find a universal approach to randomize database engines. 
As a result, previous approaches (e.g., \cite{sqlrand}) leverage a database proxy to parse a randomized query and rewrite it to a standard (i.e., derandomized) query. Implementing a robust parser for multiple database engines is challenging as shown in Section~\ref{subsubsec:advanced_sql_injection}.
Moreover, they rely on a list of known SQL keywords to randomize and derandomize, failing to prevent sophisticated attacks presented in Section~\ref{subsec:comparison_existing}.
%Unfortunately, such approaches are not able to prevent SQL queries leveraging SQL keywords that are not included in their list. % (Examples in Appendix~\ref{appendix:sqlrand-llvm}).

%For example, some SQL engines such as SQLite provide various ways to hook key operations (e.g., commit/update hooks~\cite{xxx}) while others do not have such interfaces.
%For example, in PHP, \code{mysql\_query()} sends SQL queries to a MySQL database while \code{sqlite\_query()} is used to access a SQLite database. 
%Specifically, different database engines have different APIs and internal structures. %, and execution points that can be intercepted (e.g., callback event handlers). %\MA{and therefore hooked by another call}.
%PDO: https://www.php.net/manual/en/class.pdo.php
%As a result, it is challenging to develop a generic module that can randomize diverse SQL engines. 

\begin{figure}[h]
    \centering
    %\vspace{-0.5em}
    \includegraphics[width=1\columnwidth]{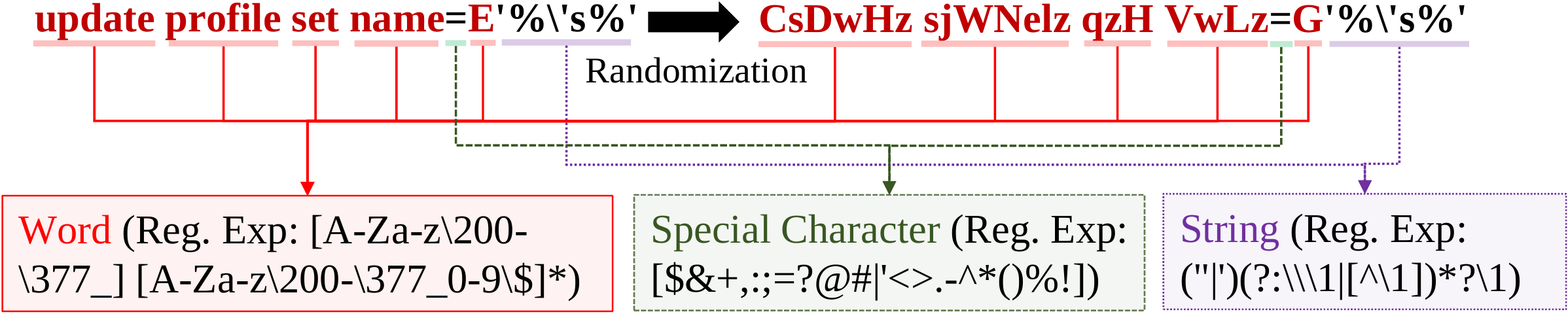}
    \vspace{-2em}
    \caption{\revised{\updated{}{Scanner Recognizing Words for Randomization}}}
    \vspace{-1em}
    \label{fig:escape_string}
\end{figure}

\noindent
\updated{}{\textbf{Bidirectional Randomization with Scanner.}}
We propose a \emph{bidirectional randomization approach} that applies the randomization scheme twice, one for randomization and another for reverse-randomization. 
\updated{}{Unlike existing techniques requiring knowledge of known SQL keywords and grammar, \sysname uses a scanner that works without such knowledge. As shown in \autoref{fig:escape_string}, \sysname only needs patterns of words, special characters, and strings.}
\updated{By doing so}{For each identified word}, it (1) derandomizes randomized intended queries and (2) randomizes (and breaks) injected malicious queries at the same time. 
\updated{It does not require a SQL parser or a list of known SQL keywords.}{} % deploy it on the SQL APIs. 
%a database front-end to randomize the database engines.
%The front-end sits between the user application and the actual database engine, like the proxies in other techniques such SQLRand~\cite{sqlrand}. 
%However, \sysname differs from existing techniques because it leverages our \emph{dual randomization approach}, which randomizes the input twice to completely prevent injected SQL queries bypass the randomization scheme. 
%Note that we present a detailed case study in Section~\ref{xxx} to demonstrate how \sysname prevents sophisticated attacks that existing randomization techniques fail to prevent.
%
Specifically, in a program that accesses a database, \sysname instruments strings that are used to compose a SQL query as shown in \autoref{fig:database_rand} \updated{}{(\blkcc{1})}. \updated{We use a scanner to identify terms in a query and then apply our randomization scheme (\blkcc{1}).}{} 
Note that untrusted inputs are not randomized \updated{}{by this instrumentation} (\blkcc{2}).
Finally, randomized trusted inputs and untrusted inputs are combined to compose a query and then passed to a SQL API such as \code{mysql\_query()}.
We hook such SQL APIs to apply our reverse-randomization (or derandomization) scheme before the query is passed to the database engine (\blkcc{3}).
\updated{Note that w}{W}e apply it for every recognized term (\emph{not only for the SQL keywords}\updated{}{ because our scanner does not have the notion of known SQL keywords}), resulting in derandomizing all the randomized terms as well as randomizing (with the reverse-randomization scheme) SQL queries from untrusted inputs.
%The second XOR cipher essentially derandomizes the keywords randomized before (at \blkcc{1}) while randomizes (and breaking) the keywords from untrusted input that are not randomized. 
To this end, if all terms in a SQL query are from the trusted sources, the resulting query can be successfully executed.
However, if some terms are from the untrusted inputs, they are randomized (via the reverse-randomization) and cannot be executed, preventing injection attacks.
%Given a mixture of randomized and standard SQL statements, it derandomizes randomized statements while randomizes standard statements. %Then, for the remaining statements (e.g., standard statements) that were not processed, we randomizes them. %and  add a randomization layer to the application. 
%Specifically, it consists of two layers. The first layer derandomizes randomized SQL statements. 
%and translates them to the original SQL statements that can be executed by an existing database engine. 
%To prevent injected (malicious) SQL statements from being executed, it also randomizes standard SQL statements. 

%We use SQLite~\cite{sqlite} to implement the dual layer randomizer.
%SQLite is one of the smallest database engines, and it is relatively easy to directly randomize the SQL grammar. 
%Specifically, we extract a SQL processor in SQLite, and create both derandomizer and randomizer.
%simply search constant strigs that define SQL keywords and replace them with randomized keywords.
%It is modified in a way that it can parse two different SQL grammar: the standard SQL grammar and our randomized grammar.
%The shim database front-end processes an input SQL statement and generates the {\it two-way translated statements} (i.e.,  standard SQL keywords $\leftrightarrow$ randomized SQL keywords). 

\updated{Note that \sysname recognizes keywords \emph{without a list of known SQL keywords}. It identifies terms with different patterns in a query, such as strings, numbers, and keywords, in a SQL grammar agnostic way. As a result, unlike existing techniques, \sysname does not fail when a new SQL keyword is added.}{
    \textit{-- Handling Escaping String Constant:}
    Note that \autoref{fig:escape_string} shows an example of PostgreSQL's unique feature of escape string constant, which is a special way of defining a string with a capital letter `E' before a string. Since it is a unique grammar for PostgreSQL, many parsers~\cite{andialbrecht/sqlparse,xwb1989/sqlparser,node-sql-parser,greenlion/PHP-SQL-Parser,moz-sql-parser} do not support it, resulting in a parsing error. \sysname considers the `E' as a word, and randomize/derandomize correctly, preserving content in the string. 
}
}

% (Details in Appendix~\ref{appendix:sqlrand-llvm}).

%First, the derandomization layer processes randomized statements, generating standard SQL statements. After the derandomization layer, all the derandomized statements are marked so that it would not be randomized again by the randomization layer.
%Next, the randomization layer transforms  standard statements to randomized statements.
%Otherwise, if the token is not randomized (i.e., standard token), the reverse process takes place. 
%To this end, intended SQL statements are randomized by applications and then derandomized by the front-end.
%By the same fashion, malicious (injected) SQL statements are not randomized by our hook module, preventing their execution.

%\YK{We hook API like mysql\_query(), and put the parser. We replace keywords in sqlite, and it works well.}

%\autoref{fig:database_rand}-(a) shows the dual layer randomizer is placed between an application and a database engine. The instrumentations are applied to API functions that execute SQL statements (e.g., \code{mysql\_query()}). 

\begin{figure}[!t]
    \centering
    %\vspace{-0.5em}
    \includegraphics[width=0.85\columnwidth]{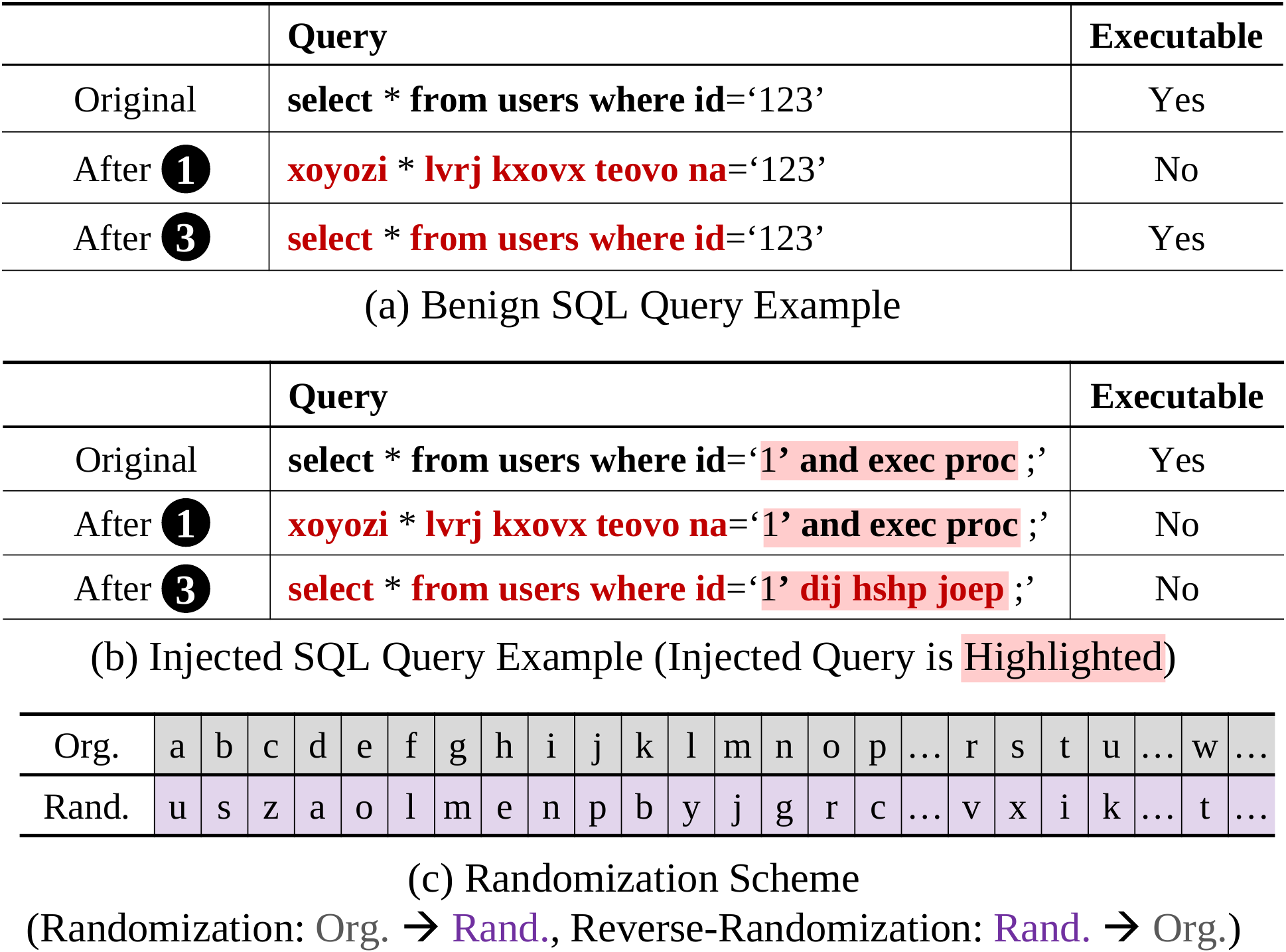}
    \vspace{-1em}
    \caption{\revised{Example of Benign and Injected Query Execution}}
    \vspace{-2em}
    \label{fig:dual_rand}
\end{figure}

\noindent
{\bf Execution of Intended SQL Queries.}
\autoref{fig:dual_rand}-(a) shows examples of how a benign SQL query is processed by \sysname with a randomization scheme shown in \autoref{fig:dual_rand}-(c), along the execution path of \autoref{fig:database_rand}.
Specifically, when an intended SQL statement is executed, it goes through the instrumentation (\blkcc{1}) hence randomized and then is passed to a SQL API. 
The randomized query is shown in the second row of \autoref{fig:dual_rand}-(a). All recognized terms, including the table name `\code{users}', are randomized.
%Assume that the dual layer randomizer uses the randomization rule shown in \autoref{fig:database_rand}-(b). An example of a randomized SQL statement is shown on the left side of \redcc{1}.
Then, in our hook function of the SQL API, we apply our reverse-randomization for all terms. Since there are no terms from untrusted input, every term is derandomized (\blkcc{3}), as shown in the third row.
The last column of \autoref{fig:dual_rand}-(a) shows whether the query can be executed without errors or not. The query after \blkcc{3} is executable.
%For example, given a query, \code{select * from table}, from trusted sources, \code{select} and \code{from} are randomized to \code{xxxxxx} and \code{yyyy} respectively.
%Then, in the API hook function, they will be derandomized back to \code{select} and \code{from}, before the query is passed to a database engine.
%To this end, the query is successfully executed.
%derandomization layer applies the randomization rule, resulting in a standard SQL statement shown next to \blkcc{2}, which is passed to the SQL engine.

\noindent
{\bf Execution of Injected SQL Queries.}
An injected SQL query is not randomized because it is not instrumented (\blkcc{1}). 
\autoref{fig:dual_rand}-(b) shows an example query with an injected query highlighted in red.
Specifically, assume that \code{select * from users where id=`\$id'} is the vulnerable SQL query, and an attacker injects a query by providing a value highlighted in \autoref{fig:dual_rand}-(b) to \code{\$id}.
%
% MA: $id corresponds to 1' and delete from users (remember that 1' is not derandomized, but the rest is)
When the \updated{SQL}{query} is passed to the SQL APIs, the beginning part of the query (from trusted inputs up until the single quote) is randomized, but the later part outside the quotes (i.e., the injected query) is not.
On the hooked API, \sysname applies the reverse-randomization in every term we recognize. %\textcolor{red}{except in tokens that are not considered from user input, determined by the system's scanner. } 
% MA: the below sentence is redundant
As a result, it effectively \emph{derandomizes} the (trusted) beginning part of the query while \emph{randomizing} the later part of the query from untrusted inputs. 
Note that the reverse-randomization applies the substitution rule in the reverse order (i.e., \textsf{\color{red}  Randomized} $\mapsto$ \textsf{Original}). For example, `\code{a} $\mapsto$ \code{\color{red} f}' is a randomization rule\updated{, where}{ of a reverse-randomization} `\code{\color{red} f} $\mapsto$ \code{a}'\updated{ is a corresponding reverse-randomization}{}.
After \blkcc{3}, \updated{due to the later part of the query that is randomized, the injected query is prevented.}{the injected query is prevented as it is reverse-randomized.}

\noindent
{\bf Randomized Table Name Translator.} 
The bidirectional randomization scheme randomizes all terms that are not originated from trusted sources. As a result, we observe that if a table name in a SQL query is originated from untrusted sources without quotes (e.g., `\code{select * from \$input}'), the table name can be randomized, resulting in a wrong query.
While using input as a table name is \updated{a bad (i.e., insecure)}{a poor} programming practice, there exist programs composing queries in that way. 
To this end, we additionally instrument \code{tbl\_derand()} to the variables that are not quoted. At runtime, it will check whether the instrumented string contains a randomized table name. \emph{If and only if it contains a single table name}, we derandomize it to the original table name. Note that it does not derandomize if the instrumented string contains multiple terms (i.e., words) to prevent injection attacks targeting the instrumented variables.
% the query will fail due to the randomized table name. 
%\YK{SQL table names can be randomized. (1) this is an insecure programming practice. (2) we handle this by translating the table back to the original.}

%\vspace{-1em}
\subsubsection{XML Processor Randomization}
An XML processor is a program or module that \updated{processes}{parses} an input XML file \updated{by parsing and executing the actions annotated in the input XML file}{and executes the annotated actions in parsed XML elements described via tags}.
%\updated{Essentially, in an XML file, each tag (i.e., entity) is an individual task to be processed.}{}

\noindent
{\bf XML External Entity (XXE) Attack.}
Among the entities, there is an XML External Entity (XXE) which refers to data from external sources (e.g., other files or networks).
The entity can refer to a sensitive password file using the following entity: ``\code{<!ENTITY xxe SYSTEM "file://FILE">}''.
An XML processor parses the entity, then it reads to include the content of `\code{FILE}' in the output. \updated{If the `\code{FILE}' is a secret file, an information leak can happen.}{}

% \begin{figure}[ht]
%     \centering
%     \includegraphics[width=0.9\columnwidth]{fig/xml_rand.pdf}
%     \vspace{-1em}
%      \caption{Randomization for XML Processors}
%      \label{fig:xmlrand}
% \end{figure}

\noindent
{\bf XML Processor Randomization.}
We randomize {\it external resources' namespaces} such as file names and network addresses at APIs that access them (e.g., file I/O APIs and network APIs). With the randomization, only the API calls with randomized file names, paths, IPs, and URLs will succeed. 
To ensure benign requests are properly handled, we analyze a target program to identify all intended XML files. 
Specifically, we identify XML files and data passed to the XML sink functions shown in \autoref{table:sinkfunctions}. If they are originated from trusted sources (e.g., constants or from configuration files), we mark them to be randomized at runtime. 
% to identify XML files and contents that can be trusted. % Hardcoded XML contents, XML file names and paths constructed through trusted sources (e.g., configuration files) are trusted. 
%In trusted XML files, we identify all entities that refer to external resources (e.g., the xxe entities). 
At runtime, when a trusted XML file is loaded, we randomize resource names/paths of XXEs in the file. 
As we randomized the namespaces in the application through APIs, intended accesses through the randomized XXEs will be successful.
For untrusted XMLs, file names, paths, and URLs in XXEs are not randomized and passed to the file I/O and network APIs, resulting in errors and preventing injection attacks.
%As our randomization scheme can only understand randomized namespaces, the access will fail (\blkcc{6}).
%Injected entities will not be randomized, failing to access the underlying file systems or network servers.
%
%
Note that when \sysname analyzes the program to identify trustable XML files, we assume all the local XML files during the offline analysis are not compromised. 
%This is because we randomize all identified XXEs from trusted sources. 
\sysname's goal is to prevent future XXE injection after the analysis. 

\subsection{\sysname Runtime Support}
\label{subsec:runtimesupport}

\subsubsection{Dynamic Randomization Support}
\sysname randomizes commands in the subsystems at runtime dynamically. We change \updated{the one-time pad}{our randomization scheme (or table)} on every command execution function invocation \updated{}{(or per input)} so that knowing previously used randomization schemes will not help subsequent attacks.
\updated{Note that \sysname's instrumentations dynamically randomize intended commands before it is passed to command execution APIs. Then, after a command execution API is called, the used one-time pad is discarded. By doing so, we minimize the possibility of attacks reusing the previously used one-time pad.}{}

\subsubsection{Randomization Primitives}
The runtime support provides two primitives: randomization and derandomization primitives.

\noindent
{\it -- Randomization Primitive} is a function that takes a string as input and returns a randomized string via a mapping between each byte in the input and randomized byte(s). 
To mitigate brute-force attacks against the randomized commands, the mapping is created per input. 
It also supports multiple randomization schemes that convert 1 byte to 2 bytes (\optmap{x}{ab}), 4 bytes (\optmap{x}{cdef}), and 8 bytes (\optmap{x}{ghijklmn}).
Details can be found in Appendix~\ref{appendix:bruteforce_attack}.

    %\sysname can extend the substitution cipher space so the search space of the possible randomized commands is large. One single character can be randomized to multiple characters while the number of length can be configured by the administrator. For example, a command \code{`kc'} can be randomized to \code{`afvycm'}, if \sysname is configured to randomize one character to four characters or other schemes as shown in \autoref{fig:brute force}-(b). \sysname will generate a new randomization table on every sink function invocation. Hence the randomization table will be different for each request.}\CJ{2)[3.2] (\#D) Explain the algorithms used to randomize variables.

%
\updated{When it randomizes,}{At runtime,} we maintain a pair of a randomized string and its \updated{one-time pad}{randomization table}, which we call {\it randomization record}. The record is later used in the derandomization function.
Note that two different strings can be randomized into the same string with two different one-time pads. For example, a one-time pad `\code{a} $\mapsto$ \code{\color{red} c}' and another one-time pad `\code{b} $\mapsto$ \code{\color{red} c}' will randomize both `\code{a}' and `\code{b}' to `\code{\color{red} c}', leading to the ambiguity in derandomization. %Specifically, when \sysname derandomizes, it looks up the randomization record with a randomized string as a key. If there are multiple records with the same randomized string, it cannot determine which one to derandomize.
To solve this problem, when it randomizes, it checks whether the \updated{new}{}randomized string exists in the existing randomization records. If it exists, it randomizes the input string again until there are no matching strings in the records.

\updated{Note that information disclosure attacks might be used to leak the stored one-time pads. However, as we always use the different one-time pads, knowing previously randomized commands does not help to launch subsequent attacks.
Also, when the randomized command is passed to command execution APIs, its corresponding one-time pad is removed. Hence, even if the attacker injects a randomized command with the previous one-time pad, the attack will not succeed.
More details can be found in Section~\ref{sec:discussion}. 
}{}

%In addition, existing memory region protection techniques, or even secure process extensions such as SGX~\cite{intel-sgx} can be used to protect the memory containing randomization records. 

\noindent
{\it -- Derandomization Primitive} takes a randomized string as input and returns the original value of the string. Given a list of randomization records, it finds a record that has a matching randomized string. Then, it leverages the record's \updated{one-time pad}{randomization scheme} to derandomize the input string.

\vspace{-0.5em}
\section{Evaluation}
\label{sec:eval}

\noindent
{\bf Objectives.} 
We evaluate \sysname on four aspects. 
First, we present analysis results on the instrumented code and its impact to show the correctness of \sysname (Section~\ref{subsec:instr}). 
Second, we run PoC exploits against a set of vulnerable programs and their \sysname instrumented versions, to show the effectiveness of \sysname in preventing command injection attacks (Section~\ref{subsec:eval_attack}).
Third, we measure the performance  overhead of \sysname (Section~\ref{subsec:eval_perf}).
Fourth, we present case studies to show the effectiveness of \sysname in advanced command injection attacks (Section~\ref{subsec:casestudy}).

\noindent
{\bf Implementation.}
%\sysname's static analysis supports diverse programming languages: C/C++, PHP, JavaScript and Lua. 
We implement our static analysis tool by leveraging LLVM~\cite{llvm} for C/C++, php-ast~\cite{php-ast} and Taint'em All~\cite{taintall} for PHP, Acorn~\cite{js-ast} for JavaScript and Lua SAST Tool~\cite{costin2017lua} for Lua.
\sysname uses \code{LD\_PRELOAD} that requires access to the shell. 
Hence, it does not support a web hosting service such as cPanel~\cite{cpanel}.

\noindent
{\bf Setup.} All the experiments were done on a machine with Intel Core i7-9700k 3.6Ghz, 16GB RAM, and 64-bit Linux Ubuntu 18.04.

\begin{table*}[t]
	\centering
	\footnotesize
	%\caption{Selected Programs for Evaluation}
	\caption{Selected Programs for Evaluation and Instrumented Results }
	\label{table:selectedprograms}
	\vspace{-1em}
	\renewcommand{\arraystretch}{0.9}	
\resizebox{1\textwidth}{!}{%	
\setlength{\tabcolsep}{3pt}

\begin{tabular}{llrccrrrrrrrrrrrrrrrr} 
	\toprule
	\multirow{4}{*}{\bf ID} &
	\multirow{4}{*}{\bf Name} & 
	\multicolumn{1}{c}{\multirow{4}{*}{\bf Size}} &    %<=====FIX 
	\multirow{3}{*}{\bf Vulner-} &
	\multirow{3}{*}{\bf Lang-}&
	
	\multicolumn{5}{c}{\bf \# Instrumentations} &
	\multicolumn{3}{c}{\bf \# Instr. Affecting} &
	\multicolumn{6}{c}{\bf \# Affected Vars./Funcs.} &
	\multicolumn{2}{c}{\bf Dep. Analysis}

	\\ \cmidrule(lr){6-10} \cmidrule(lr){11-13} \cmidrule(lr){14-19} \cmidrule(rr){20-21}
	
	& & &
	\multirow{2}{*}{\bf ability} & 
	\multirow{2}{*}{\bf uage(s)}
	&
	\multirow{2}{*}{\bf Const.} &
	\multicolumn{3}{c}{\bf Dynamic } &
	\multirow{2}{*}{\bf Sinks}& 
	\multirow{2}{*}{\bf BB$^1$} &
	\multirow{2}{*}{\bf Fn$^2$ } &
	\multirow{2}{*}{\bf Fns$^3$} &
	\multicolumn{2}{c}{\multirow{1}{*}{\bf Local$^4$}} &
	\multicolumn{2}{c}{\multirow{1}{*}{\bf Global$^5$}} &
	\multicolumn{2}{c}{\multirow{1}{*}{\bf Funcs$^6$}} &
	\multirow{2}{*}{\bf For-} &
	\multirow{2}{*}{\bf Back-} 
    %\multirow{2}{*}{\bf Local} &
	%\multirow{2}{*}{\bf Global} &
	%\multirow{2}{*}{\bf Func} 
	
	\\ 
	\cmidrule{7-9}
	& & & & & &
	{\bf 1-5}&{\bf 6-10}&{\bf $>$11} &
	& & &
	& \multicolumn{2}{c}{(Total)} & \multicolumn{2}{c}{(Total)} & \multicolumn{2}{c}{(Total)} & {\bf ward} & {\bf ward}
	\\
	
%	\midrule
	\cmidrule(lr){1-5} 	\cmidrule(lr){6-10}	\cmidrule(lr){11-13}	\cmidrule(lr){14-19} \cmidrule(lr){20-21} %\cmidrule(lr){17-18}
	\rowcolor{gray!40}
	{\tt s1}&
	WordPress~\cite{wordpress} &
	42.60 MB  & 
	Cmd.$^7$~\cite{CVE-s1} &
	PHP &
	38 & 279 &  127 &  18  &  7 & 3 & 1 & 458 &	\avgtotal{11.39}{178} & \avgtotal{2.95}{15} & \avgtotal{7.04}{90} & \cellcolor{red!25} 10.2$^\alpha$ & \cellcolor{red!25} 6.9$^\beta$
	%& & &
	
	\\ 
	%\cmidrule(lr){1-5} 	\cmidrule(lr){6-10}	\cmidrule(lr){11-13}	\cmidrule(lr){14-16} %\cmidrule(lr){17-18}
	%\rowcolor{gray!40}
    {\tt s2}&
    Activity Monitor~\cite{PlainviewActivityMonitor} &
	0.99 MB &Cmd.$^7$~\cite{CVE-s2} &
	PHP &
	6 & 12 & 9 & 0  &  6 & 2 & 4 & 21 & \avgtotal{9.89}{27} & \avgtotal{2.77}{2} & \avgtotal{7.53}{34} & 7.7 & \cellcolor{red!25} 6.3$^\gamma$
	%& & &
	
%	\hline	
	\\ 
%\cmidrule(lr){1-5} 	\cmidrule(lr){6-10}	\cmidrule(lr){11-13}	\cmidrule(lr){14-16} %\cmidrule(lr){17-18}
	\rowcolor{gray!40}	
	{\tt s3}&
	AVideo-Encode~\cite{videoencoder} &
	8.93 MB &
	Cmd.$^7$~\cite{videoencodercve} &
	PHP &
	2 & 48 & 8 & 3  &  27 & 3 & 37 & 21 & \avgtotal{0.98}{63} & \avgtotal{0}{0} & \avgtotal{1.36}{79} & 1.7 & 6.7
	%& & &
	
	\\ 
	%\cmidrule(lr){1-5} 	\cmidrule(lr){6-10}	\cmidrule(lr){11-13}	\cmidrule(lr){14-16} %\cmidrule(lr){17-18}
	%\rowcolor{gray!40}
    {\tt s4}&
	Pepperminty-Wiki~\cite{Pepperminty-Wiki} &
	23.00 MB &
	XXE$^8$~\cite{CVE-wiki} &
	PHP$^\dagger$ & % , XML 
	0 & 2 & 0 & 0  &  2 & 0 & 2 & 0 & \avgtotal{0}{2} & \avgtotal{0}{0} & \avgtotal{1}{2} & 1 & 1
	%& & &
	
	\\ 
	%\cmidrule(lr){1-5} 	\cmidrule(lr){6-10}	\cmidrule(lr){11-13}	\cmidrule(lr){14-16} %\cmidrule(lr){17-18}
	\rowcolor{gray!40}
	{\tt s5}&
	PHPSHE~\cite{PHPSHE} &
	11.91 MB &
	XXE$^8$~\cite{CVE-shopping} &
	PHP$^\ddag$ & % , XML, SQL 
	54 & 183 & 26 & 7  &  5 & 54 & 0 & 236 & \avgtotal{3.82}{96} & \avgtotal{0}{0} & \avgtotal{3.76}{67} & 5.7 & 3.6
	% &  &  &
	
	\\ 
	%\cmidrule(lr){1-5} 	\cmidrule(lr){6-10}	\cmidrule(lr){11-13}	\cmidrule(lr){14-16} %\cmidrule(lr){17-18}
	%\rowcolor{gray!40}
	{\tt s6}&
	Pie Register~\cite{pieregister} &
	5.51 MB &
	SQL$^9$~\cite{CVE-s5} &
	PHP$^*$ & % , SQL
	0 & 68 & 5 & 0  &  2 & 0 & 0 & 73 & \avgtotal{3.06}{26} & \avgtotal{3}{3} & \avgtotal{4.28}{27} & 6.3 & \cellcolor{red!25} 7.2$^\delta$
	% &  &  &
	
	\\ 
	%\cmidrule(lr){1-5} 	\cmidrule(lr){6-10}	\cmidrule(lr){11-13}	\cmidrule(lr){14-16} %\cmidrule(lr){17-18}
	\rowcolor{gray!40}	
	{\tt s7}&
	Lighttpd~\cite{lighttpd} &
	17.40 MB &
	SQL$^9$~\cite{CVE-s6} &
	C &
	5 & 4 & 0 & 1 &  10 & 5 & 5 & 0 & \avgtotal{0.5}{5} & \avgtotal{0}{0} & \avgtotal{0.5}{5} & 1.6 & 7.3
	% &  &  &
	
	\\ 
	%\cmidrule(lr){1-5} 	\cmidrule(lr){6-10}	\cmidrule(lr){11-13}	\cmidrule(lr){14-16} %\cmidrule(lr){17-18}
	%\rowcolor{gray!40}
	{\tt s8}&
	Leptonica~\cite{Leptonica} &
	24.10 MB &
	Cmd.$^7$~\cite{CVE-s7} &
	C  &
	0 & 0 & 0 & 2  &  2 & 0 & 2 & 0 & \avgtotal{2}{2} & \avgtotal{0}{0} & \avgtotal{2}{2} & 2.4 & 12.1
	%& & &
	
	\\ 
	%\cmidrule(lr){1-5} 	\cmidrule(lr){6-10}	\cmidrule(lr){11-13}	\cmidrule(lr){14-16} %\cmidrule(lr){17-18}
	\rowcolor{gray!40}
	{\tt s9}&
	GNU-Patch~\cite{Patch} &
	4.96 MB &
	Cmd.$^7$~\cite{CVE-s8} &
	C  &
	0 & 0 & 0 & 7  &  2 & 0 & 1 & 6 & \avgtotal{1.00}{6} &  \avgtotal{0.86}{5} & \avgtotal{3.57}{1} & 4.9 & 10.2
	%& & &
	
	\\ 
	%\cmidrule(lr){1-5} 	\cmidrule(lr){6-10}	\cmidrule(lr){11-13}	\cmidrule(lr){14-16} %\cmidrule(lr){17-18}
	%\rowcolor{gray!40}
	{\tt s10}&
	Goahead~\cite{goahead} &
	18.20 MB &
	Cmd.$^7$~\cite{CVE-s9} &
	C  &
	0 & 0 & 0 & 1  &  1 & 0 & 1 & 0 & \avgtotal{1}{1} & \avgtotal{0}{0} & \avgtotal{1}{1} & 3 & 9
	%& & &
    
	\\ 
	%\cmidrule(lr){1-5} 	\cmidrule(lr){6-10}	\cmidrule(lr){11-13}	\cmidrule(lr){14-16} %\cmidrule(lr){17-18}
	\rowcolor{gray!40}
	{\tt s11}&
	LuCI~\cite{LuCI} &
	43.10 MB &
	Cmd.$^7$~\cite{CVE-s10} &
	C, Lua  &
	19 & 102 & 17& 2  &  52 & 19 & 37 & 84 & \avgtotal{2.24}{136} & \avgtotal{0}{0} & \avgtotal{1.96}{132} & 2.4 & 6.4
	% &  &  &
	
	\\ 
	%\cmidrule(lr){1-5} 	\cmidrule(lr){6-10}	\cmidrule(lr){11-13}	\cmidrule(lr){14-16} %\cmidrule(lr){17-18}

	%\rowcolor{gray!40}
	{\tt s12}&
	jison~\cite{jison} &
	1.25 MB &
	Cmd.$^7$~\cite{CVE-jison} &
	JS$^{\mathsection}$ &
	0 & 2 & 0 & 0  &  2 & 2 & 0 & 0 & \avgtotal{0}{0} & \avgtotal{0}{0} & \avgtotal{0}{0} & 1 & 3
	%& & &
	
	\\ 
	%\cmidrule(lr){1-5} 	\cmidrule(lr){6-10}	\cmidrule(lr){11-13}	\cmidrule(lr){14-16} %\cmidrule(lr){17-18}
	\rowcolor{gray!40}
	{\tt s13}&
	Kill-port~\cite{killportprocesses} &
	34.80 KB &
	Cmd.$^7$~\cite{CVE-s12} &
	JS$^{\mathsection}$ &
	0 & 5 & 0 & 0 &  2 & 5 & 0 & 0 & \avgtotal{0}{0} & \avgtotal{0}{0} & \avgtotal{0}{0} & 1 & 4.5
	%& & &
	
	\\ 
	%\cmidrule(lr){1-5} 	\cmidrule(lr){6-10}	\cmidrule(lr){11-13}	\cmidrule(lr){14-16} %\cmidrule(lr){17-18}
	%\rowcolor{gray!40}
	{\tt s14}&
	egg-scripts~\cite{eggscripts} &
	58.40 KB &
	Cmd.$^7$~\cite{CVE-s13} &
	JS$^{\mathsection}$ &
	2 & 1 & 0 & 0  &  3 & 3 & 0 & 0 & \avgtotal{0}{0} & \avgtotal{0}{0} & \avgtotal{0}{0} & 1 & 3.3
	%& & &
	
	\\ 
	%\cmidrule(lr){1-5} 	\cmidrule(lr){6-10}	\cmidrule(lr){11-13}	\cmidrule(lr){14-16} %\cmidrule(lr){17-18}
	\rowcolor{gray!40}
	{\tt s15}&
	node-df~\cite{node-df} &
	40.00 KB &
	Cmd.$^7$~\cite{CVE-node-df} &
	JS$^{\mathsection}$ &
	 0 & 1 & 0 & 0 & 1  & 0 & 1 & 0 & \avgtotal{1}{1} & \avgtotal{0}{0} & \avgtotal{1}{1} & 1 & 2
	%& & &
	
	\\ 
	%\cmidrule(lr){1-5} 	\cmidrule(lr){6-10}	\cmidrule(lr){11-13}	\cmidrule(lr){14-16} %\cmidrule(lr){17-18}
	%\rowcolor{gray!40}
	{\tt s16}&
	PM2~\cite{PM2} &
	4.42 MB &
	Cmd.$^7$~\cite{CVE-s15} &
	JS$^{\mathsection}$ &
	7 & 25 & 2 & 1  &  34 & 21 & 23 & 2 & \avgtotal{0.77}{21} & \avgtotal{0}{0} & \avgtotal{1.95}{31} & 1.3 & 5.3
	%& & &
	
	\\ 
	%\cmidrule(lr){1-5} 	\cmidrule(lr){6-10}	\cmidrule(lr){11-13}	\cmidrule(lr){14-16} %\cmidrule(lr){17-18}
	\rowcolor{gray!40}
	{\tt s17}&
	fs-git~\cite{fs-git} &
	130.00 KB &
	Cmd.$^7$~\cite{CVE-s16} &
	JS$^{\mathsection}$ &
	0 & 1 & 0 & 0  &  1 & 0 & 0 & 1 & \avgtotal{1}{1} & \avgtotal{0}{0} & \avgtotal{2}{2} & 2 & 4
	%& & &
	
	\\ 
	%\cmidrule(lr){1-5} 	\cmidrule(lr){6-10}	\cmidrule(lr){11-13}	\cmidrule(lr){14-16} %\cmidrule(lr){17-18}
	%\rowcolor{gray!40}
	{\tt s18}&
	Meta-git~\cite{meta-git} &
	262.00 KB &
	Cmd.$^7$~\cite{CVE-s17} &
	JS$^{\mathsection}$ &
	0 & 1 & 0 & 0  &  3 & 0 & 0 & 1 & \avgtotal{2}{2} & \avgtotal{0}{0} & \avgtotal{2}{2} & 1 & 5
	%& & &
	
	\\ 
	%\cmidrule(lr){1-5} 	\cmidrule(lr){6-10}	\cmidrule(lr){11-13}	\cmidrule(lr){14-16} %\cmidrule(lr){17-18}
	\rowcolor{gray!40}
	{\tt s19}&
	Listening Process~\cite{listening-processes} &
	131.00 KB &
	Cmd.$^7$~\cite{CVE-s18} &
	JS$^{\mathsection}$ &
	0 & 3 & 0 & 0  &  3 & 3 & 0 & 0 & \avgtotal{0}{0} & \avgtotal{0}{0} & \avgtotal{0}{0} & 1 & 3
	%& & &
	
	\\ 
	%\cmidrule(lr){1-5} 	\cmidrule(lr){6-10}	\cmidrule(lr){11-13}	\cmidrule(lr){14-16} %\cmidrule(lr){17-18}
	%\rowcolor{gray!40}
	{\tt s20}&
	NPM lsof~\cite{lsof} &
	18.00 KB &
	Cmd.$^7$~\cite{CVE-s19}  &
	JS$^{\mathsection}$ &
	0 & 3 & 0 & 0  &  3 & 3 & 0 & 0 & \avgtotal{0}{0} & \avgtotal{0}{0} & \avgtotal{0}{0} & 1 & 2
	%& & &
    
    \\ 
    %\cmidrule(lr){1-5} 	\cmidrule(lr){6-10}	\cmidrule(lr){11-13}	\cmidrule(lr){14-16} %\cmidrule(lr){17-18}	
	\rowcolor{gray!40}
	{\tt s21}&
	NPM opencv~\cite{opencv} &
	22.60 MB &
	Cmd.$^7$~\cite{CVE-s20} &
	JS$^{\mathsection}$ &
	1 & 2 & 0 & 0  &  3 & 3 & 0 & 0 & \avgtotal{0}{0} & \avgtotal{0}{0} & \avgtotal{0}{0} & 1 & 2.3
	%& & &
    
	\\ 
	%\cmidrule(lr){1-5} 	\cmidrule(lr){6-10}	\cmidrule(lr){11-13}	\cmidrule(lr){14-16} %\cmidrule(lr){17-18}	
	%\rowcolor{gray!40}
	{\tt s22}&
	logkitty~\cite{logkitty} &
	514.00 KB &
	Cmd.$^7$~\cite{CVE-logkitty} &
	JS$^{\mathsection}$ &
	0 & 2 & 0 & 0  & 2 & 0 & 0 & 2 & \avgtotal{2}{3} & \avgtotal{0}{0} & \avgtotal{2.5}{4} & 3 & 3
	%& & &
    
	\\ 
	%\cmidrule(lr){1-5} 	\cmidrule(lr){6-10}	\cmidrule(lr){11-13}	\cmidrule(lr){14-16} %\cmidrule(lr){17-18}	
	\rowcolor{gray!40}	
	{\tt s23}&
	gitpublish~\cite{git-publish} &
	32.00 KB &
	Cmd.$^7$~\cite{CVE-s22} &
	JS$^{\mathsection}$  &
	0 & 9 & 0 & 0  &  3 & 2 & 0 & 7 & \avgtotal{2}{8} & \avgtotal{0}{0} & \avgtotal{2}{8} & 1 & 5.7
	%& & &
    
	\\ 
	%\cmidrule(lr){1-5} 	\cmidrule(lr){6-10}	\cmidrule(lr){11-13}	\cmidrule(lr){14-16} %\cmidrule(lr){17-18}
	%\rowcolor{gray!40}	
	{\tt s24}&
	codecov~\cite{codecov} &
	290.00 KB &
	Cmd.$^7$~\cite{CVE-s23} &
	JS$^{\mathsection}$ &
	4 & 0 & 2 & 0  &  6 & 4 & 2 & 0 & \avgtotal{0.5}{3} & \avgtotal{0}{0} & \avgtotal{0.5}{3} & 1 & 7.4
	% & &  &

		\\ 
		%\cmidrule(lr){1-5} 	\cmidrule(lr){6-10}	\cmidrule(lr){11-13}	\cmidrule(lr){14-16} %\cmidrule(lr){17-18}
	\rowcolor{gray!40}
	{\tt s25}&
	pdfinfojs~\cite{pdfinfojs} &
	77.00 KB &
	Cmd.$^7$~\cite{CVE-s24} &
	JS$^{\mathsection}$ &
	0 & 3 & 0 & 0  &  3 & 3 & 0 & 0 & \avgtotal{0}{0} & \avgtotal{0}{0} & \avgtotal{0}{0} & 1 & 4.3
	%& & &
    
	\\ 
	%\cmidrule(lr){1-5} 	\cmidrule(lr){6-10}	\cmidrule(lr){11-13}	\cmidrule(lr){14-16} %\cmidrule(lr){17-18}
	%\rowcolor{gray!40}
	{\tt s26}&
	libnmap~\cite{libnmap} &
	157.00 KB &
	Cmd.$^7$~\cite{CVE-libnmap} &
	JS$^{\mathsection}$ &
	0 & 1 & 0 & 0  &  1 & 0 & 0 & 1 & \avgtotal{4}{4} & \avgtotal{1}{1} & \avgtotal{4}{4} & 3 & 4
	%& & &
    
	\\ 
	%\cmidrule(lr){1-5} 	\cmidrule(lr){6-10}	\cmidrule(lr){11-13}	\cmidrule(lr){14-16} %\cmidrule(lr){17-18}
	\rowcolor{gray!40}	
	{\tt s27}&
	pdf-image~\cite{PDF-image} &
	14.00 KB &
	Cmd.$^7$~\cite{CVE-pdfimage} &
	JS$^{\mathsection}$ &
	0 & 1 & 1 & 0  &  2 & 0 & 0 & 2 & \avgtotal{0}{2} & \avgtotal{0}{0} & \avgtotal{2}{4} & 2 & 7.5
	%& & &
	\\ 
	\bottomrule 
	\vspace{-0.5em}
	\\
	\multicolumn{21}{l}{1: Basic block. 2: Function. 3: Multiple Functions. 4: Local variable (Avg.). 5: Global/member variable (Avg.). 6: Functions (Avg.). 7: Shell Command Injection. 8: XXE Injection.} \\
	\multicolumn{21}{l}{ 9: SQL Injection. $\dagger$: PHP and XML. $\ddagger$: PHP, XML, and SQL. $*$: PHP and SQL.  ${\mathsection}$: JavaScript. $\alpha$: 4 FN (False negative) cases. $\beta$: 24 FN cases. $\gamma$: 3 FN cases. $\delta$: 2 FN cases.} \\
	\multicolumn{21}{l}{($\alpha,\beta, \gamma, \delta$): No FN cases when we apply the bidirectional analysis. FN cases are caused when only forward or backward analysis is applied alone.}
	%7: Execution prevented with ``Command not found''
	%\multicolumn{16}{l}{8: Execution prevented with ``File not found''. 9: Execution prevented with a time out error. 10: Average.}
	%\multicolumn{15}{l}{1: Execution prevented with an error ``Command not found''. 2: Execution prevented with an error ``File not found''. 3: Execution prevented with a time out error.} \\
\end{tabular}
}
\vspace{-1em}
\end{table*}

\noindent
{\bf Program Selection.}
%To evaluate the effectiveness of \sysname in preventing \updated{command}{input} injection attacks (Section~\ref{subsec:eval_perf}), 
We search publicly known \updated{command}{input} injection vulnerabilities (including SQL and XXE\updated{(XML external entity)}{} injections) in recent five years. Among them, we reproduced 27 vulnerabilities and used the vulnerable programs as shown in \autoref{table:selectedprograms}. Note that the versions of the evaluated programs can be found in Appendix~\ref{appendix:versions} (\autoref{table:versions}).
%shows the programs including names, sizes, languages written in, and brief descriptions. 
The selected programs are diverse, including popular programs such as WordPress~\cite{wordpress} and OpenCV~\cite{opencv}.
They are also written in diverse programming languages such as PHP, C/C++, Lua, and JavaScript. 
%LuCI~\cite{LuCI} is an Embedded Configuration Interface widely used in embedded devices. 
The programs and vulnerabilities\updated{ used in the evaluation}{} are on \cite{csr-tool}\updated{ (\autoref{table:vulnanddescription})}{}.

%\label{subsec:eval_perf}
\noindent
{\bf Input Selection.}
%We use the above 27 programs in \autoref{table:selectedprograms} to measure the performance of \sysname (Section~\ref{subsec:eval_perf}). 
To obtain realistic test input (or test data) that can cover diverse aspects of the program, we leverage publicly available input data sources. %\mw{added methods for test}
For instance, Leptonica~\cite{Leptonica} provides 278 test cases with 192 images. 
Other programs also have developer provided test cases: NPM-opencv~\cite{opencv}, fs-git~\cite{fs-git}, PM2~\cite{PM2} and codecov~\cite{codecov}. For the programs with less than 100 test cases, we extend them on different inputs with around 100 cases. 
%XXX and XXX use Mocha~\cite{Mocha}, a test automation framework, to manually write unit test cases for them. 
%
For the programs that accept PDFs (e.g., pdfinfojs), videos (e.g., Avideo-Encoder), and patches (e.g., GNU-Patch), we crawl more than 100 samples for each type from public websites~\cite{tpn-pdf, ted-video}.
%Some programs have standard input like \code{pdfinfojs} for pdf files, \code{Cool Video Gallery} for videos, \code{Gnu-Patch} for patch, \code{Ruby Resolv Module} for host name. So we generate inputs that are various both in size and contents to test the programs. 
%For the cases that have limited execution options. We use Mocha~\cite{Mocha}, a test framework, to manually write unit test cases for them. 
%Overall, we obtain at least 100 test cases for all programs selected for the performance evaluation.
% \YK{add later...}
To run the programs for the performance evaluation (Section~\ref{subsec:eval_perf}), we leverage Apache Jmeter~\cite{apachejmeter} and Selenium~\cite{selenium} for web applications. We also use Selenium scripts provided by \cite{sec19-selenium} to simulate requests and interactions for web services such as WordPress. 
OLPTBench~\cite{difallah2013oltp}, which aims to conduct extensive tests on relational database systems, is used to test diverse SQL queries. 
In addition, a large publicly available XML data-set (1026 MB total)~\cite{xmldata} is used.
We also include popular web servers~\cite{apache,lighttpd,OpenLightSpeed,Cherokee}, SQL engines~\cite{sqlite,Mysql}, and XML libraries~\cite{libxml,simplexml,libxmljs,luaexpat} in our evaluation. % to show the performance of \sysname in various aspects.

\subsection{Instrumentation Results and Correctness}
\label{subsec:instr}
\autoref{table:selectedprograms} presents the results of our instrumentation in detail.

\noindent
{\bf Statistics\updated{ (\# Instrumentation)}{}.}
The ``Const.'' and ``Dynamic'' columns  represent the number of completely constant commands\updated{ (e.g., `\code{\$cmd = {\color{blue}"rm *.bak"}}')}{}  and the number of dynamically composed commands with other values\updated{ (e.g., `\code{\$cmd = {\color{red}\$base\_dir}.{\color{blue}"/ping"}}')}{} respectively. 
There are three groups based on the number of variables involved in creating a command or query dynamically. 
The first group includes cases where 1$\sim$5 variables are involved, that are trivial to verify that they do not break benign functionalities. Most cases belong to this group. 
The second and third groups indicate 6$\sim$10 and more than 11 variables are involved respectively.  
We checked them all that they do not break benign functionalities. Examples and details can be found in Appendix~\ref{appendix:manual_analysis_instr_detail}.
%
%They are more complex to analyze. 
%We find that the number of cases decreases as the number of variables involved increases.  
%in which only the path name is constant but the arguments are from the users' inputs. 
The ``Sinks'' column represents the number of sink functions identified by \sysname. 
Note that while there are applications that require many instrumentations (e.g., 462 for WordPress), most of them are constants or dynamic cases with only a few variables are involved. 
%In other words, they are simple cases that do not make the instrumentation complicated.
%There are three exceptions: {\tt s6} (Pie Register), {\tt s11} (LuCI) and {\tt s16} (PM2).
%
WordPress is a content management system stores/retrieves contents from databases, PHPSHE is a website builder, and Pie Register is a user registration form service. 
LuCI is a web interface for configuring OpenWrt~\cite{openwrt} that runs various commands in nature. 
These programs include many SQL queries, leading to a large number of instrumentations.
However, patterns of queries in those programs  are simple and similar to each other.
Further, we analyze the dynamically composed commands.
Most cases are appending file names to base folders to compose paths and  adding table names in queries. 

\noindent
{\bf Correctness.}
We run test cases and analyze all the instrumented code to show the instrumented programs' correctness. 

-- \textit{Testing Instrumented Programs:}
    To empirically show that our instrumentation does not break the original functionalities, we run test cases that can cover instrumented code and other parts of the program code affected by the instrumentation. 
    We leverage test cases provided by developers of the target applications. If there are no provided test cases or test cases are not sufficient, we manually extend test cases to cover those. All the test cases are presented in \autoref{table:selectedprograms_cov}. % where the number of test cases we additionally created is shown in the ``Added'' column.
    In total, we run 15,916 test cases for the 27 programs, achieving the average code coverage of 78.17\%. For the code that is not covered by the test cases, we manually checked that they are not affected by our instrumentation. %More details can be found in \cite{csr-tool}.
    %we use real inputs and unit test cases to run the instrumented programs. For those applications not having test cases including AVideo-Encode, Pepperminty-Wiki, PHPSHE, Pie-register, and LuCI, we create test cases to simulate the realistic tasks. We also added test cases to the applications having unit test cases in the source code to improve the code coverage. Multiple code test and code coverage report tools are used in the experiments. We leverage PHPUnit~\cite{PHPUnit} and Xdebug~\cite{Xdebug} for PHP applications, Mocha~\cite{Mocha}, Jest~\cite{Jest} and nyc~\cite{nyc} for JavaScript applications, LuaCov~\cite{LuaCov} for Lua applications and Gcov~\cite{Gcov} for C/C++ applications. In those applications, there are functions and modules not supported by our test environment. For example, Leptonica, kill-port and PM2 have code to support operating systems other than Linux. LuCI has functions that need hardware support like LTE network. These parts of code are removed. The average code coverage is 78.17\%. Although there is code not covered by our test cases, we checked that those code is independent from our instrumentation. The details of the results are shown in Appendix[XXXXX].

-- \textit{Manual Analysis of Instrumentation:} 
We analyze the impact of our instrumentations and categorize them into three types: instrumentations that affect (1) a single basic block (the BB$^1$ column), (2) a single function (the Fn$^2$ column), and (3) multiple functions (the Fns$^3$ column).
The first category only affects statements within its basic block. 
Mostly, they are the cases where a constant string is instrumented and directly passed to a sink function. For this case, it is trivial to prove that it does not impact the correctness of the program as the impact of the instrumentation is contained within the current basic block.
%In this case, the instrumentation does not break the program's benign functionality, as long as our command target randomization scheme is correct. 
%Note that we do not observe errors in our command target randomization scheme during our evaluation.
For the second category (i.e., single function), the instrumented values are stored into local variables, but it does not affect other functions (i.e., they are not returned or passed to other functions). Hence, the impact of instrumentation is limited within the function. 
The last category (i.e., multiple functions) means that the instrumentation affects multiple functions because the instrumented value is stored to a variable shared between functions (e.g., global variable) or passed to other functions as arguments. 
We verify all the cases in the three categories that they do not break the original functionalities of the target programs by tracing dependencies caused by our instrumentations.
%Note that there are four programs containing instrumentations affecting many functions: WordPress, PHPSHE, Pie Register, and LuCI. 
Details with example code for the three categories are in Appendix~\ref{appendix:manual_analysis_instr_detail}.

\updated{{\it -- Affected Variables and Functions by Instrumentation:}}{}
We also inspect local and global variables and functions affected by the instrumentations.
In the next three columns, the average number of variables/functions affected by each instrumentation is presented, followed by the total number of variables/functions affected in the entire program.
The average number of variables and functions affected per instrumentation is not large: less than 12 local variables, 2 global variables, and 8 functions. 
%We use static analysis to trace and verify the complexity of the instrumentation and its impact automatically.
We verify all of them and \sysname does not break the benign functionality.

\vspace{-1em}
\subsection{Effectiveness}
\label{subsec:eval_attack}

\subsubsection{Against PoC (Proof of Concept) Exploits}
\label{subsec:poc_eval}
We reproduce 27 PoC exploits on \sysname instrumented programs, as shown in \autoref{table:selectedprograms}.
%shows the v we have reproduced on vanilla applications.
%Then, we apply \sysname to the applications and rerun the exploits to check whether \sysname successfully prevents the attacks.
%The first column shows the identifier (ID) of the programs. 
%\textsl{Vulnerability} column shows the reported vulnerability IDs or names in public (e.g., CVE and Hackerone). %and functions or elements (XML element for XXE attack in {\tt s4})  respectively.
%The fourth column shows the functions (or XML element for XXE attack in {\tt s4}) that are exploited. 
The ``Vulnerability'' column shows attack type (e.g., Command injection, XXE injection, and SQL injection) with citations.
%\textsl{Result} column shows the results when the PoC exploits are launched on the \sysname protected programs.
All the PoC (Proof of Concept) attacks are successful in the vanilla versions, while prevented in the \sysname protected programs.
%As shown in the last column, there are three different results: 
%{\it Command not found}, {\it File not found}, and {\it Time out}.
%The first two essentially mean that the exploit failed to execute the command as the command is already randomized. Time out happens because the program waits for the completion of the command while the command will never be completed as it fails to execute.
%to Meng, Also this should be fixed with superscript in table 2.
 %, meaning that it is essentially the command not found error.
%command execution is finished, which will not be satisfied if the program fails to execute the command. 
%Hence, it is essentially equivalent to the command not found error. 
%\YK{Why crash on exeSync?} \mw{I did not give the correct input for the program. There must be a process to kill, so the vulnerability can be triggered. But I used the wrong PID, so it crahsed} \YK{is this your fault? need to run exp again?}

%\vspace{-1em}
\subsubsection{Against Automated Vulnerability Discovery Tools} 
To see whether \sysname can prevent \emph{diverse malicious commands}, in addition to the tested CVEs in Section~\ref{subsec:poc_eval}, we leverage three automated vulnerability discovery tools, Commix~\cite{commix}, sqlmap~\cite{sqlmap}, and xcat~\cite{xcat}, to test a diverse set of known malicious commands.
We launch 102 command injection attacks, 97 SQL injection attacks, and 24 XXE injection attacks, leveraging the three tools.
They essentially brute-force the target programs' inputs using the known malicious commands. Then, they check whether it is vulnerable to command injection attacks. 
The result shows that \sysname successfully prevents all 223 tested attacks. Details are in Appendix~\ref{appendix:vulnerability_discovery_tools}.

%Then, we test the identified payloads to show the effectiveness of \sysname.
\revised{
\updated{}{
\subsubsection{Bidirectional Analysis Compared to Backward and Forward Analysis}

We apply forward and backward data flow analysis alone to the programs and presented the average length of dependency chain obtained from each analysis in the last two columns of \autoref{table:selectedprograms}. 
We observe that data-flow analysis accuracy, including forward and backward analyses, decreases as the data dependency chain's length becomes larger than 10 in general, causing false negatives. We find such cases in WordPress~\cite{wordpress}, Activity Monitor~\cite{PlainviewActivityMonitor}, and Pie Register~\cite{pieregister}, marked with $\alpha$, $\beta$, $\gamma$, and $\delta$ with the red cell background color. 
%When we use forward or backward analysis alone for them, we find false negative cases from the data flow analysis results. 
%Note that with our bidirectional analysis, we observe no false negative. 
We manually verified that all the results from our analysis are true-positives. In particular, we run other static/dynamic taint-analysis techniques~\cite{taintless,psalm,pecltaint} and manually verify that the dependencies identified by the existing techniques but not by ours are false-positives. 
Appendix~\ref{appendix:bidirectional_effectiveness} and \ref{appendix:bidir_accuracy} provide more details including examples and accuracy of the bidirectional analysis.
}
}

\vspace{-1em}
\subsection{Performance Evaluation}
\label{subsec:eval_perf}
\begin{figure}[h]
    \centering
    %\vspace{-0.5em}
    \includegraphics[width=1\columnwidth]{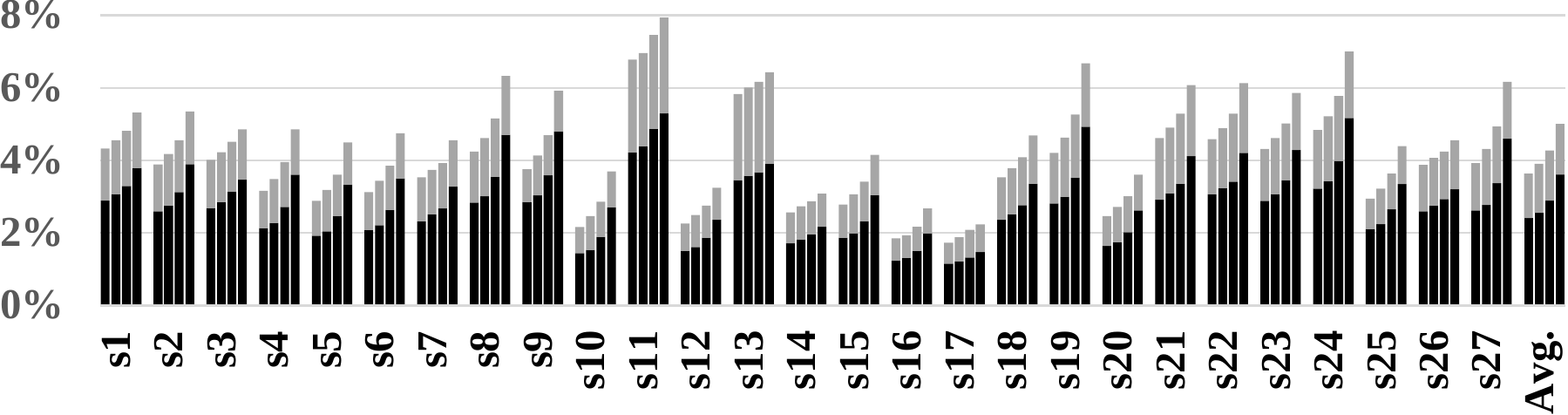}
    \vspace{-2.5em}
    \caption{\revised{Runtime Overhead}}
    \vspace{-1em}

    \label{fig:overhead}
\end{figure}

\vspace{-0.5em}
\noindent
{\bf Runtime Overhead (Overhead: $\approx$5\%).} 
We measure the runtime overhead of \sysname on the 27 programs\updated{}{ in Table~\ref{table:selectedprograms} as shown in \autoref{fig:overhead}}.
\updated{}{Note that each application has 4 measures as we use 4 different randomization schemes mapping 1 byte to 1, 2, 4, and 8 bytes}.
\updated{}{In each bar, the bottom black portion represents the overhead caused by creating randomization tables, including those for rerandomization, while the top gray portion is the overhead from the computations for randomization.} 
For each program, we use 100 typical benign test inputs that cover instrumented statements. % to evaluate performance overhead.
For each input, we run ten times and take the average. 
%We modify the sample programs provided by the developer to make sure all the sink APIs are executed in the test. 
%\updated{To obtain realistic inputs,}{For data files used for inputs,} we use non-trivial size data for inputs. For example, we use a large XML file whose size is around 100MB obtained from a publicly available XML data-set maintained by the University of Washington~\cite{xmldata} and we obtain 100 large graphs from \code{Leptonica}~\cite{Leptonica}.
\updated{\autoref{fig:Perf_all} shows the result.}{} 
The average overhead is 3.64\%\updated{}{, 3.91\%, 4.28\% and 5.01\% for 1, 2, 4, and 8 bytes randomization schemes respectively}. 
%\CJ{10) [8] (\#E) Evaluation of the randomization approach in different applications (e.g., overhead from managing randomization tables).}
% s1 s10 s12 s23
%\updated{Among the programs,}{} \code{s11} (LuCI) has slightly higher overhead than others, i.e., 7\%$\sim$8\%. 
%Our manual inspection reveals that they execute more commands (e.g., 150 commands while a typical program executes 50 commands) during the evaluation. %\updated{}{Managing larger randomization table is the main contribution of the increasement of the run-time overhead in different kinds of schemes. Storage overhead for one single randomization table is 54 bytes, 106 bytes, 209 bytes and 417 bytes respectively when \sysname is configured to randomize one character to 1, 2, 4, or 8 characters.}

\noindent
{\bf \updated{Overhead on}{Throughput of} Full Stack Web \updated{Services (Overhead: 3.69\%)}{Servers (WordPress)}.}
We measure the overhead on \updated{}{throughput of} full-stack web services to understand the performance overhead of realistic deployment of \sysname.
\updated{Specifically,}{}We applied \sysname to four different web servers: Apache 2.4.41, Lighttpd 1.4.55, Cherokee 1.2.102, and Openlightspeed 1.5.11. 
We also apply \sysname to SQLite 3.31.0, PHP 7.2, and WordPress 4.9.8, along with the listed vulnerable plugins. %We also applied \sysname to The instrumented database used in this part is Sqilte version 3.31.0. 
\updated{To test the web services, }{}Apache Jmeter~\cite{apachejmeter} is used to request 10,000 concurrent webpages, covering various functionalities of WordPress, including posting blogs, changing themes, and activating/deactivating/configuring plugins. 
\updated{}{The average overhead on throughput is 3.69\% (4.33\%, 3.76\%, 3.18\%, and 3.47\% overhead on Apache, Lighttpd, OpenLightSpeed, and Cherokee respectively).
}
%for the plugins that need extra inputs. For example, the input of the Advanced XML Reader is a 100 MB XML file, and the inputs of Cool Video Gallery are 100 different short videos. 

% \input{tex/table_server_overhead}

\noindent
{\bf Overhead on Database Engines and XML Parsers.} 
%To understand the performance overhead on database engines, 
We apply \sysname to SQLite and MySQL and run various SQL queries using data-sets from OLTP-Bench~\cite{difallah2013oltp}. % to run various SQL queries.
The result shows that the overhead with SQLite is 4.9\% and MySQL is 5.3\%. 
We also measure the overhead on four XML parsers~\cite{libxml, simplexml, libxmljs,luaexpat}. % using large datasets~\cite{xmldata}. % (i.e., 1GB in total). % on four XML parsers %Libxml~\cite{libxml}, SimpleXML~\cite{simplexml}, libxmljs~\cite{libxmljs} and LuaExpat~\cite{luaexpat}, 
%with instrumentations. 
The average overhead is 1.4\% (Details in Appendix~\ref{appendix:microbenchmarks_xml}).

%Table~\ref{table:OverheadDB} shows the result. Overall, the overhead is less than 5.5\%. SQLite is slightly faster than MySQL due to the complexity of MySQL that requires complicated instrumentations.

\revised{
\noindent
\updated{}{\textbf{Memory Overhead.} \sysname needs to maintain randomization tables on memory during execution. 
Memory overhead for one randomization table is 54 bytes, 106 bytes, 209 bytes and 417 bytes respectively when \sysname is configured to randomize 1 to 1, 2, 4, and 8 bytes. 
At runtime, the memory overhead is $\sim$1MB on large programs such as WordPress, with 8 bytes randomization scheme.
}
}
%\begin{figure}[h]
 %   \centering
 %   \includegraphics[width=0.7\columnwidth]{fig%/db_perf.pdf}
%     \caption{Overhead of \sysname on Database %Engines}
%     \label{fig:perf_db}
%\end{figure}

%\noindent
%{\bf Microbenchmarks.}
%We conduct additional microbenchmarks on each instrumented APIs in PHP, C, Lua, and JavaScript.
%We break down the APIs into three different types: Command, XML, and SQL APIs for APIs that execute commands, XML queries, and SQL queries, respectively.
%
%The results show that the overhead is reasonable (from 1.26\% to 5.71\%). %We present details in Appendix~\ref{appendix:microbenchmarks_apis}.
%In addition, we conduct microbenchmarks on the instrumented OpenWrt firmware's uHTTPd~\cite{uHTTPd} and LuCI web configuration interface~\cite{LuCI}. The average overhead is less than 6\%~\cite{csr-tool}. % (Details in Appendix~\ref{appendix:embedded_devices_perf}).  
%More details can be found on \cite{csr-tool}.

%%%%%%%%%%%%%%%%%%%%%%%%%%%%%%
%\vspace{-1em}
\subsection{Case Study}
\label{subsec:casestudy}
%\vspace{-0.5em}
\updated{We present two case studies in this section and two additional case studies in Appendix~\ref{appendix:case_study}.}{}

%\vspace{-1em}
\subsubsection{Advanced SQL Injections Exploiting Parsers}
\label{subsubsec:advanced_sql_injection}
We present a few sophisticated injection attacks exploiting flaws of 11 popular parsers~\cite{andialbrecht/sqlparse, xwb1989/sqlparser, client9/libinjection, greenlion/PHP-SQL-Parser, moz-sql-parser, hyrise/sql-parser, alibaba/nquery, phpmyadmin/sql-parser, flora-sql-parser, node-sql-parser, druid-sql-parser}, showing the weaknesses of the parser-based for randomization approaches~\cite{sqlrand, autorand, diglossia}.
All of the cases are successfully prevented by \sysname, demonstrating the effectiveness of \sysname.
%
%\autoref{fig:sqlinjection_examples} shows 3 types of SQL injection attacks (A1, A2, and A3 for Attack 1, 2, and 3 respectively) that existing parsers fail to recognize.
%Note that existing randomization techniques such as SQLRand~\cite{sqlrand} and AutoRand~\cite{autorand} use parsers to randomize and derandomize SQL queries (keywords). 
%Hence, if parsers fail to recognize certain structures, keywords in the structures are not randomized, and attackers leverage them to inject malicious queries.
%
%\sysname  \emph{handles (i.e., prevents) all of them}.

\begin{figure}[h]
    \centering
    %\vspace{-0.5em}
    \includegraphics[width=0.95\columnwidth]{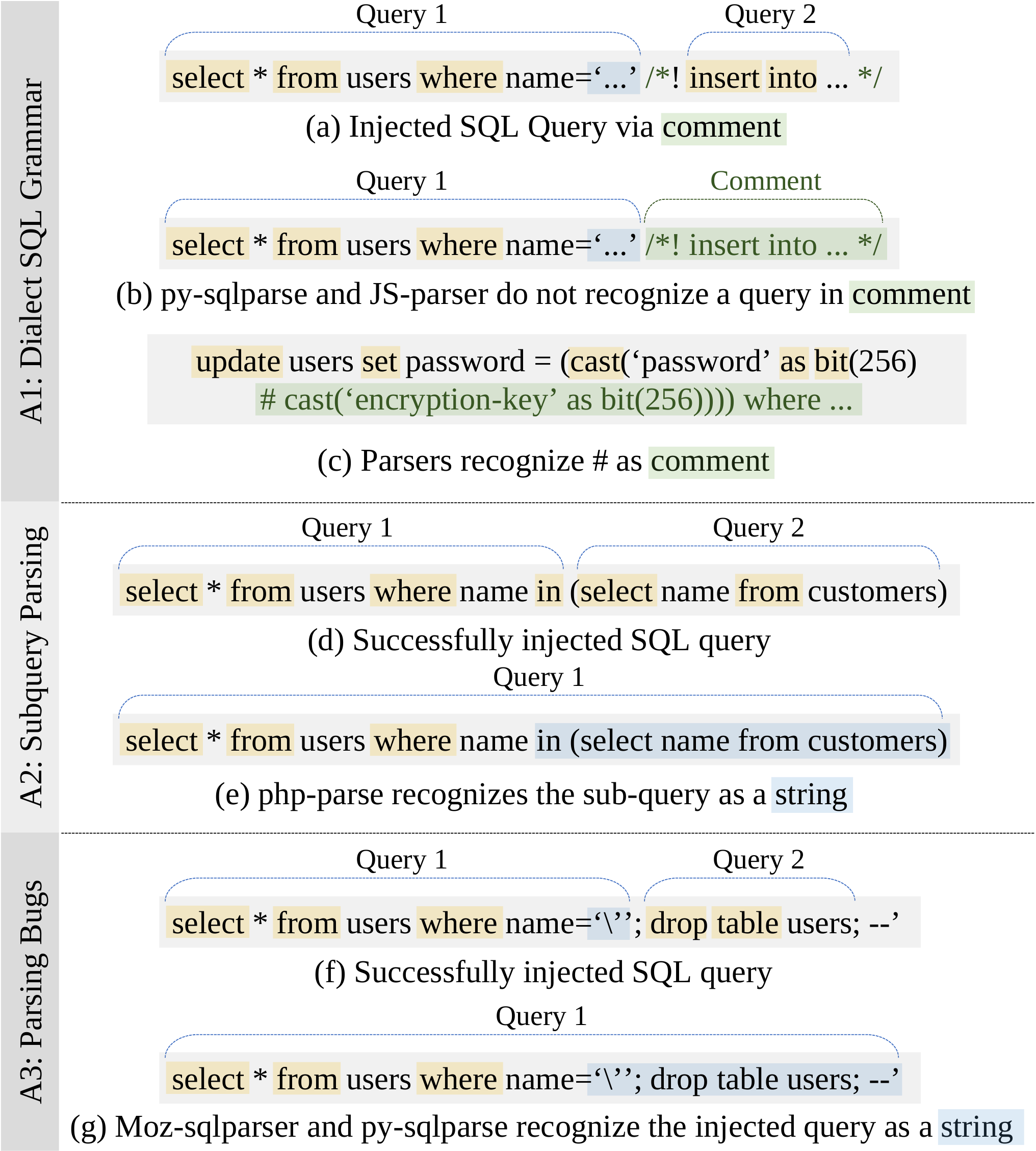}
    \vspace{-1em}
     \caption{SQL Injections that Parsers Fail to Recognize (Yellow: keywords, Blue: strings)}
    \label{fig:sqlinjection_examples}
     %\vspace{-1em}
\end{figure}

\noindent
{\bf A1: Dialect SQL Grammar ((a), (b), and (c)).} 
The dialect SQL attack shows that using a parser is an insecure design choice of the existing techniques. 
For instance, MySQL supports a SQL dialect: if a query in a comment starts with ``\code{/*!}'', it can be executed, as shown in \autoref{fig:sqlinjection_examples}-(a).
However, many parsers do not support this dialect.
An attacker can inject a malicious payload inside the comment, exploiting parsers that cannot recognize queries in a comment.
We confirmed py-sqlparse~\cite{andialbrecht/sqlparse} and JS-parser~\cite{node-sql-parser} fail to recognize injected queries in a comment as shown in \autoref{fig:sqlinjection_examples}-(b).
In addition, as shown in \autoref{fig:sqlinjection_examples}-(c), PostgreSQL~\cite{PostgreSQL} considers the `\#' symbol as an XOR operator, while others typically consider it as a single line comment operator. 
An attacker can also inject a malicious query with `\#'. 
Note that some techniques may automatically remove queries after `\#', removing injected queries. However, this will break benign queries using \# is an XOR operator as shown in as shown in \autoref{fig:sqlinjection_examples}-(c): doing a simple XOR encryption on a password.

\noindent
{\bf A2: Sub-query Parsing Error ((d) and (e)).} 
Attackers can inject malicious queries as a subquery to exploit approaches relying on parsers that cannot parse sub-queries correctly.
For instance, \autoref{fig:sqlinjection_examples}-(d) shows a SQL statement including two queries where the second query is a sub-query. 
As shown in \autoref{fig:sqlinjection_examples}-(e), php-parse~\cite{phpmyadmin/sql-parser} parses the entire sub-query as a string. %Hence, techniques using php-parse cannot prevent attacks that inject a malicious query into the sub-query.
%Highlighted terms are the keywords (to be randomized and derandomized by a parser).
Note that we present a specific case study for this attack type in Section~\ref{subsec:comparison_existing}.

\noindent
{\bf A3: String Parsing Error ((f) and (g)).} 
Moz-sqlparser~\cite{moz-sql-parser} and py-sqlparse~\cite{andialbrecht/sqlparse} have a bug in parsing a string~\cite{sqlparsebug}, allowing injected queries to be considered as a string that is not a randomization target.
For example, \autoref{fig:sqlinjection_examples}-(f) shows two SQL queries where the Query 2 is an injected malicious query. 
\cite{moz-sql-parser, andialbrecht/sqlparse} mistakenly consider the entire second query as a part of a string (blue marked).
%Note that if previous techniques use such parsers, they will not be able to prevent the injected query.

%\autoref{fig:sqlinjection_examples}-(f) shows two SQL queries where the Query 2 is an injected malicious query. 
%Moz-sqlparser~\cite{moz-sql-parser} and py-sqlparse~\cite{andialbrecht/sqlparse} fail to recognize the second query, as shown in \autoref{fig:sqlinjection_examples}-(g).
%This is because of a bug in parsing a string~\cite{sqlparsebug}, and they consider the entire second query as a part of a string (as highlighted with blue). Note that if previous techniques use such parsers, they will not be able to prevent the injected query.

%\vspace{-1em}
\revised{
\subsubsection{Comparison with Existing Techniques}
\label{subsec:comparison_existing}
We compare \sysname with two state-of-the-art techniques~\cite{diglossia, sqlrand-llvm}.

\begin{figure}[h]
    \centering  
    %\vspace{-1em}
     \includegraphics[width=1.0\columnwidth]{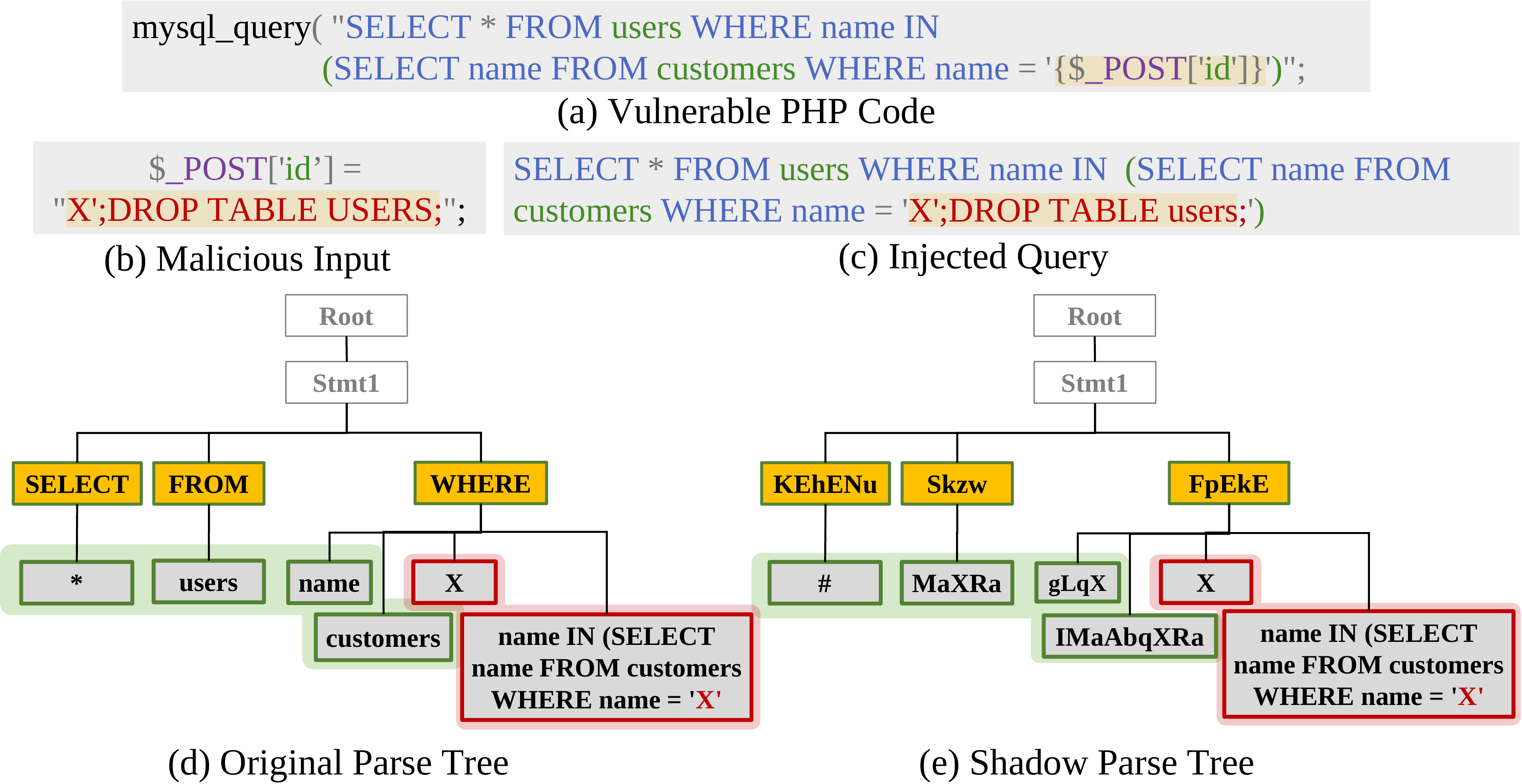}
     \vspace{-2em}
     \caption{Diglossia with php-parse~\cite{phpmyadmin/sql-parser}}
     \vspace{-1em}
    \label{fig:diglossia_failure_2} 
\end{figure}

\noindent
\textbf{Diglossia~\cite{diglossia} vs \sysname.}
%Diglossia uses an additional shadow parser that uses a different language from the original parser. % to detect injected code.
Diglossia runs two parsers, the original parser and the shadow parser, together.
The shadow parser is created to use a different language than the original parser and its input is obtained by translating the original input into the other language. 
At runtime, it obtains two parse trees from the parsers.
Different nodes between the trees indicate the parts originated from untrusted sources. 
If identical nodes are representing keywords (not strings/numbers), it detects an injection attack.
In this experiment, we implement our own version of Diglossia using php-parse~\cite{phpmyadmin/sql-parser} and \sysname's randomization scheme for the translation, since Diglossia does not provide its source code.

\autoref{fig:diglossia_failure_2}-(a) shows a vulnerable PHP code.
Given the malicious input shown in \autoref{fig:diglossia_failure_2}-(b), the malicious query is injected as shown in \autoref{fig:diglossia_failure_2}-(c).
As explained in Section~\ref{subsubsec:advanced_sql_injection}, the parser failed to parse the subquery after the \code{IN} keyword, resulting in an incorrect tree as shown in \autoref{fig:diglossia_failure_2}-(d). 
The last two children nodes (with red borders) of \code{WHERE} are unknown type nodes. 
When the malicious input is injected, both parse trees have the injected query as unknown nodes, resulting in a broken trees. As a result, it failed to recognize injected query. 
%Note that `\code{X}' appears twice in \autoref{fig:diglossia_failure_2}-(d).
%In other words, 
Note that \autoref{fig:diglossia_failure_2}-(d) and (e) show that they have identical nodes, marked with red borders. However, they are considered as literal nodes, hence not considered as an injected code. 
Worse, while the parser fails to process the query, it does not show error messages but silently suppresses the errors, missing the opportunities to detect the attack.
%To this end, the injected query is allowed. 
On the other hand, \sysname successfully prevents the injected SQL query \code{DROP TABLE users} from being executed.
Note that the performance of \sysname (about 5\%) is slightly better than and Diglossia (7.54\%).

\noindent
\textbf{sqlrand-llvm~\cite{sqlrand-llvm} vs \sysname.}
sqlrand-llvm~\cite{sqlrand-llvm} is an implementation of SQLRand using LLVM. %While this aims to implement SQLRand, there are a few different design choices made by sqlrand-llvm. 
It hooks \code{mysql\_query()} to tokenize a randomized input query and compare each token with a list of randomized SQL keywords. % via a string operation. 
It then derandomizes the matched tokens and then pass the deranomized query to the SQL engine.

\begin{figure}[ht]
    \centering
    %\vspace{-0.5em}
    \includegraphics[width=0.9\columnwidth]{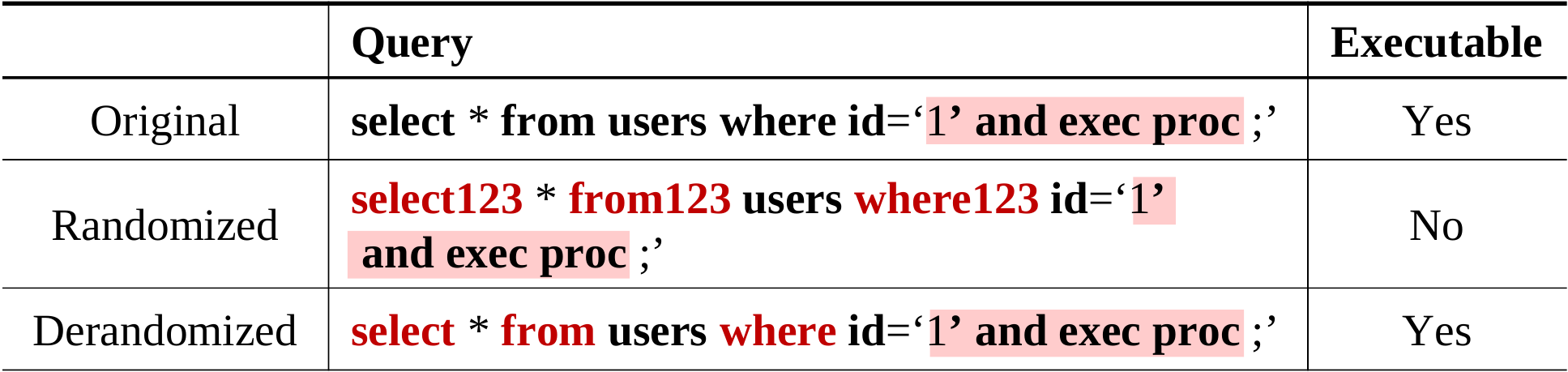}
    \vspace{-1em}
     \caption{SQL Injection Example with sqlrand-llvm~\cite{sqlrand-llvm}}
     \vspace{-1em}
     \label{fig:sqlrand-injected}
\end{figure}

{\it 1) SQL Injection with New Keywords:} 
sqlrand-llvm maintains a list of known SQL keywords and another list of randomized keywords. 
If a query with a keyword that is not included in the list are injected, it cannot prevent.
For instance, as shown in \autoref{fig:sqlrand-injected}, SQL keywords in the original query are randomized by appending \code{123} to the keywords.
During the derandomization process, if it encounters a SQL keyword that is not randomized, it considers the keyword is injected. 
However, it does not support \code{exec} and \code{proc} keywords according to the sqlrand-llvm's source code~\cite{sqlrand-llvm}.
%However, for a keyword that is not included in the list of known SQL keywords, it allows the query.
As shown in \autoref{fig:sqlrand-injected}, \code{exec} and \code{proc} in the injected query (highlighted) are not detected.
Note that \code{exec proc} can execute a stored procedure called \code{proc}.
\sysname prevents the attack with a similar performance: \sysname (5\%) and sqlrand-llvm (4.13\%).
%Since \sysname does not leverage a list of known SQL keywords, all terms go through the dual randomization scheme, \updated{resulting in randomized injected SQL query (i.e., \code{\color{red} ; hshp joep})}{breaking injected SQL queries}.

{\it 2) Breaking Benign Queries:} 
sqlrand-llvm uses \code{strtok()} to randomize/derandomize keywords even if they are a part of a string.
This results in an error if a string contains a SQL keyword. 
Consider a query ``\code{select * from users where name=`\$name'}'', where the value of \code{\$name} is `\code{grant}'. Its randomized query is ``\code{select123 * from123 users where123 name=`grant'}''. 
The value \code{grant} is from the user at runtime, hence not randomized. Unfortunately, \code{grant} is one of the known SQL keywords used in sqlrand-llvm, meaning that it will detect an injection attack (false positive) because the \code{grant} is not randomized.
\sysname does not have this issue as it does not randomize string type values.
}

\vspace{-0.5em}
\section{Discussion}
\label{sec:discussion}

\noindent
{\bf Prepared Statements.}
Prepared statements~\cite{prepared_statements} aim to prevent SQL injections by separating input data from a SQL query during the query construction.
While effective, they have limitations. First, some SQL keywords are not supported in prepared statements such as \code{PASSWORD} and \code{DESC} (5 more in Appendix~\ref{appendix:enc_sql_statements}). Second, changing existing SQL queries to prepared statements requires manual effort. 
\revised{Note that we manually check 866 SQL queries from all our target programs, and none of them is a prepared statement, showing the needs of \sysname in practice  (Appendix~\ref{appendix:enc_sql_statements}). %Third, SQL injections with prepared statements are still possible~\cite{phpdelusions_net}. 
}
%Third, they may incur higher overhead. 

\revised{
\noindent
{\bf Memory Disclosure on Randomization Records.}
\updated{After \sysname randomizes subsystems, it stores the used keys so that a randomized command name can be derandomized later.}{\sysname maintains randomization records that contain previously used randomization schemes.}
\updated{While the randomization record is deleted soon after a command execution API is called, there is still a short time window that the records exist on the memory.}{}
Attackers who can leak the memory pages containing the records may obtain \sysname's previously used keys. 
However, \sysname chooses a new randomization key on every new input. Hence, knowing previous randomization keys does not help in launching subsequent attacks.  
%\updated{xx}{}
%\CJ{2) [2.1] (\#E) How easy to evade detection (e.g., learning attacker)?}
%Note that in command injection attacks, it is difficult to overwrite already injected commands.
%
Also, existing memory protection techniques~\cite{intel-sgx} can be used to protect the records. 

\noindent 
{\bf Limitations.}
    When a target application is updated, one needs to run \sysname to analyze and instrument the updated target application. 
    Typically, this simply requires re-running \sysname on the updated application. 
    However, if an update significantly changes program code relevant to the trusted commands, it requires manual efforts to redefine the trusted command specifications. Further, we analyze updates of 42 applications, including popular programs, to check whether updates in practice lead to changes in trusted command specifications. The results show that they do not change trusted command specifications. Details are in Appendix~\ref{appendix:software_update_impact}. 
    If an application runs a completely dynamic command (e.g., \code{system(``\$\_GET[`cmd']'')}, \sysname blocks it and notifies users to fix the program. % (e.g., using constant commands selected by inputs can be a fix).
    %\CJ{@MW, fix XXX here}
    %\mw{Once the protection is applied, if a new module or plugin is added to the program, the protection of \sysname will be broken. We need to re-apply \sysname to the target application.}}\CJ{14) limitation}
}
%\noindent
\updated{{\bf Additional Discussions.}
In Appendix~\ref{appendix:additional_discussion}, we present more discussions on an alternative approach to \sysname and preventing the command injections with multiple commands.
}{}
%\vspace{-1em}
\section{Related Work}
\label{sec:related}

\noindent
{\bf Runtime Protection of Web Application.}
There have been many researchers that have proposed runtime protection systems against command and SQL injection attacks~\cite{nguyen-tuong-sql,  Haldar_2005, chin_2009, csse, sqlcheck, sqlrand, halfond06fse, halfond_wasp, sekar_ndss, AMNESIA, CANDID}. 
Taint tracking techniques track untrusted user inputs in server-side applications at runtime~\cite{nguyen-tuong-sql,  Haldar_2005, chin_2009, csse, sqlcheck}.
\cite{sekar_ndss,  AMNESIA} leverage static analysis to infer possible benign commands and use them to detect injection attacks. 
CANDID~\cite{CANDID}  employs dynamic analysis to extract and model an accurate structure of SQL queries. % so that they can be used for malicious queries.
\cite{halfond06fse} proposes positive tainting that dynamically tracks trusted inputs.  Unlike them, \sysname focuses on randomizing trusted commands, which is more lightweight than existing approaches (e.g., up to 19\% overhead in \cite{halfond06fse}).
Diglossia~\cite{diglossia} proposes a dual parsing technique that uses different languages during the parsing to detect injected SQL queries. However, it relies on parsers which can be exploited as shown in Sections~\ref{subsubsec:advanced_sql_injection} and \ref{subsec:comparison_existing}. 

Among the existing approaches, SQLRand~\cite{sqlrand, autorand} is the closest work to our approach. It randomizes SQL keywords and uses a proxy that can parse and derandomize the randomized SQL statements. %It then passes the derandomized SQL statements to the existing database engines.
%
%{\color{red}
Compared to SQLRand, \sysname does not rely on parsers which can be attacked and exploited as presented in Section~\ref{subsubsec:advanced_sql_injection}. 
%that require more generic system design, while SQLRand only prevents SQL injection. 
%Specifically, SQLRand does not consider commands defined by configurations files from users, which are commonly observed in command injection attacks. 
%Note that we also analyzed a version of SQLRand, sqlrand-llvm~\cite{sqlrand-llvm}, to show \sysname outperforms it (Section~\ref{subsec:comparison_existing}). 
%
There are also randomization based techniques such as Instruction Set Randomization~\cite{isr, isr2, isr3}. 
While sharing the randomization idea, \sysname's design provides solutions for preventing advanced attacks exploiting ambiguous grammars~\cite{sql_dialect}, as shown in Section~\ref{subsubsec:advanced_sql_injection}.
\cite{isr2} randomizes a programming language, leveraging a similar method to SQLRand. However, it is vulnerable to attacks exploiting language specification changes across different versions. %, which can be mitigated by our design. We leave this as future work.

% because it leverages a scanner (which is grammar agnostic) and the dual randomization scheme.
%\sysname supports dynamic randomization meaning that our randomization scheme is changing at runtime while  SQLRand's randomization is static. 
%AutoRand: https://dl.acm.org/doi/10.1007/978-3-319-40667-1_3 
% Java Bytecode, only for java
% https://link.springer.com/chapter/10.1007%2F978-3-319-40667-1_3
%}

\noindent
{\bf Security Analysis of Web Applications\updated{}{/Randomization}.}
Researchers have proposed various techniques to analyze vulnerabilities in web applications~\cite{webssari, xie_aiken, pixy, wassermann_2007,wassermann_2008,minamide_2005, noxes, saner_2008}. 
\cite{webssari} uses static analysis to identify vulnerabilities in PHP applications. 
Xie et al.~\cite{xie_aiken} propose a symbolic execution based program analysis technique to find SQL injection vulnerabilities. %Pixy~\cite{pixy} is an open source vulnerability analysis tool for PHP applications. 
String-taint analysis~\cite{wassermann_2007,wassermann_2008,minamide_2005} tracks untrusted substrings from user inputs to prevent information leak attacks.
%make sure no trusted scripts can be included in SQL queries and web pages generated by PHP programs.
\cite{saner_2008} combines dynamic and static analysis to find vulnerabilities in input sanitizers.
\sysname also uses static taint analysis and data flow analysis. % for randomizing commands and SQL queries. 
\updated{}{\cite{ahmed2020} studies the impact of timing of rerandomization. \sysname rerandomizes subsystems per input event, following the paper's recommendation.}

\noindent
{\bf Security Testing for Web Applications.}
Security testing aims to identify inputs that can expose input validation vulnerabilities in web applications~\cite{survey_bau, huang_2003, mcallister_2008, doupe_2012, martin_2008, kiezun_2009, saxena_2010a, saxena_2010b, saner_2008}. 
\cite{huang_2003} is a pioneer of web application testing by injecting XSS and SQL attacks. Mcallister et al.~\cite{mcallister_2008} propose a guided and stateful fuzzing technique to improve the performance. % to speed up the vulnerability searching.
%
%There are many techniques that enhance the effectiveness and efficiency of security testing.
Doup\'{e} et al.~\cite{doupe_2012} propose incrementally building a state machine during crawling to understand the internal structure of the web applications for better web application fuzzing. % the web application better.
To enhance input generation efficiency, Martin et al.~\cite{martin_2008} leverage model checking and static analysis, ARDILLA~\cite{kiezun_2009} applies symbolic execution, and Saxena et al.~\cite{saxena_2010a, saxena_2010b} use both dynamic taint analysis and symbolic execution for input mutation space pruning.
\updated{}{\cite{10.1145/3319535.3363195} systematically measures security issues in the payment card industry's webservices.}
\sysname aims to provide runtime protection. They are orthogonal to \sysname and are complementary. 
\section{Conclusion}
\label{sec:conclusion}
\vspace{-0.5em}
%Though many techniques have been applied to prevent the command injections, the attackers can still execute malicious commands by leveraging the advance-with-time methods or bugs in the software. 
In this paper, we introduce \sysname, a randomization based \updated{command}{input} injection prevention technique. \sysname is more robust than state-of-the-art randomization techniques\updated{, preventing advanced attacks exploiting subtle language dialects}{}.
Our extensive evaluation results show that \sysname successfully prevents advanced attacks with low overhead (\updated{3.69\% on web servers and 5.83\% on embedded devices}{$<$4\%}).  We release our tool's source code and result to public~\cite{csr-tool}.

\begin{acks}
  We thank the anonymous referees and our shepherd 
Giancarlo Pellegrino for their constructive feedback. The authors gratefully acknowledge the support of NSF under awards 1916499, 1908021, and 1850392. Any opinions, findings, and conclusions or recommendations expressed in this material are those of the authors and do not necessarily reflect the views of the sponsor.
\end{acks}

%\clearpage
\bibliographystyle{ACM-Reference-Format}
\bibliography{paper}

%\clearpage
\section{Appendix}
\subsection{Supplementary Text and Experiment}
\subsubsection{Sink Functions}
\label{appendix:sinkfunctions}
In addition to \autoref{table:sinkfunctions}, \autoref{table:additional_sink_functions} provides additional sink functions for XML and database subsystems.

\begin{table}[h]
	\centering
	\caption{Sink Functions}
	\vspace{-1em}
	\label{table:additional_sink_functions}
	
\resizebox{1.0\columnwidth}{!}{%	

\begin{tabular}{l c c} % 4 columns
	\toprule
	{\bf Sink Functions} & 
	{\bf Subsystem} &
	{\bf Language}
	\\ 
	\midrule
	
	{\tt mysqli::multi\_query()}, {\tt mysqli::prepare()},  &
	\multirow{3}{*}{MySQL} &
	\multirow{3}{*}{PHP} \\ 
	
	{\tt mysqli::real\_query()}, {\tt mysqli::select\_db()}, &
	&
	 \\ 
	
	{\tt mysqli::send\_query()} &
	&
	\\ \midrule

	{\tt mysql\_create\_db()}, {\tt mysql\_drop\_db()}, &
	\multirow{3}{*}{MySQL} &
	\multirow{3}{*}{C/C++} \\
	{\tt mysql\_query()}, {\tt mysql\_real\_query()}, &
	 &
	 \\
	{\tt mysql\_select\_db()}  &
	 &
	 \\ \midrule

	{\tt sqlite\_array\_query()}, {\tt sqlite\_exec()}, &
	\multirow{4}{*}{SQLite} &
	\multirow{4}{*}{PHP} \\ 
	{\tt sqlite\_open()}, {\tt sqlite\_query()}, & & \\ 
	{\tt sqlite\_popen()}, {\tt sqlite\_single\_query()}, & & \\ 
	{\tt sqlite\_unbuffered\_query()} & & \\ \midrule
	
	{\tt sqlite3\_get\_table()}, {\tt sqlite3\_exec()}, &
	\multirow{3}{*}{SQLite} &
	\multirow{3}{*}{C/C++} \\ 
	{\tt sqlite3\_prepare()$^1$}, {\tt sqlite3\_prepare16()$^2$},  & & \\ 
	{\tt sqlite3\_open()$^3$} & & \\ \midrule

	{\tt libxml.parseXmlString()}, {\tt parser.parseString()}, & \multirow{2}{*}{XML} & \multirow{2}{*}{JavaScript} \\ 
	{\tt parser.push()}, {\tt element.find()}, {\tt element.get()} & & \\ \midrule

	{\tt callbacks.StartDoctypeDecl()}, {\tt parser:parse()} & XML & Lua \\

	\bottomrule

	\multicolumn{3}{l}{$1$: inlcuding \code{sqlite3\_prepare\_v2()},  \code{sqlite3\_prepare\_v3()}.} \\
	
	\multicolumn{3}{l}{$2$: inlcuding \code{sqlite3\_prepare16\_v2()},  \code{sqlite3\_prepare16\_v3()}.} \\
	
	\multicolumn{3}{l}{$3$: inlcuding \code{sqlite3\_open16()}, \code{sqlite3\_open\_v2()}.}

\end{tabular}

}
%\caption{Database Sink Functions}
\vspace{-1em}
\end{table}

% \subsection{Details of Selected Programs and Vulnerability}
% \label{appendix:details_selected_programs}

% \autoref{table:vulnanddescription} presents
% detailed descriptions of the target applications and vulnerabilities we used in the evaluation in Section~\ref{sec:eval}.
% The first column shows the program ID that corresponds to the first column of \autoref{table:selectedprograms}.
% The third column shows the vulnerabilities that we used in our evaluation with citations. The last column presents brief descriptions of the applications.

%As shown in the last column, there are three different results: 
%{\it Command not found}, {\it File not found}, and {\it Time out}.
%The first two essentially mean that the exploit failed to execute the command as the command is already randomized. Time out happens because the program waits for the completion of the command while the command will never be completed as it fails to execute.

% \input{tex/table_vulnAndDescrption}

\subsubsection{Automated Vulnerability Discovery Tools}
\label{appendix:vulnerability_discovery_tools}
Commix~\cite{commix} is an automated testing tool that aims to find command injection vulnerabilities on web server-side applications.
We test all programs except for \code{s4} and \code{s5} which do not have functions executing OS/shell commands. 
%Note that they are tested for SQL injection and XXE injection attacks later in this section.
%Among them, \code{Commix} were not able to find command injection vulnerability in \code{s5}, \code{s6} 
%In this experiment, we choose five different applications: . \code{s2} is WordPress Plainview Activity Monotor plugin which represents the typical PHP program.
%\code{s7} is \code{Leptonica} image processing library and it is the C program with the most instrument statements. 
%\code{s10} is the only program running on the embedded devices, and at the same time, it has the most instrument statements of all the programs. \code{s20} is the largest JavaScript program in our cases.
%
%To use Commix, we define all the input interfaces for each target program. 
%Then, it automatically injects malicious commands to the target programs.
Commix identified vulnerabilities shown in \autoref{table:selectedprograms}, and successfully executed 102 malicious commands, while it failed to do so for \sysname protected programs. %. The result shows that \sysname protects all the injected commands.
sqlmap~\cite{sqlmap} is a penetration testing tool for SQL injection vulnerability testing. We apply sqlmap to the applications that use SQL database engines: \code{s1} (WordPress), \code{s5} (Pie Register), and \code{s6} (Lighttpd).
We instruct sqlmap to inject the SQL statements through typical input channels (e.g., \code{GET} and \code{POST} requests).
sqlmap supports various types of injection attack payloads, including boolean blind SQL injection, error-based SQL injection, stacked queries SQL injection, and time blind SQL injection, just to name a few.
\sysname mitigates all the injected statements. %failed to be executed, while they were successfully injected in vanilla versions.
%(e.g., \code{PUT} 
%We installed the vulnerable \code{s5} (Pie Register) on the server and defines the \code{sqlmap} use \code{PUT} to send a generated payload to the program. 
%And it succeeded in finding time-based sql injection vulnerability. Then we tested it with our \sysname, the \code{Sqlmap} fails to exploit the vulnerability.\mw{add data} 
%In the test, we use various methods to try to exploit the vulnerability including boolean blind sql injection, error based sql injection, inline query sql injection, stacked queries sql injection, time blind sql injection and union query sql injection. In all there are 97 different kinds of payloads.
% http://sqlmap.org/
% blackhat europe 2009 https://www.blackhat.com/presentations/bh-europe-09/Guimaraes/Blackhat-europe-09-Damele-SQLInjection-whitepaper.pdf
%We use \code{xcat} \cite{xcat} to test our \sysname on XML injection. 
xcat~\cite{xcat} is a command line tool to exploit and investigate XML injection vulnerabilities. We tested \code{s4} and \code{s5}, which are vulnerable to the XXE injection. 
%
%Note that while the program is a plugin of WordPress. As it has compatible issues with the current version of \code{WordPress}, we fix the issues for this experiment.
%
xcat successfully discovers XXE injection vulnerability in the original programs while it failed with the \sysname protected application.
%Because the \code{Advanced XML Reader} is a project not maintained for 6 years, it has server compatible problem with the current \code{Wordpress} and \code{PHP}. 
%So we modify the part of its source code to make it can work on the newer version environment. 
%We use \code{Xcat} set a built-in out of bound server to test if it exists XXE injection vulnerability. We define the vulnerable parameter and xpath query for the \code{xcat}. It succeeded in finding the XXE injection vulnerability but after applyintg \sysname to the program. \code{xcat} cannot find any potential vulnerability. 
%https://github.com/orf/xcat

\subsubsection{Overhead on Database Engines}
\label{appendix:overhead_database_engines}

%In Section~\ref{subsec:eval_perf}, we apply \sysname to SQLite and MySQL to evaluate performance overhead of \sysname on database engines.
We use OLTP-Bench~\cite{difallah2013oltp}, which is an extensible testbed for benchmarking relational databases. 
It provides 15 data-sets. However, when we test the data-sets on the vanilla MySQL and SQLite, \emph{only three data-sets (TPC-C, Wikipedia, and Twitter) were successfully completed} while all others lead to crashes. 
Hence, we select the three working data-sets.
%\autoref{table:OverheadDB} shows the result. 
The average overheads are 4.9\% and 5.3\% for SQLite and MySQL respectively. % less than 5.5\%. %SQLite is slightly faster than MySQL due to the complexity of MySQL leading to additional instrumentations.

\subsubsection{Overhead on XML Library}
\label{appendix:microbenchmarks_xml}
%In Section~\ref{subsec:eval_perf}, we evaluate the performance overhead of \sysname on four popular XML parsers.
We use Libxml~\cite{libxml}, SimpleXML~\cite{simplexml}, libxmljs~\cite{libxmljs}, and LuaExpat~\cite{luaexpat}. 
For XML test-data, we download a data-set (1GB in total) from the University of Washington~\cite{xmldata}.
The average overheads are 1.5\%, 1.38\%, 1.43\%, and 1.29\% for Libxml, SimpleXML, LuaExpat, and libxmljs respectively.

\subsubsection{Overhead on OpenWrt}
\label{appendix:embedded_devices_perf}
We applied \sysname to the OpenWrt firmware's uHTTPd~\cite{uHTTPd} web server and LuCI web configuration interface~\cite{LuCI}.
%LuCI runs instrumented functions such as \code{os.execute()}, and we confirm that the test cases cover the instrumented functions.
%As they do not run on the authors' workstation, which is x64, 
We use QEMU~\cite{QEMU} to run OpenWrt ARM firmware with 256MB RAM, which represents the standard router hardware specification. 
%
%uhttpd is the official webserver used by Openwrt project to run the Luci Configuration Interface. 
We use Apache Jmeter to generate 1,000 concurrent requests to visit the LuCI interface to get system status information. 
Note that 1,000 parallel requests are sufficient to exhaust the test system's resources and the test workload is more intensive than the common usage.
The average overhead is 5.83\%.
%to retrieve system information or status from the router and present it on the web page. 
%Our test's pressure is much higher than the common usage, but the overhead is still around 3\%.

\subsubsection{Versions of the Evaluated Programs}
\label{appendix:versions}
\autoref{table:versions} shows the versions of all the evaluated programs including those in \autoref{table:extra_programs}.

%\addtolength{\tabcolsep}{-5pt}
\setlength{\tabcolsep}{3pt}
\begin{table}[h]
	\centering
	\caption{Versions of the Evaluated Programs }
	\vspace{-1em}
	\label{table:versions}
	\resizebox{0.95\columnwidth}{!}{
\begin{tabular}{l r l r l r l r } 
    \toprule
    {\bf ID} &
    {\bf Version} &
    {\bf ID } & 
    {\bf Version} &
    {\bf ID } & 
    {\bf Version} &
    {\bf ID } & 
    {\bf Version} 
	\\
	\midrule

	\gcell \code{s1} &
	5.3.2 &
	\gcell \code{s12} & 
	0.4.18	 &
	\gcell \code{s23} & 
	0.2.4-beta	 &
	\gcell \code{s34} & 
	5.1.3
 \\

	\gcell \code{s2} &
	20161228$^1$ &
	\gcell \code{s13} & 
	1.1.0	 &
	\gcell \code{s24} & 
	3.6.1	 &
	\gcell \code{s35} & 
	2.0.4
 \\

	\gcell \code{s3} &
	2.3 &
	\gcell \code{s14} & 
	2.6.0	 &
	\gcell \code{s25} & 
	0.3.6 &
	\gcell \code{s36} & 
	6.2.3 \\

	\gcell \code{s4} &
	0.15 &
	\gcell \code{s15} & 
	0.1.4	 &
	\gcell \code{s26} & 
	0.4.13	 &
	\gcell \code{s37} & 
	3.0.2 \\

	\gcell \code{s5} &
	1.7 &
	\gcell \code{s16} & 
	3.5.0 &
	\gcell \code{s27} & 
	1.0.2 &
	\gcell \code{s38} & 
	3.36.0 \\

	\gcell \code{s6} &
	3.0.9	 &
	\gcell \code{s17} & 
	1.0.1	 &
	\gcell \code{s28} & 
	5.2.2 &
	\gcell \code{s39} & 
	3.22.1 \\

	\gcell \code{s7} &
	1.4.35	 &
	\gcell \code{s18} & 
	1.1.2 &
	\gcell \code{s29} & 
	15.2 &
	\gcell \code{s40} & 
	0.7.2 \\

	\gcell \code{s8} &
	1.74.4 &
	\gcell \code{s19} & 
	1.2.0 &
	\gcell \code{s30} & 
	4.1.7 &
	\gcell \code{s41} & 
	6.3.0 \\

	\gcell \code{s9} &
	2.7.6	 &
	\gcell \code{s20} & 
	0.1.0 &
	\gcell \code{s31} & 
	3.0.12 &
	\gcell \code{s42} & 
	1.0.0-pre.45
 \\

	\gcell \code{s10} &
	3.6.5 &
	\gcell \code{s21} & 
	6.0.0 &
	\gcell \code{s32} & 
	0.7 &
	& \\

	\gcell \code{s11} &
	0.10 &
	\gcell \code{s22} & 
	0.7.0	 &
	\gcell \code{s33} & 
	5.1.8 &
	& \\

	\bottomrule 

	\multicolumn{8}{l}{1: This project does not have an explicit version number. This is} \\
	\multicolumn{8}{l}{ the date of the last commit.}

\end{tabular}
}
%\vspace{-2em}
\end{table}

\subsubsection{Trusted Command Specification (TCS) Generation Tool}
\label{appendix:tcstool}
We provide an automated trusted command specification generation tool~\cite{csr-tool} that takes a list of sink-functions (e.g., \autoref{table:sinkfunctions}) and trusted-folders (e.g., \code{/var/www/}) as input. It derived all the TCSs used in the paper without significant domain-knowledge and completed the analysis in less than four minutes. 
%Our project website~\cite{csr-tool} also presents the inputs for this tool. 
It can also detect incomplete specifications (e.g., untrusted commands passed to sink-functions). Note that we did not observe incomplete specifications.

\noindent
{\bf Performance of TCS Generator.}
\autoref{table:tcs_time} shows the time required to generate the TCS by our TCS generator. We generate the same TCS used in our evaluation. Note that generating TCS for Leptonica (s8) took the longest time: 217.75 seconds, which is 3 min 37.75 seconds.

%\addtolength{\tabcolsep}{-5pt}
\setlength{\tabcolsep}{3pt}
\begin{table}[h]
	\centering
	\caption{Time to generate TCS for each application}
	\vspace{-1em}
	\label{table:tcs_time}
	\resizebox{0.7\columnwidth}{!}{
\begin{tabular}{l r l r l r} 
    \toprule
    {\bf ID} &
    {\bf Time (s)} &
    {\bf ID } & 
    {\bf Time (s)} &
    {\bf ID } & 
    {\bf Time (s)}
	\\
	\midrule

	\gcell \code{s1}  & 	165.59 & 	\gcell \code{s15} & 	1.25	 & 	\gcell \code{s29} & 	31.96	  \\
	\gcell \code{s2}  & 	7.82 & 	\gcell \code{s16} & 	22.29	 & 	\gcell \code{s30} & 	4.13	  \\
	\gcell \code{s3}  & 	5.18 & 	\gcell \code{s17} & 	1.18	 & 	\gcell \code{s31} & 	29.24	  \\
	\gcell \code{s4}  & 	7.99 & 	\gcell \code{s18} & 	1.59	 & 	\gcell \code{s32} & 	2.88	  \\
	\gcell \code{s5}  & 	14.73 & 	\gcell \code{s19} & 	1.14	 & 	\gcell \code{s33} & 	4.71	  \\
	\gcell \code{s6}  & 	6.23 & 	\gcell \code{s20} & 	1.22	 & 	\gcell \code{s34} & 	7.31	  \\
	\gcell \code{s7}  & 	100.32 & 	\gcell \code{s21} & 	0.83	 & 	\gcell \code{s35} & 	28.64	  \\
	\gcell \code{s8}  & 	217.75 & 	\gcell \code{s22} & 	0.75	 & 	\gcell \code{s36} & 	18.38	  \\
	\gcell \code{s9}  & 	12.28 & 	\gcell \code{s23} & 	0.74	 & 	\gcell \code{s37} & 	9.23	  \\
	\gcell \code{s10}  & 	54.02 & 	\gcell \code{s24} & 	1.17	 & 	\gcell \code{s38} & 	15.19	  \\
	\gcell \code{s11}  & 	146.36 & 	\gcell \code{s25} &     1.35	 & 	\gcell \code{s39} & 	8.28	  \\
	\gcell \code{s12}  & 	9.76 & 	\gcell \code{s26} & 	0.58	 & 	\gcell \code{s40} & 	0.98	  \\
	\gcell \code{s13}  & 	1.18 & 	\gcell \code{s27} & 	1.28	 & 	\gcell \code{s41} & 	1.19	  \\
	\gcell \code{s14}  & 	1.61 & 	\gcell \code{s28} & 	9.68	 & 	\gcell \code{s42} & 	0.87	  \\

 \bottomrule 

%	\multicolumn{6}{l}{1: This project does not have an explicit version number. This is} \\
%	\multicolumn{6}{l}{ the date of the last commit.}

\end{tabular}
}
%\vspace{-2em}
\end{table}

\subsection{Effectiveness of \sysname}

\revised{
%%%%%%%%%%%%%%%%%%%%%%%%%%%%%%%%%%%%%%%%%%%%%%%%%%%%%%%%%%%%%%%%%%%%%%%%%%%%
%%%%%%%%%%%%%%%%%%%%%%%%%%%%%%%%%%%%%%%%%%%%%%%%%%%%%%%%%%%%%%%%%%%%%%%%%%%%
%%%%%%%%%%%%%%%%%%%%%%%%%%%%%%%%%%%%%%%%%%%%%%%%%%%%%%%%%%%%%%%%%%%%%%%%%%%%
\subsubsection{Applicability of \sysname} 
To understand whether \sysname can be a generic solution for various applications, we additionally collect the five most popular applications from three well-known open-source package managers (NPM~\cite{npmpackage}, Packagist~\cite{packagist} and WordPress Plugin~\cite{wordpressplugins}) as shown in \autoref{table:extra_programs}. 
We prune out programs that are not meant to be deployed such as a unit-test framework~\cite{PHPUnit}. 
%We apply \sysname to them and the results are shown in \autoref{table:extra_programs}. 
\sysname successfully handled them without errors 
}

%\CJ{4) [2.4] (\#C) How general or application-specific is the solution?}

%%%%%%%%%%%%%%%%%%%%%%%%%%%%%%%%%%%%%%%%%%%%%%%%%%%%%%
%%%%%%%%%%%%%%%%%%%%%%%%%%%%%%%%%%%%%%%%%%%%%%%%%%%%%%
%%%%%%%%%%%%%%%%%%%%%%%%%%%%%%%%%%%%%%%%%%%%%%%%%%%%%%

\revised{
\subsubsection{Correctness of Instrumentation}
\label{appendix:correctness_instrumentation}
\updated{}{
\autoref{table:selectedprograms_cov} shows the number of test cases. Our additional test cases to cover all the instrumented code and increase code coverage are shown in the ``Added'' column. 
}
}
%The experiment results in \autoref{table:selectedprograms_cov} show that our instrumentation does not break the functionalities so \sysname instrumented the applications in the correct position.} 
%\CJ{1) [1] (\#C) Correctness verification: verify the instrumentation does not break the application.}

\begin{table}[tp]
	\centering
	%\caption{Selected Programs for Evaluation}
	\caption{\updated{}{Popular Applications From Pakcage Managers}}
	\label{table:extra_programs}
	\vspace{-1em}
	\renewcommand{\arraystretch}{1.3}

\resizebox{1.0\columnwidth}{!}{%	

\begin{tabular}{c l r l r r r r r} 
	\toprule
	\multirow{3}{*}{\bf ID} & 
	\multirow{3}{*}{\bf Name} & 
	\multicolumn{1}{c}{\multirow{3}{*}{\bf Size}} &
	\multicolumn{1}{l}{\multirow{3}{*}{\bf Source}} &
	
	\multicolumn{5}{c}{\bf \# Instrumentation} 
	\\ \cmidrule{5-9}

	&
 	& &
 	&
	\multirow{2}{*}{\bf Const.} &
	\multicolumn{3}{c}{\bf Dynamic } &
	\multirow{2}{*}{\bf Sinks}
    %\multirow{2}{*}{\bf Local} &
	%\multirow{2}{*}{\bf Global} &
	%\multirow{2}{*}{\bf Func} 
	
	\\ 
	\cmidrule{6-8}
	 & & & & &
	{\bf 1-5}&{\bf 6-10}&{\bf $>$11} 
	
	\\
	
%	\midrule
	\midrule
	\rowcolor{gray!40}
	
	\code{s28} &
	Contact-Form-7~\cite{ContactForm7} &
	744.00 KB  & 
	WordPress &    
    1 & 4 &  
    0 & 0 & 5 
	
	\\ 
	%\cmidrule(lr){1-4} 	\cmidrule(lr){5-9}	\cmidrule(lr){10-12}	\cmidrule(lr){13-15} %\cmidrule(lr){17-18}

	\code{s29} &
    Yoast SEO~\cite{YoastSEO} &
    13.70 MB &
    WordPress & 
    6 & 12 & 
    9 & 0 & 6 
	
	\\ 
	%\cmidrule(lr){1-4} 	\cmidrule(lr){5-9}	\cmidrule(lr){10-12}	\cmidrule(lr){13-15} %\cmidrule(lr){17-18}
	\rowcolor{gray!40}
	\code{s30} &
	Akismet Spam Protection~\cite{AkismetSpamProtection} &
	288.00 KB &
	WordPress & 
    0 & 17 & 
    0 & 0 & 17 
	
	\\ 
	%\cmidrule(lr){1-4} 	\cmidrule(lr){5-9}	\cmidrule(lr){10-12}	\cmidrule(lr){13-15} %\cmidrule(lr){17-18}
	\code{s31} &	
	Elementor Website Builder~\cite{ElementorWebsiteBuilder} &
	18.00 MB &
	WordPress &  
	2 & 21 &
	0 & 0 & 23 
    
	%& & &
	
	\\ 
	%\cmidrule(lr){1-4} 	\cmidrule(lr){5-9}	\cmidrule(lr){10-12}	\cmidrule(lr){13-15} %\cmidrule(lr){17-18}
	\rowcolor{gray!40}
	\code{s32} &
	WordPress Importer~\cite{WordPressImporter} &
	100.00 KB &
	WordPress &  
    0 & 
    2 & 0 & 0 & 
    2 
	% &  &  &
	
	\\ \midrule
	%\cmidrule(lr){1-4} 	\cmidrule(lr){5-9}	\cmidrule(lr){10-12}	\cmidrule(lr){13-15} %\cmidrule(lr){17-18}
	\code{s33} &
	Symfony Console~\cite{console} &
	584.00 KB &
	Packagist& 
    8 & 
    10 & 0 & 0 &
    15 
	% &  &  &
	
	\\ 
	%\cmidrule(lr){1-4} 	\cmidrule(lr){5-9}	\cmidrule(lr){10-12}	\cmidrule(lr){13-15} %\cmidrule(lr){17-18}
	\rowcolor{gray!40}
	\code{s34} &
	Environment~\cite{envrionment} &
	49.00 KB &
	Packagist& 
    3 &
    0 & 0 & 0 & 
    3 
	% &  &  &
	
	\\ 
	%\cmidrule(lr){1-4} 	\cmidrule(lr){5-9}	\cmidrule(lr){10-12}	\cmidrule(lr){13-15} %\cmidrule(lr){17-18}
	\code{s35} &
	Composer~\cite{Composer} &
	120.00 KB &
	Packagist&  
    4 & 
    4 & 0 & 0 & 
    8 
	%& & &
	
	\\ 
	%\cmidrule(lr){1-4} 	\cmidrule(lr){5-9}	\cmidrule(lr){10-12}	\cmidrule(lr){13-15} %\cmidrule(lr){17-18}
	\rowcolor{gray!40}
	\code{s36} &
	Swiftmailer~\cite{swiftmailer} &
	2.08 MB &
	Packagist& 
    0 & 
    1 & 0 & 0 & 
    1 
	%& & &
	
	\\ 
	%\cmidrule(lr){1-4} 	\cmidrule(lr){5-9}	\cmidrule(lr){10-12}	\cmidrule(lr){13-15} %\cmidrule(lr){17-18}
	\code{s37} &
	Version~\cite{yalc} &
	20.00 KB &
	Packagist& 
    1 & 
    0 & 0 & 0 & 
    1 
	%& & &
    
	\\ \midrule
	%\cmidrule(lr){1-4} 	\cmidrule(lr){5-9}	\cmidrule(lr){10-12}	\cmidrule(lr){13-15} %\cmidrule(lr){17-18}
	
	\rowcolor{gray!40}
	\code{s38} &
	Ghost~\cite{ghost} &
	58.80 MB &
	 NPM &
    1 & 
    0 & 0& 0 &  
    1 
	% &  &  &
	
	\\ 
	%\cmidrule(lr){1-4} 	\cmidrule(lr){5-9}	\cmidrule(lr){10-12}	\cmidrule(lr){13-15} %\cmidrule(lr){17-18}

	\code{s39} &
	Lerna~\cite{Lerna} &
	12.10 MB &
	NPM & 
    1 & 
    0 & 0& 0 &  
    1 
	%& & &
	
	\\ 
	%\cmidrule(lr){1-4} 	\cmidrule(lr){5-9}	\cmidrule(lr){10-12}	\cmidrule(lr){13-15} %\cmidrule(lr){17-18}
	
	\rowcolor{gray!40}
	\code{s40} &
	Npkill~\cite{npkill} &
	7.69 MB &
	NPM & 
    2 & 
    7 & 0 & 0 & 
    9 
	%& & &
	
	\\ 
	%\cmidrule(lr){1-4} 	\cmidrule(lr){5-9}	\cmidrule(lr){10-12}	\cmidrule(lr){13-15} %\cmidrule(lr){17-18}
	
	\code{s41} &
	Release~\cite{release} &
	900.00 KB &
	NPM &
    0 & 
    3 & 0 & 0 &  
    3 
	%& & &
	
	\\ 
	%\cmidrule(lr){1-4} 	\cmidrule(lr){5-9}	\cmidrule(lr){10-12}	\cmidrule(lr){13-15} %\cmidrule(lr){17-18}
	
	\rowcolor{gray!40}
	\code{s42} &
	Yalc~\cite{yalc} &
	704.00 KB &
	NPM & 
    0 & 
    4 & 0 & 0 &
    4 
	%& & &

	\\ 
	\bottomrule 
	%7: Execution prevented with ``Command not found''
	%\multicolumn{16}{l}{8: Execution prevented with ``File not found''. 9: Execution prevented with a time out error. 10: Average.}
	%\multicolumn{15}{l}{1: Execution prevented with an error ``Command not found''. 2: Execution prevented with an error ``File not found''. 3: Execution prevented with a time out error.} \\
\end{tabular}
}
\vspace{-1.5em}
\end{table}

\begin{table}[h]
	\centering
	\caption{\revised{Test Cases and Code Coverages}}
	\vspace{-1em}
	\label{table:selectedprograms_cov}
	\resizebox{1.0\columnwidth}{!}{
\begin{tabular}{l r r r r  l r r r r } 
    \toprule
    \multirow{2}{*}{\bf ID} &
    \multirow{2}{*}{\bf Line of} &
    \multicolumn{2}{c}{\bf \# of Test Cases} &
	\multicolumn{1}{c}{\multirow{2}{*}{\bf Cov--} } &
	\multirow{2}{*}{\bf ID} &
    \multirow{2}{*}{\bf Line of} &
    \multicolumn{2}{c}{\bf \# of Test Cases} &
	\multicolumn{1}{c}{\multirow{2}{*}{\bf Cov--} }
	\\ \cmidrule{3-4} \cmidrule{8-9}
	
	& \multicolumn{1}{c}{\bf Code} & {\bf Added} & {\bf Total}& \multicolumn{1}{c}{\bf   erage} & & \multicolumn{1}{c}{\bf Code} & {\bf Added} & {\bf Total}& {\bf   erage}
	\\ \midrule

	\rowcolor{gray!40}
	\code{s1} &
	116,356    &
	200 & 11,677&
	58.54\% &
	\code{s15}  &
	146 &
	39 & 63 &
	76.14\% \\

    \code{s2}  &
    9,881 &
    100 & 142 &
	75.76\% &
	\code{s16}  &
	12,281 &
	50 & 332 &
	72.78\% \\

	\rowcolor{gray!40}
	\code{s3}  &
	67,560 &
	404 & 404 & 
	70.49\%  &
	\code{s17}  &
	402 &
	50 & 60 &
	88.33\% \\

    \code{s4}  &
	6,124 &
	161 & 161 &
	82.30\%  &
	\code{s18}  &
	69 &
	49 & 53 &
	86.21\% \\

	\rowcolor{gray!40}
	\code{s5}  &
	12,872 &
	219 & 219 &
	71.56\% &
	\code{s19}  &
	78 &
	51 & 54 &
	89.74\% \\

	\code{s6}  &
	9,944 &
	188 & 188 &
	73.92\% &
	\code{s20}  &
	76 &
	50 & 58 &
	83.85\% \\

	\rowcolor{gray!40}
	\code{s7}  &
	42,840 &
	100 & 459 &
	64.00\% &
	\code{s21} &
	1,912 &
	49 & 141 &
	86.00\% \\

	\code{s8}  &
	86,668 &
	97 & 443 &
	65.38\% &
	\code{s22}  &
	999 &
	40 & 53 &
	79.17\% \\
	
	\rowcolor{gray!40}
	\code{s9}  &
	30,011 &
	100 & 107 &
	71.00\% &
	\code{s23}  &
	301 &
	60 & 62 &
	79.46\% \\

	\code{s10}  &
	58,994 &
	233 & 289 &
	68.96\% &
	\code{s24}  &
	792 &
	38 & 122 &
	83.22\% \\
    
	\rowcolor{gray!40}
	\code{s11}  &
	34,237 &
	396 & 396 &
	72.42\% &
	\code{s25}  &
	145 &
	38 & 48 &
	91.19\% \\

	\code{s12}  &
	1,431 &
	10 & 122 &
	78.23\% &
	\code{s26}  &
	284 &
	60 & 64 &
	74.11\% \\

	\rowcolor{gray!40}
	\code{s13}  &
	174 &
	50 & 53 &
	83.73\% &
	\code{s27}  &
	143&
	55 & 57 &
	87.50\% \\

	\code{s14}  &
	281 &
	55 & 89 &
	96.65\% \\

	\bottomrule 
	%7: Execution prevented with ``Command not found''
	%\multicolumn{16}{l}{8: Execution prevented with ``File not found''. 9: Execution prevented with a time out error. 10: Average.}
	%\multicolumn{15}{l}{1: Execution prevented with an error ``Command not found''. 2: Execution prevented with an error ``File not found''. 3: Execution prevented with a time out error.} \\
\end{tabular}
}
\vspace{-1em}
\end{table}

%%%%%%%%%%%%%%%%%%%%%%%%%%%%%%%%%%%%%%%%%%%%%%%%%%%%%%
%%%%%%%%%%%%%%%%%%%%%%%%%%%%%%%%%%%%%%%%%%%%%%%%%%%%%%
%%%%%%%%%%%%%%%%%%%%%%%%%%%%%%%%%%%%%%%%%%%%%%%%%%%%%%

\subsubsection{Supporting Polymorphic Objects}

\begin{figure}[!h]
    \centering
    \vspace{1em}
    \includegraphics[width=1.0\columnwidth]{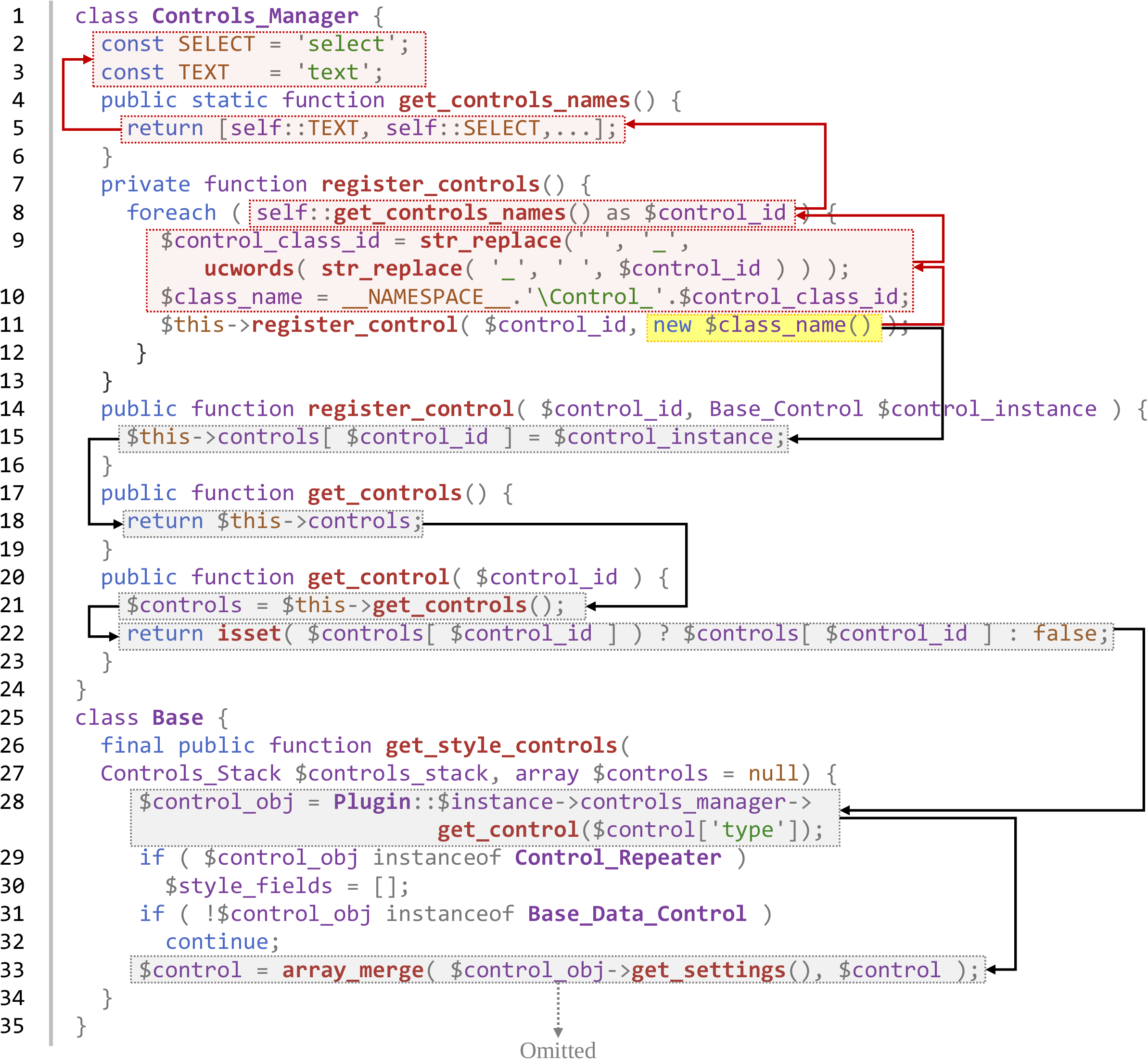}
    \vspace{-2em}
    \caption{\updated{}{Handling Polymorphic Objects (Red and black arrows represent backward and forward analysis respectively)}}
    \vspace{-1em}
    \label{fig:oop_case}
\end{figure}

\sysname supports complex real-world applications including OOP programs such as WordPress in \autoref{table:selectedprograms}.
In this example, we show that our analysis handles polymorphism and dynamic bindings. 
In particular, \autoref{fig:oop_case} shows how \sysname analyzes polymorphic objects in a WordPress plugin called Elementor~\cite{ElementorWebsiteBuilder}.
From line 11, we identify an instantiation of an object with a string (\code{\$class\_name}). We backtrace the string variable (annotated via red arrows), identifying that the class name starts with ``\code{Control\_}''. However, as the return value of \code{get\_control\_names()} can be updated at runtime, we conservatively assume that any class that has name starting with \code{Control\_} (i.e., \code{Control\_*}) can be created at line 11.

Then, we conduct a forward analysis to identify the object's usage (annotated through black arrows). \autoref{fig:oop_case} shows only a few of the forward flows due to space. 
%Note that our analyses (both forward and backward) are conservative, meaning that it may have false positives but no false negatives. 
We check all the omitted flows and they are not relevant to command execution. %, meaning that false positives do not cause wrong instrumentations.

%Note that our analysis is conservative which may cause false positive cases. However, with the false positives, since they are not relevant to the command execution APIs, \sysname is not affected by this limitation during our evaluation.

%%%%%%%%%%%%%%%%%%%%%%%%%%%%%%%%%%%%%%%%%%%%%%%%%%%%%%%%%%%%%%%
%%%%%%%%%%%%%%%%%%%%%%%%%%%%%%%%%%%%%%%%%%%%%%%%%%%%%%%%%%%%%%%
%%%%%%%%%%%%%%%%%%%%%%%%%%%%%%%%%%%%%%%%%%%%%%%%%%%%%%%%%%%%%%%

\revised{
\subsubsection{Effectiveness of Bidirectional Analysis}
\label{appendix:bidirectional_effectiveness}
%\sysname uses bidirectional analysis to mitigate inaccurate analysis results caused by false negative of the forward and backward analysis alone, as described in Section~\ref{sec:design}. 
This section provides an example of the effectiveness of bidirectional analysis. % mitigating false negative issues of the forward and backward analysis alone.

\begin{figure}[h]
    \centering
    \includegraphics[width=1.0\columnwidth]{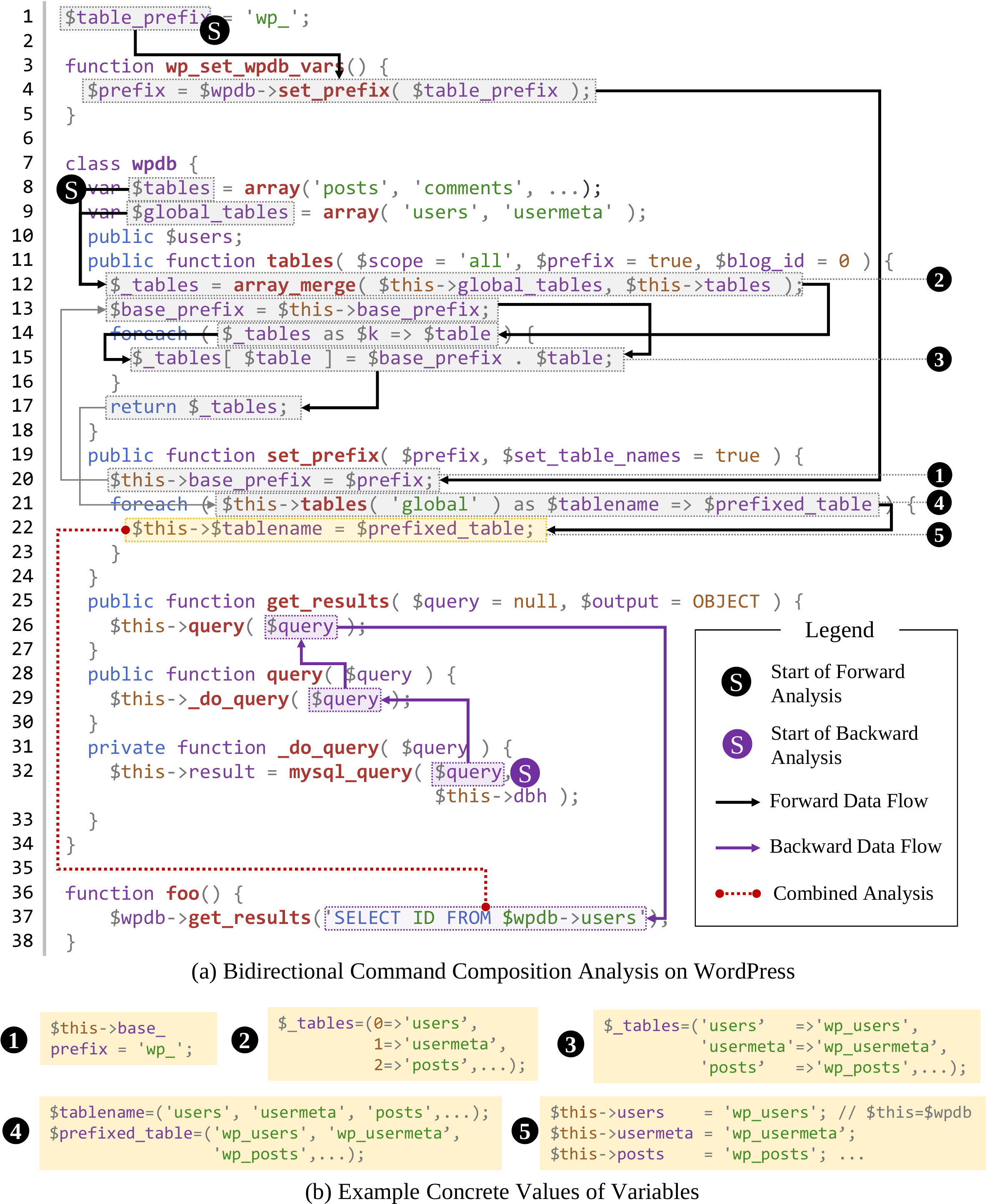}
    \vspace{-2.5em}
    \caption{\updated{}{Bidirectional Analysis on WordPress}}
    \vspace{-1em}
    \label{fig:false_negative}
\end{figure}

\textit{1) Backward Analysis:} 
The backward flow analysis begins from the sink function \code{mysql\_query()} at line 32. Following the backward data flow (depicted as purple arrows), it reaches to line 38, which is a SQL query. However, the backward analysis alone is not able to identify the original of \code{\$wpdb->users} to determine whether this is from a trusted source (hence requires instrumentation) or not.
Note that the value of \code{\$wpdb->users} is assigned by dynamic construct, which is unreachable by the backward analysis. 

\textit{2) Forward Analysis:} Our forward analysis starts from trusted sources such as constant strings at lines 1, 8, and 9. \code{\$table\_prefix} (global variable) is assigned to \code{\$this->base\_prefix} at line 20 (\blkcc{1}), which will be used at line 13 in \code{tables()}. \autoref{fig:false_negative}-(b) shows values of variables at the lines marked by circled numbers.
From lines 8 and 9, the two arrays are merged at line 12 (\blkcc{2}), resulting in an array shown in \autoref{fig:false_negative}-(b). At line 15, it constructs an array consisting of pairs of table names and table names with \code{wp\_} prefix (\blkcc{3}). The composed new table (\code{\$\_table}) is returned by \code{tables()}, which is called at line 21 in \code{set\_prefix()}.
Hence, we further analyze \code{set\_prefix()} which iterates arrays shown in \autoref{fig:false_negative}-(b)-\blkcc{4}. 
Note that PHP allows a string variable to be used to specify a member variable's name in an object (line 22). To this end, the line 22 essentially executes statements shown in \autoref{fig:false_negative}-(b)-\blkcc{5}.
Note that the first statement define \code{\$this->users}, where \code{\$this} is essentially \code{\$wpdb}.

\textit{3) Combining Forward and Backward Analysis:}
Recall that the backward analysis stops at line 37, as it was unable to resolve \code{\$wpdb->users}, while the forward analysis can resolve the value. 
Bidirectional analysis successfully finds out that the variable in the query (\code{\$wpdb->users}) is a constant string from a trusted source.
}

\subsubsection{Evaluation of Bidirectional Analysis's Accuracy}
\label{appendix:bidir_accuracy}
% https://github.com/cmd-spinner/commandrandom-spinner-php/tree/master/Supplementary/Evaluation
In this section, we explain the details of how we evaluate \sysname's bidirectional analysis's effectiveness and correctness. 
Note that since obtaining the ground-truth is challenging, we try to evaluate the bidirectional analysis's accuracy as follows.
We manually verified that all the results from our analysis are true-positives. We also run other static/dynamic taint-analysis tools~\cite{taintall, psalm, pecltaint} and compare the results (i.e., dependencies) from them with the result from \sysname. 
As shown in \autoref{table:num_detected}, we observe that \sysname covers the majority of the dependencies chains that are generated by the other tools. For dependencies not covered by \sysname, we manually check them and find that they are false-positives (hence we are not missing anything covered by other tools).

Note that during this process, we have updated and implemented a few tools. First, we update the AST parser of taintless~\cite{taintless} to the new version and add extra rules to help it handle WordPress's callback function hook.
Second, we add additional plugins to psalm~\cite{psalm} to enhance its ability on tracing data flow on object inheritance. Third, we add additional sinks to PECL taint~\cite{pecltaint} for tainting WordPress.

\noindent
{\bf Procedure of the Evaluation.}
We do our evaluation as follows. 

\begin{itemize}[leftmargin=*]
    \item [\it 1.] Run the bi-directional analysis and manually verify the dependency chains identified by the analysis. a) Manually check the propagation rules applied by the bi-directional analysis (both forward and backward analyses). b) Verify that all the dependencies identified by the bi-directional analysis are true-positives.
    
    \item [\it 2.] Run other static/dynamic analysis tools to get dependency chains (Note that static/dynamic analysis tools suffer from over and under-approximations). a) If other tools find more dependencies, then they might be potential false-negatives of the bi-directional analysis. We manually verify them all, and the result shows that they are all false-positives, meaning that we did not find false-negatives from the bi-directional analysis. b) If other tools find lesser dependencies, then they might be potential false-positives of the bi-directional analysis. We manually verify them all, and the result shows that they are all false-negatives, meaning that we did not find false-positives from the bi-directional analysis.
\end{itemize}

\noindent
{\bf Procedure and Method for Manual Analysis.}
Our manual analysis leverages existing static/dynamic analysis techniques. While they are inaccurate, we only apply them for a single dependency chain and reason about the result. Since we only reason a single dependency at a time, the task was manageable even though it is a time-consuming task. 
We conduct inter-procedural manual analysis, meaning that we follow through the callee functions' arguments if values propagate through the functions.
The analysis finishes when the data reaches a trusted/untrusted source.
In addition to the static/dynamic taint analysis techniques, we manually run the programs and observe how the concrete values are propagated by changing inputs and checking output differences. Note that if an output value is changed from the above testing due to the input change, there is a dependency.

%\addtolength{\tabcolsep}{-5pt}
\setlength{\tabcolsep}{3pt}
\begin{table}[h]
	\centering
	\caption{Effectiveness of \sysname's bidirectional analysis compared with existing techniques}
	\vspace{-1em}
	\label{table:num_detected}
	\resizebox{1.0\columnwidth}{!}{
\begin{tabular}{l r r r r} 
    \toprule
    {\bf Testbeds} &
    {\bf \sysname } &
    {\bf Taintless } &
    {\bf Psalm } &
    {\bf PECL taint}
    \\
    \midrule

	\rowcolor{gray!25} 
	WordPress*  & 	462 & 413 & 426 & 537 \\
	\rowcolor{gray!25}
	Activity Monitor*  & 	27 & 16 & 17 & 27 \\
	\rowcolor{gray!25}
	Avideo Encoder*	 & 	61 & 66 & 61 & 61 \\
	\rowcolor{gray!25}
	PHPSHE*  & 	270 & 301 & 266 & 223 \\
	\rowcolor{gray!25}
	Pie Register*  & 	73 & 79 & 77 & 73 \\
	Pepperminty WiKi  & 	2 & 2 & 2 & 2 \\
	Contact-Form-7  & 	5 & 5 & 5 & 5 \\
	Yoast SEO  & 	27 & 27 & 27 & 27 \\
	Akismet Spam Protection  & 	17 & 17 & 17 & 17 \\
	Elementor Website Builder  & 	23 & 23 & 23 & 23 \\
	WordPress Importer  & 	2 & 2 & 2 & 2 \\
	Symfony Console  & 	18 & 18 & 18 & 18 \\
	Environment  & 	3 & 3 & 3 & 3 \\
	Composer  & 	8 & 8 & 8 & 8 \\
	Swiftmailer  & 	1 & 1 & 1 & 1 \\
	Version  & 	1 & 1 & 1 & 1 \\

 \bottomrule 

	\multicolumn{5}{l}{*: Except for these 5 applications, there is no difference between the tools.} \\
%	\multicolumn{6}{l}{ the date of the last commit.}

\end{tabular}
}
%\vspace{-2em}
\end{table}

%* Except for these 6 applications, there is no difference between the tools.
To make sure \sysname's bi-directional analysis does not miss anything, we compared the results with existing techniques (Taintless, Psalm, and PECL taint). We manually analyzed them and verified that all the results from bi-directional analysis are true-positives. 
Details on the notable cases are as follows.

\begin{itemize}[leftmargin=*]
    \item [\it 1.] {\it WordPress:} 
    Compared to Taintless, Taintless has 49 false negatives. Among them, 24 false negatives are caused as described in \autoref{fig:false_negative}. 5 false negatives are caused by handling PHP dynamic function call (e.g., \code{call\_user\_func\_array()}). 20 false negatives are caused by handling WordPress \code{apply\_filter} which invokes a function by the nickname registered by \code{add\_filter}.
    Compared to Psalm, Psalm has 24 false negatives as described in \autoref{fig:false_negative}. Psalm is not accurate in handling object inheritance. It will miss the data dependencies from subclass methods to base class methods in 36 cases.
    Compared to PECL taint, PECL taint has 35 false positives caused by handling WordPress \code{do\_action} dynamic function hook.  PECL taint has 40 false positives caused by string array filtering operation.
    
    \item [\it 2.] {\it Activity Monitor:}
     Compared to Taintless, Taintless has 11 false negatives. Among them, 3 false negatives are caused as shown in \autoref{fig:false_negative}. 8 false negatives are caused by not supporting WordPress \code{apply\_filter} which invoke a function registered by \code{add\_filter} dynamically. The data flow will be broken when it goes into such APIs.
    Compared to Psalm, Psalm has 14 false negative and 4 false positive cases. Among them, 3 false negatives are caused as shown in \autoref{fig:false_negative}. 8 false negatives are caused by \code{add\_filter} and \code{apply\_filter}. 3 false negatives are caused by mishandling object inheritance. Variables defined in base class will not be recognized in subclass. 4 false positive cases are caused by mishandling regex matching API \code{preg\_match}.
    
    \item[\it 3.] {\it Avideo-Encoder:}
    Compared to Taintless, Taintless has 2 false negatives and 7 false positives. Among them, 2 false negatives are caused by unsupported API \code{DateTime()} which should be considered as trusted. 7 false positives are caused by mishandling regex API \code{preg\_match}. 
    
    \item[\it 4.] {\it PHPSHE:}
    Compared to Taintless, Taintless has 16 false negatives and 47 false positives. Among them, 16 false negatives are caused by parsing error on one PHP file. Internal bug on an old version of PHP-Parser. 47 false positives are caused by history upgrading scripts.
    Compared to Psalm, Psalm has 15 false negatives and 11 false positives. Among them, 3 false negatives are caused by \code{time()} API. 12 false negatives are caused by class object inheritance. 11 false positives are caused by SQL keywords in arguments used matching pattern of \code{preg\_match} functions. 
    Compared to PECL taint, PECL taint has 47 false negatives because of PHP fatal error in executing database update script
    
    \item[\it 5.] {\it Pie-register:}
    Compared to Taintless, Taintless has 2 false negatives and 8 false positives. Among them, 2 false negatives are caused by the case shown in \autoref{fig:false_negative}. 8 false positives are caused by SQL keywords in the embedded HTML while they are not SQL statements.
    Compared to Psalm, Psalm has 2 false negatives caused by the case shown in \autoref{fig:false_negative}.

\end{itemize}

%%%%%%%%%%%%%%%%%%%%%%%%%%%%%%%%%%%%%%%%%%%%%%%%%%%%%%%%%%%%%%%%%%%%%%%%%%%%
%%%%%%%%%%%%%%%%%%%%%%%%%%%%%%%%%%%%%%%%%%%%%%%%%%%%%%%%%%%%%%%%%%%%%%%%%%%%
%%%%%%%%%%%%%%%%%%%%%%%%%%%%%%%%%%%%%%%%%%%%%%%%%%%%%%%%%%%%%%%%%%%%%%%%%%%%

\subsubsection{Impact Analysis for Instrumentated Code}
\label{appendix:manual_analysis_instr_detail}

\begin{figure}[h]
    \centering
    \includegraphics[width=0.9\columnwidth]{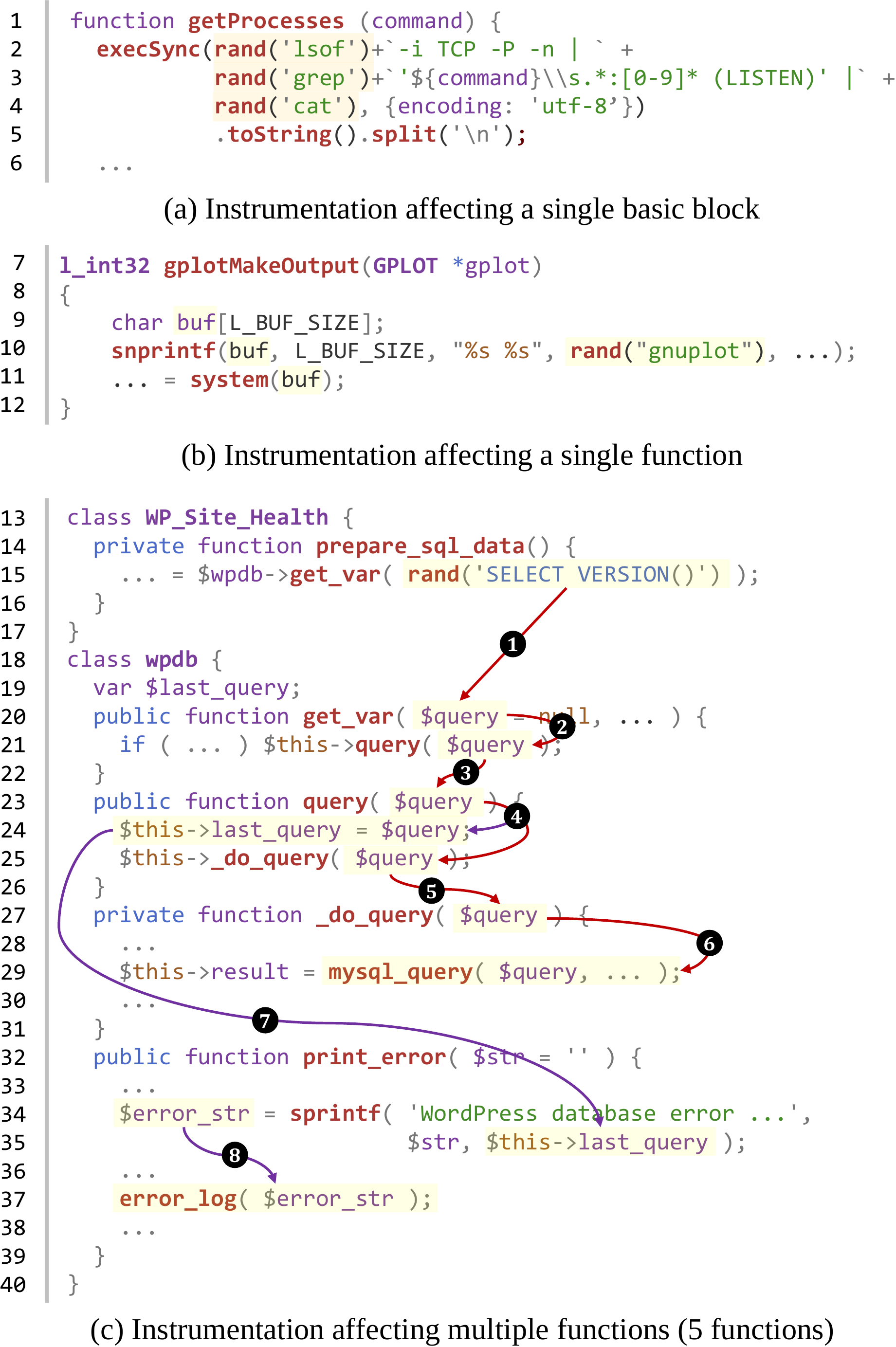}
    \vspace{-1em}
     \caption{Examples of Impact of Instrumentation}
     \label{fig:instr_sample}
     \vspace{-1em}
\end{figure}

\autoref{fig:instr_sample} shows examples of instrumentations impacting a single basic block (a), a single function (b), and multiple functions (c).
%We explain each example as follows.

\noindent
{\bf Single Basic Block (the BB column in \autoref{table:selectedprograms}).}
This is the simplest type of instrumentation. As shown in \autoref{fig:instr_sample}-(a), all the instrumented commands (i.e., \code{lsof}, \code{grep}, and \code{cat}) are directly fed into the sync function (\code{execSync} at line 2). The instrumented commands are not saved and transferred to other functions. %This does not affect any other parts of the target program.

\noindent
{\bf Single Function (the Fn column in \autoref{table:selectedprograms}).}
Instrumented commands can affect or stored in local variables. However, they only affect statements within the same function and do not propagate to other functions. 
In \autoref{fig:instr_sample}-(b), the instrumentation (\code{rand()} at line 10) affects a local variable \code{buf}. However, the local variable does not affect any other statements nor passed/returned to other functions.
Note that it is relatively easy to verify the impact of instrumentation since it only requires analysis within the function. %Our analysis is capable of identifying all the affected variables and code to make sure it does not break the target program's functionalities.

\noindent
{\bf Multiple Functions (the Fns column in \autoref{table:selectedprograms}).}
In this type, an instrumentation affects multiple functions through function calls and global/member variables. 
\autoref{fig:instr_sample}-(c) shows an example. 
The instrumented SQL query is shown at line 15. The randomized query is passed to \code{get\_var()} (\blkcc{1}). The query is then used to call \code{query()} function (\blkcc{2}) and passed to the function again (\blkcc{3}). In the \code{query()} function, it is stored to the \code{\$last\_query} member variable (at line 24, \blkcc{4}) and passed to the \code{\_do\_query()} function (\blkcc{4}). Finally, in the \code{\_do\_query()} function, the query is used to call a sink function which is \code{mysql\_query()}.
Note that the \code{\$last\_query} variable that stores the randomized query is used later in the \code{print\_error()} function at lines 34 and 37 (\blkcc{7} and \blkcc{8}). 
In this example, the instrumentation at line 15 affects 5 functions (\code{prepare\_sql\_data()}, \code{get\_var()}, \code{query()}, \code{\_do\_query()}, and \code{print\_error()}).

%We verified all the instrumentations belong to this category are correct by tracing variables affected by the instrumentations. 
%To check whether this type of instrumentation breaks the benign functionalities or not, we trace all the variables that are affected by the instrumentation. %The next three columns present the number of local and global variables as well as functions. 
%Details are presented in the next section.

%%%%%%%%%%%%%%%%%%%%%%%%%%%%%%%%%%%%%%%%%%%%%%%%%%%%%%%%%%%%%%%
%%%%%%%%%%%%%%%%%%%%%%%%%%%%%%%%%%%%%%%%%%%%%%%%%%%%%%%%%%%%%%%
%%%%%%%%%%%%%%%%%%%%%%%%%%%%%%%%%%%%%%%%%%%%%%%%%%%%%%%%%%%%%%%

\vspace{-0.5em}
\subsection{Additional Discussions}
\label{appendix:additional_discussion}
\revised{
\subsubsection{Alternative Approach: Screening Unintended Commands}
\label{appendix:whitelist_approach}
One can develop an approach that only allows intended commands identified.
For instance, given a function call ``\code{system("rm file \$opt")}'', the approach will only allow the ``\code{rm}'' command. 
Such an approach (i.e., allowlist method) is fundamentally different from \sysname since it cannot distinguish \textit{different instances of commands} and enforce the same rule for every commands on an API. %, failing to identify the injected commands having the same name.
%\updated{}{\mw{The mechanism of allowlist and how we use our analysis results are different. We use static analysis to identify the intended commands from the trusted sources. Allowlist will allow the execution of commands even if it is not the developers' intention. So the attacker can inject the same command to compromise the system. However, we use static analysis to identify the benign commands from the trusted sources and those benign commands can only be used in the developers' intended conditions.}}\CJ{8) [6] (\#A) Clear justification of complex design choices (e.g., motivation for designing the Spinner instead of using a simple allowlist or sanitization).}
\updated{Unfortunately,}{For example,} it cannot prevent if an attacker injects the same command (e.g., ``\code{rm}'' in this case). % while \sysname can prevent such an attack. % because only randomized commands will be properly executed while the injected command is not randomized.
\updated{}{\sysname randomizes the first ``\code{rm}'' and leaves the second ``\code{rm}'' command, which is injected, preventing the attack.}
For SQL injections, approaches relying on known/allowed SQL keywords cannot prevent attacks leveraging keywords that are not considered (e.g., Section~\ref{subsec:comparison_existing}) while \sysname can prevent them.
}
% http://g2pc1.bu.edu/~qzpeng/manual/MySQL%20Commands.htm list of some sql statements

%%%%%%%%%%%%%%%%%%%%%%%%%%%%%%%%%%%%%%%%%%%%%%%%%%%%%%%%%%%%%%%%%%%%%%%%%%%%
%%%%%%%%%%%%%%%%%%%%%%%%%%%%%%%%%%%%%%%%%%%%%%%%%%%%%%%%%%%%%%%%%%%%%%%%%%%%
%%%%%%%%%%%%%%%%%%%%%%%%%%%%%%%%%%%%%%%%%%%%%%%%%%%%%%%%%%%%%%%%%%%%%%%%%%%%
% \subsubsection{Command Line including Multiple Commands}
% A command line containing multiple commands can be passed to a command execution API. For instance, two commands can be chained by using a pipe operator: `\code{ls | grep arg}'. To randomize them, \sysname's scanner identifies the four terms (\code{ls}, \code{|},  \code{grep}, and \code{arg}), and randomizes \code{ls} and \code{grep} as they are commands. 
% All commands are executed by the randomized shell process (by \sysname) as shown in Section~\ref{subsubsec:shellcommand_rand}.
% %
% In addition, a program may accept another binary program as an argument and can execute at runtime (e.g., `\code{exec open ls}'). In such case, the first program (i.e., \code{exec}) calls a command execution API to execute the second program (i.e., \code{open}) which again calls a command execution API to run the third command (i.e., \code{ls}). In such a case, all the programs except for the last program should be hardened by \sysname to prevent command injection attacks.

\subsubsection{Prepared Statements in Practice}
\label{appendix:enc_sql_statements}
%The most often recommended way to prevent SQL injection is to make use of prepared statements. However, 
As mentioned in Section~\ref{sec:discussion}, prepared statements are not well adopted in practice. We analyze all the SQL queries in the applications used in our evaluation. We find that 866 SQL queries from WordPress~\cite{wordpress} (459 queries), Pie Register~\cite{pieregister} (70 queries), PHPSHE~\cite{PHPSHE} (277 queries), AVideo-Encode~\cite{videoencoder} (39 queries), and Plainview Activity Monitor~\cite{PlainviewActivityMonitor} (21 queries).
None of them use the prepared statements.

\noindent
{\bf Unsupported Keywords.}
The following SQL keywords are not supported: \code{DESCRIBE} (or \code{DESC}), \code{ALTER DATABASE}, \code{LOAD DATA}, \code{LOAD XML}, \code{RENAME USER}, and \code{SHOW TABLES LIKE}.
In particular, WordPress (\code{s1}) is using the unsupported keywords, i.e., \code{DESC}, \code{SHOW TABLES LIKE}, in their queries, making it challenging to convert.
%We analyze all the SQL queries (866 queries) in the applications used in our evaluation, and there are XXX programs using them. 

\begin{figure}[h]
    \centering
    \includegraphics[width=1\columnwidth]{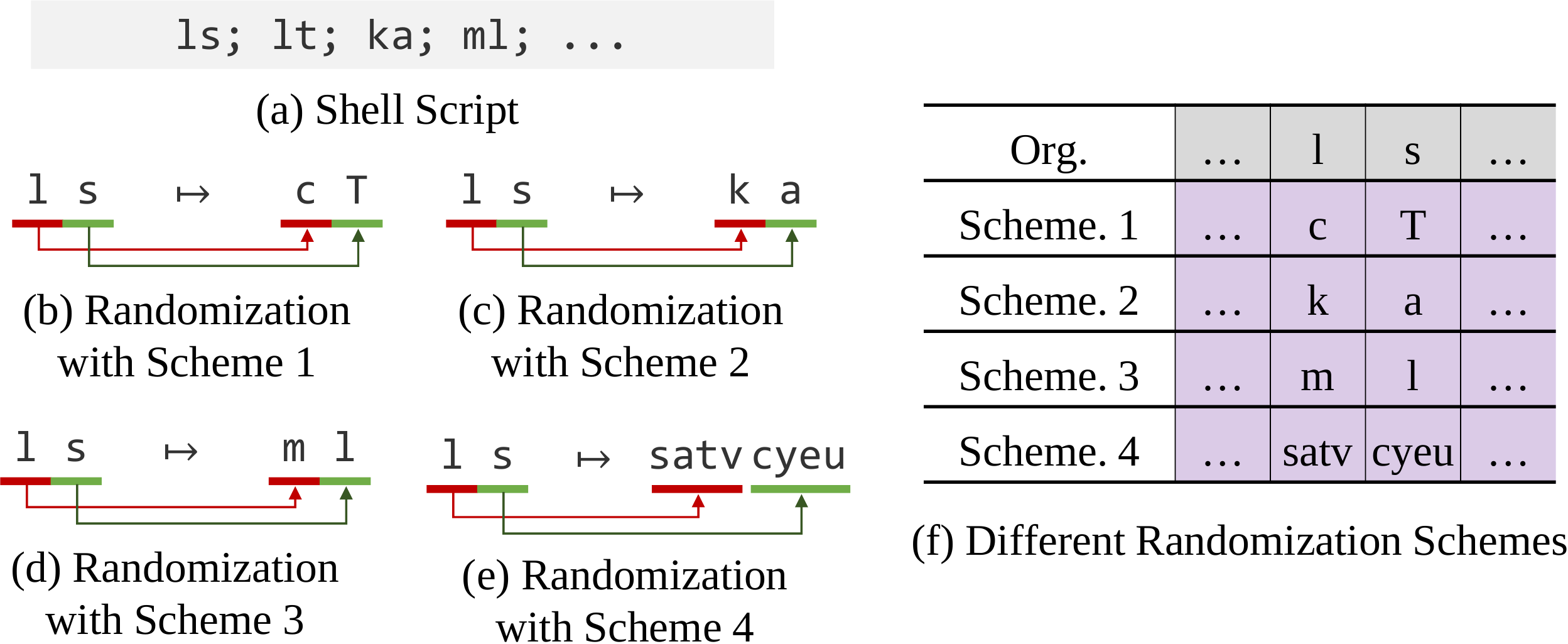}
    \vspace{-2em}
    \caption{\updated{}{Randomization Schemes Used for Each Command}}
    \label{fig:bruteforce}
\end{figure}

\revised{
\subsubsection{Brute-force Attack \sysname.}
\label{appendix:bruteforce_attack}
Attackers may inject multiple commands (or a shell script file containing multiple commands) to try out a number of guesses of randomization schemes.
From the attacker's perspective, if any of the guesses lead to the successful execution of the command, the attack is successful.
\autoref{fig:bruteforce}-(a) shows such a shell script containing multiple commands. 
We find that the Linux shell process handles individual commands separately, causing multiple command execution API invocations for each command. 
Recall that \sysname uses different randomizations on command execution API invocations. To this end, we randomize the subsystems differently, as shown in 
\autoref{fig:bruteforce}-(b-e). 
The first command failed because we randomize \optmap{ls}{cT}. The second attempt also failed as `\code{ka}' is expected. Even if an attacker learned this previous randomized command and injects \code{ka} next time, as shown in this example, it still fails as \sysname changes the randomization scheme to \optmap{ls}{ml}. Finally, one may try to inject a large number of the same command (e.g., millions of \code{sl}), waiting for our randomization scheme to become \optmap{ls}{sl}. Unfortunately, \sysname allows can be configured to use multiple bytes translation rules. For example, the randomization scheme 4 translates a single byte to 4 bytes. With this, searching space is practically too large to brute-force. 
Specifically, assume our randomization schemes use all printable ASCII characters (94 of them) to substitute, two-byte commands such as `\code{ls}', can be randomized to 8,741 (=$P(94,2)-1$) different two-byte characters. For 4 bytes commands, the space becomes extremely large: $P(94^4,2)-1$.
%Assuming that an attacker knows that his target is protected by \sysname and uses the basic one-to-one randomization, he will try to inject commands \code{`ls'} using a script. However, as shown in \autoref{fig:bruteforce}-(b), the shell will handle those commands one by one and each of them will have a different randomization table so the brute-force attack will fail.
}
%\CJ{2) [2.2] (\#A+C) Is the randomization method for a simple command easy to be attacked (e.g., `ls’ to `sl’ )?}

\noindent
{\bf Effectiveness of Dynamic Randomization.}
\sysname dynamically changes randomization scheme on every command, which we call dynamic randomization. 
To understand the effectiveness of dynamic randomization compared with the static randomization which uses a single randomization scheme during the entire execution, we tried brute-force attacks on both static and dynamic randomization approaches. 
In general, attackers need to try twice more attacks to break the dynamic approach than the static approach. 
For instance, using the dynamic approach (1-to-1 mapping) for three characters-long commands requires 70,191 more attempts to succeed the attack (which we believe quite effective) than the static approach.
% Details can be found on \cite{technical_report}.

\noindent
{\bf Experiment Results.} 
We conduct brute-force attack experiments. Specifically, we brute-force four different randomization schemes to show the effectiveness of the dynamic randomization scheme. 

%\addtolength{\tabcolsep}{-5pt}
\setlength{\tabcolsep}{3pt}
\begin{table}[h]
	\centering
	\caption{Brute force attacks on static and dynamic randomization schemes}
	\vspace{-1em}
	\label{table:brute_force_attack}
	\resizebox{0.85\columnwidth}{!}{
\begin{tabular}{l r r r r} 
    \toprule
    {\bf } &
    {\bf 1 to 1} &
    {\bf 1 to 2 } & 
    {\bf 1 to 3} &
	{\bf 1 to 4} \\
	\midrule

	{\bf Static randomization}  &  71.1K  & 9.8M   &  1,389T$^*$ & 195Q$^*$  	\\
	\rowcolor{gray!30} 
	{\bf Dynamic randomization} &  141.3K & 19.7M  &  2,779T$^*$ &	391Q$^*$    \\
	\bottomrule 
	% quintillion
	
	\multicolumn{5}{l}{T: Trillion.  Q: Quintillion. $^*$: Estimated value.}
\end{tabular}
}
%\vspace{-2em}
\end{table}

\autoref{table:brute_force_attack} shows the number of failed attempts before the first correct guess, leading to a successful attack. For instance, using the 1 to 1 mapping scheme, the static method prevents 71.1K attempts successfully. With the dynamic randomization scheme, the attack has to run 141.3K commands until the first successful guess.
Note that we decided to use the estimation for 1 to 3 and 1 to 4 randomization schemes because the experiment did not finish within 10 hours. We observe this result follows the distribution (i.e., static randomization approach follows the uniform distribution and dynamic randomization approach follows the geometric distribution). According to this observation, we put the expected value through the statistical method.
For the case of 1 to 2 scheme, using dynamic approaches for this command requires 9,807,906,470 more attempts to succeed the attack than static randomization.

\setlength{\tabcolsep}{3pt}
\begin{table}[h]
	\centering
	\caption{Update History of All Evaluated Programs }
	\vspace{-1em}
	\label{table:history_release}
	\resizebox{0.95\columnwidth}{!}{
\begin{tabular}{l l l r r c r} 
    \toprule
    {\bf ID} &
    {\bf Trusted Src.$\textsuperscript{1}$} &
    {\bf Language } & 
    {\bf \# S$\textsuperscript{2}$} &
	{\bf \# V$\textsuperscript{3}$} &
	{\bf Timeline (dd/mm/yyyy)} &
	{\bf Dur.$^4$ } 
	\\
	\midrule

	\code{s1}&
	Const.$^5$ , Conf.$^6$   &
	PHP  &  7 & 16  & 11/16/2017 $\sim$  10/29/2020 & 35 
	\\

	\code{s2} &
    Const.$^5$ , Conf.$^6$  &
    PHP & 6 & 17 & 05/11/2014 $\sim$ 08/26/2018 & 51
     \\

	\code{s3} &
	Const.$^5$  &
	PHP & 27 & 3  & 08/12/2017 $\sim$ 01/13/2020 & 29
	  \\

	\code{s4} &
	Const.$^5$  &
	PHP & 2 & 20  & 11/25/2014 $\sim$ 09/11/2020 & 69
	  \\
	
	\code{s5} &
	Const.$^5$ , Conf.$^6$  &
	PHP & 5 & 3  & 01/01/2017 $\sim$ 09/05/2018 & 20
	 \\
	
	\code{s6} &
	Const.$^5$ , Conf.$^6$  &
	PHP & 2 & 19  & 10/04/2011 $\sim$ 10/22/2020 & 108
	\\

	\code{s7} &
	Const.$^5$ , Path &
	C & 10 & 25  & 01/02/2016 $\sim$ 10/25/2020 & 57
	 \\
	
	\code{s8} &
	Const.$^5$  &
	C & 2 & 20  & 01/14/2016 $\sim$ 07/28/2020 & 54
	 \\

	\code{s9} &
	Const.$^5$  &
	C & 2 & 7  & 09/12/2012 $\sim$ 02/06/2018 & 64
	 \\

	\code{s10} &
	Const.$^5$ , Path &
	C & 1 & 3  & 12/22/2018 $\sim$ 07/15/2020 & 18
	 \\

	\code{s11} &
	Const.$^5$  & 
	Lua & 52 & 13  & 10/09/2014 $\sim$ 09/28/2020 & 71
	 \\

	\code{s12} &
	Const.$^5$  &
	JavaScript & 2 & 54  & 12/28/2009 $\sim$ 06/18/2012 & 29
	 \\

	\code{s13} &
	Const.$^5$  &
	JavaScript & 2 & 9  & 01/05/2018 $\sim$ 10/01/2019 & 20
	 \\

	\code{s14} &
	Const.$^5$  &
	JavaScript & 3 & 23  & 08/02/2017 $\sim$ 02/24/2020 & 30
	 \\

	\code{s15} &
	Const.$^5$  &
	JavaScript & 1 & 4  & 06/03/2014 $\sim$ 02/16/2018 & 44
	 \\

	\code{s16} &
	Const.$^5$  &
	JavaScript & 34 & 20  & 09/15/2016 $\sim$ 09/29/2020 & 48
	 \\

	\code{s17} &
    Const.$^5$    &
	JavaScript & 1 & 14  & 09/20/2014 $\sim$ 06/01/2017 & 32
	 \\

	\code{s18} &
	Const.$^5$  &
	JavaScript & 3 & 6  & 03/03/2017 $\sim$ 11/26/2019 & 32
	 \\

	\code{s19} &
    Const.$^5$  &
	JavaScript & 3 & 5  & 08/12/2017 $\sim$ 08/18/2017 & $<$1$^8$
	 \\

	\code{s20} &
	Const.$^5$  &
	JavaScript & 3 & 2  & 05/23/2014 $\sim$ 01/06/2020 & 67
	 \\

	\code{s21} &
	Const.$^5$  &
	JavaScript & 3 & 4  & 12/23/2013 $\sim$ 05/16/2020 & 76
	 \\

	\code{s22} &
	Const.$^5$  &
	JavaScript & 2 & 15  & 01/25/2019 $\sim$ 01/10/2020 & 11
	 \\

	\code{s23} &
	Const.$^5$  &
	JavaScript & 3 & 6  & 04/07/2016 $\sim$ 03/23/2017 & 11
	 \\
		
	\code{s24} &
	Const.$^5$  &
	JavaScript & 6 & 23  & 10/16/2015 $\sim$ 05/09/2017 & 18
	 \\
	
	\code{s25} &
	Const.$^5$  &
	JavaScript & 3 & 11  & 02/22/2013 $\sim$ 06/28/2018 & 64
	 \\

	\code{s26} &
	Const.$^5$  &
	JavaScript & 1 & 25  & 11/23/2016 $\sim$ 10/30/2019 & 35
	 \\

	\code{s27} &
	Const.$^5$  &
	JavaScript & 2 & 4  & 06/30/2015 $\sim$ 01/30/2016 & 7
	 \\ \midrule

	\code{s28} &
	Const.$^5$ , Conf.$^6$  &
	PHP & 5 & 35  & 05/06/2013 $\sim$ 10/21/2020 & 89
	\\

	\code{s29} &
	Const.$^5$ , Conf.$^6$  &
	PHP & 6 & 28  & 09/03/2019 $\sim$ 10/15/2020 & 13
	\\

	\code{s30} &
	Const.$^5$ , Conf.$^6$  &
	PHP & 17 & 27  & 03/04/2016 $\sim$ 10/15/2020 & 55
	\\

	\code{s31} &
	Const.$^5$ , Conf.$^6$  &  PHP& 2 & 23  & 05/30/2016 $\sim$ 10/20/2020 & 52
	\\ 

	\code{s32} &
	Const.$^5$ , Conf.$^6$  & PHP & 2 & 11  & 10/25/2010 $\sim$ 04/04/2020 & 113
	\\ 

	\code{s33} &
	Const.$^5$  & PHP & 15 & 32  &  01/07/2015 $\sim$ 10/04/2020 & 68
	\\ 

	\code{s34} &
	Const.$^5$  & PHP & 3 & 10  & 02/18/2014 $\sim$ 09/28/2020 & 29
	\\ 

	\code{s35} &
	Const.$^5$ , Env.$^7$  & PHP & 8 & 14  & 04/15/2016 $\sim$ 10/24/2020 & 54
	\\ 

	\code{s36} &
	Const.$^5$  & PHP & 1 & 13  & 12/19/2016 $\sim$ 11/12/2019 & 34
	\\ 

	\code{s37} &
	Const.$^5$  & PHP & 1 & 10  & 03/03/2017 $\sim$ 09/28/2020 & 42
	\\
	
	\code{s38} &
	Const.$^5$  &
	JavaScript & 1 & 38  & 03/26/2014 $\sim$ 10/20/2020 & 78
	 \\

	\code{s39} &
	Const.$^5$  &
	JavaScript & 1 & 32  & 12/04/2015 $\sim$ 06/08/2020 & 54
	 \\

	\code{s40} &
	Const.$^5$  &
	JavaScript & 7 & 20  & 07/29/2019 $\sim$ 01/20/2020 & 5
	 \\

	\code{s41} &
	Const.$^5$  &
	JavaScript & 3 & 21  & 12/28/2016 $\sim$ 07/28/2020 & 43 
	 \\

	\code{s42} &
	Const.$^5$  &
	JavaScript & 2 & 27  & 11/22/2017 $\sim$ 10/22/2020 & 35
	 \\
	 
	\bottomrule 
	\multicolumn{7}{l}{1: Trusted Sources. 2: Versions. 3: Sinks. 4: Duration in months. 5: Constant String. }\\
	\multicolumn{7}{l}{ 6: Configuration File. 7: Environment Variable. 8: Less than 1 month.}
	%7: Execution prevented with ``Command not found''
	%\multicolumn{16}{l}{8: Execution prevented with ``File not found''. 9: Execution prevented with a time out error. 10: Average.}
	%\multicolumn{15}{l}{1: Execution prevented with an error ``Command not found''. 2: Execution prevented with an error ``File not found''. 3: Execution prevented with a time out error.} \\
\end{tabular}
}
%\vspace{-2em}
\end{table}

%%%%%%%%%%%%%%%%%%%%%%%%%%%%%%%%%%%%%%%%%%%%%%%%%%%%%%%%%%%%%%%%%%%%%%%%%%%%
%%%%%%%%%%%%%%%%%%%%%%%%%%%%%%%%%%%%%%%%%%%%%%%%%%%%%%%%%%%%%%%%%%%%%%%%%%%%
%%%%%%%%%%%%%%%%%%%%%%%%%%%%%%%%%%%%%%%%%%%%%%%%%%%%%%%%%%%%%%%%%%%%%%%%%%%%
\revised{
\subsubsection{Impact of Software Updates on \sysname}
\label{appendix:software_update_impact}
As discussed in Section~\ref{sec:discussion}, if software updates of a target application cause changes in the trusted command specification, manual analysis of the target application is required. 
To understand how prevalent such cases are in practice, we study the update history of 42 applications (27 applications in Table~\ref{table:selectedprograms} and 15 programs in Table~\ref{table:extra_programs}).
As shown in \autoref{table:history_release}, we track major version updates from the first stable version to the most recent major update until November 2020. We analyze each major update to understand whether the trusted command specification of an old version should be updated for a new version to use \sysname.
The result shows that none of the trusted command specifications are changed between versions.

\autoref{table:history_release} shows the results. 
All 42 applications use constant strings as a trusted source. 
%An example of trusted command specification of such applications is shown in \autoref{fig:trusted_specifications}.
There are 9 programs that have both configuration files and constant strings as trusted sources. 
Their trusted command specification is similar to \autoref{fig:spec_example}-(a). 
\texttt{s7} and \texttt{s10} have folder paths and constant strings as trusted sources and can be defined as shown in \autoref{fig:spec_example}-(c).
\texttt{s35} requires the environment variable as a trusted source. For this program, to prevent attacks that attempt to compromise environment variables, we hook \code{setenv} and \code{getenv}. % functions to detect unauthorized modifications. %\CJ{@MW, fix XXX and missing link here.}
%source and the file is not exposed to the remote-user. One application Composer uses the environment variable as a trusted source. There is only one application monolog that allows the trusted source to be modified by remote users. However, it is a log framework for developers that are not meant to be exposed. 2 applications lighttpd and Goahead use file path a trusted source because they are both web servers. Examples of trusted command specifications are shown in \autoref{fig:trusted_specifications}
}

%\addtolength{\tabcolsep}{-5pt}
\setlength{\tabcolsep}{3pt}
\begin{table}[h]
	\centering
	\caption{Performance of \sysname}
	\vspace{-1em}
	\label{table:perf_corr}
	\resizebox{0.95\columnwidth}{!}{
\begin{tabular}{l l l} 
    \toprule
    {\bf Program} &
    {\bf Version used in Section~\ref{sec:eval} } &
    {\bf Latest version } 
    \\

	\midrule

	WordPress  & 	4.33\% (released in 12/18/19) & 4.41\% (released in 2/22/21) \\
	Leptonica  & 	4.25\% (released in 6/11/17) & 4.21\% (released in 7/28/20) \\

 \bottomrule 

%	\multicolumn{3}{l}{*: Estimated value (unit: trillion): We decided to use the estimation} \\
%	\multicolumn{6}{l}{ the date of the last commit.}

\end{tabular}
}
%\vspace{-2em}
\end{table}

\noindent
{\bf \sysname on Different Versions of Target Programs.}
To understand the impact of software updates on the performance of \sysname, we applied \sysname to WordPress (v5.6.2; released on Feb 22, 2021) and Leptonica (v1.8; released in July 28, 2020) in addition to the versions we have evaluated in Section~\ref{sec:eval}, as shown in \autoref{table:perf_corr}. 
The resulting protected programs are correct where we observe a similar average overhead of 4.31\%.

%The intended commands of one application might be changed in the new version. But if the trusted sources remain the same, we only need to re-apply \sysname again. 
%We checked the trusted sources of the history releases of the applications evaluated in this paper. For the frequently updated applications like WordPress, we checked the major update of the releases. 
%We find that the trusted sources remain unchaged in different versions so we can easily apply \sysname in different versions of the applications. The detailed results are shown in \autoref{table:history_release}.} \CJ{7) [5] (\#B+C) Need discussion on the system’s scalability and reproducibility}

%%%%%%%%%%%%%%%%%%%%%%%%%%%%%%%%%%%%%%%%%%%%%%%%%%%%%%%%%%%%%%%%%%%%%%%%%%%%
%%%%%%%%%%%%%%%%%%%%%%%%%%%%%%%%%%%%%%%%%%%%%%%%%%%%%%%%%%%%%%%%%%%%%%%%%%%%
%%%%%%%%%%%%%%%%%%%%%%%%%%%%%%%%%%%%%%%%%%%%%%%%%%%%%%%%%%%%%%%%%%%%%%%%%%%%
\revised{
\subsubsection{Preventing Trusted Sources from Being Compromised}
\sysname often trusts configuration files that cannot be modified by remote attackers. However, if our analysis is incomplete or the system has other vulnerabilities that allow attackers to compromise the trusted configuration files, \sysname's protection can be affected. 
As a mitigation, we implement a kernel module that denies any modifications to the configuration files. We also tried secure file systems~\cite{li2004secure} to prevent unauthorized modifications to the configuration files. 
We enabled them during our evaluation, and we do not observe any errors caused by them, meaning that users may also use such approaches to protect \sysname.
}
\subsection{Diglossia~\cite{diglossia} vs \sysname}
\label{appendix:diglossia_comparison}

\begin{figure}[h]
    \centering  
    %\vspace{-1em}
     \includegraphics[width=1.0\columnwidth]{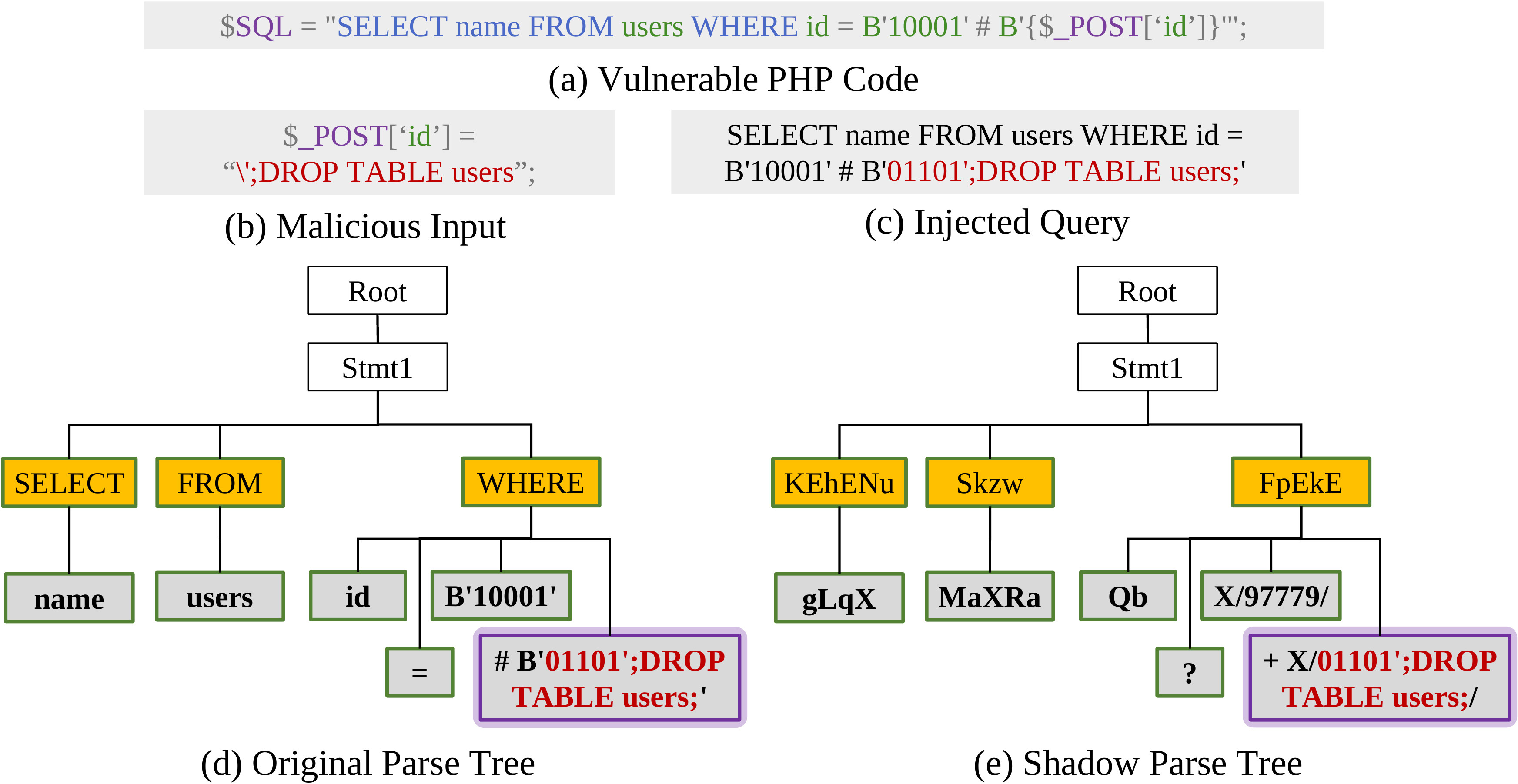}
     %\vspace{-2em}
     \caption{\updated{}{Failure Case of Diglossia with PHP-SQL-Parser}}
     %\vspace{-2em}
    \label{fig:diglossia_failure_1} 
\end{figure}

\revised{
%\subsubsection{Diglossia~\cite{diglossia} vs \sysname}
%\label{appendix:diglossia_comparison}
In addition to Section~\ref{subsec:comparison_existing}, we compare \sysname with another our own implementation of Diglossia~\cite{diglossia} using PHP-SQL-Parser~\cite{greenlion/PHP-SQL-Parser}, which is the most popular SQL Parser for PHP in GitHub. 
\autoref{fig:diglossia_failure_1}-(a) shows a vulnerable PHP program's code.
Given the malicious input shown in \autoref{fig:diglossia_failure_1}-(b), the malicious query is injected as shown in \autoref{fig:diglossia_failure_1}-(c).
\autoref{fig:diglossia_failure_1}-(d) and (e) show parse trees from the original parser and the shadow parser. Nodes with yellow backgrounds represent keywords while nodes with gray backgrounds represent strings or numbers which are allowed to be injected.
Nodes with green borders are correctly translated in the shadow parser, meaning that they are intended nodes.
Nodes with the violet borders are those that are \emph{not fully translated}, meaning that some values (i.e., the first 2 characters) are translated and some are not.
%Note that they have hence having the same value between the original parser and the shadow parser.
Note that Diglossia detects an injected input by identifying nodes with the same values between the two parse trees. In this case, we do not have such nodes, meaning that Diglossia will miss the attack. 
%They are essentially injected input. Diglossia will raise an alarm if the injected input is a keyword, which is yellow background boxes in \autoref{fig:diglossia_failure_1}.
The injected code is not properly parsed due to the bug of the parser. It fails to recognize SQL grammar after the \code{\#} symbol, an XOR operator in PostgreSQL. 
%Since the entire injected query is not considered as a keyword, the exploitation is successful.

\sysname uses a scanner and applies reverse-randomization scheme to the injected query, preventing the attack.
}

\subsubsection{Preventing XXE Injection}
XML External Entity (XXE) injection allows attackers to inject an XML external entity in an XML file. XML external entity is a custom XML tag that allows an entity to be defined based on the content of a file path or URL.
An attacker can abuse the external entity to leak the content of arbitrary files.
In this case, we use a WordPress plugin, Advanced XML Reader~\cite{AdvancedXMLReader}, to demonstrate how \sysname prevents the XXE injection attack.

\begin{figure}[ht]
    \centering
    \includegraphics[width=0.8\columnwidth]{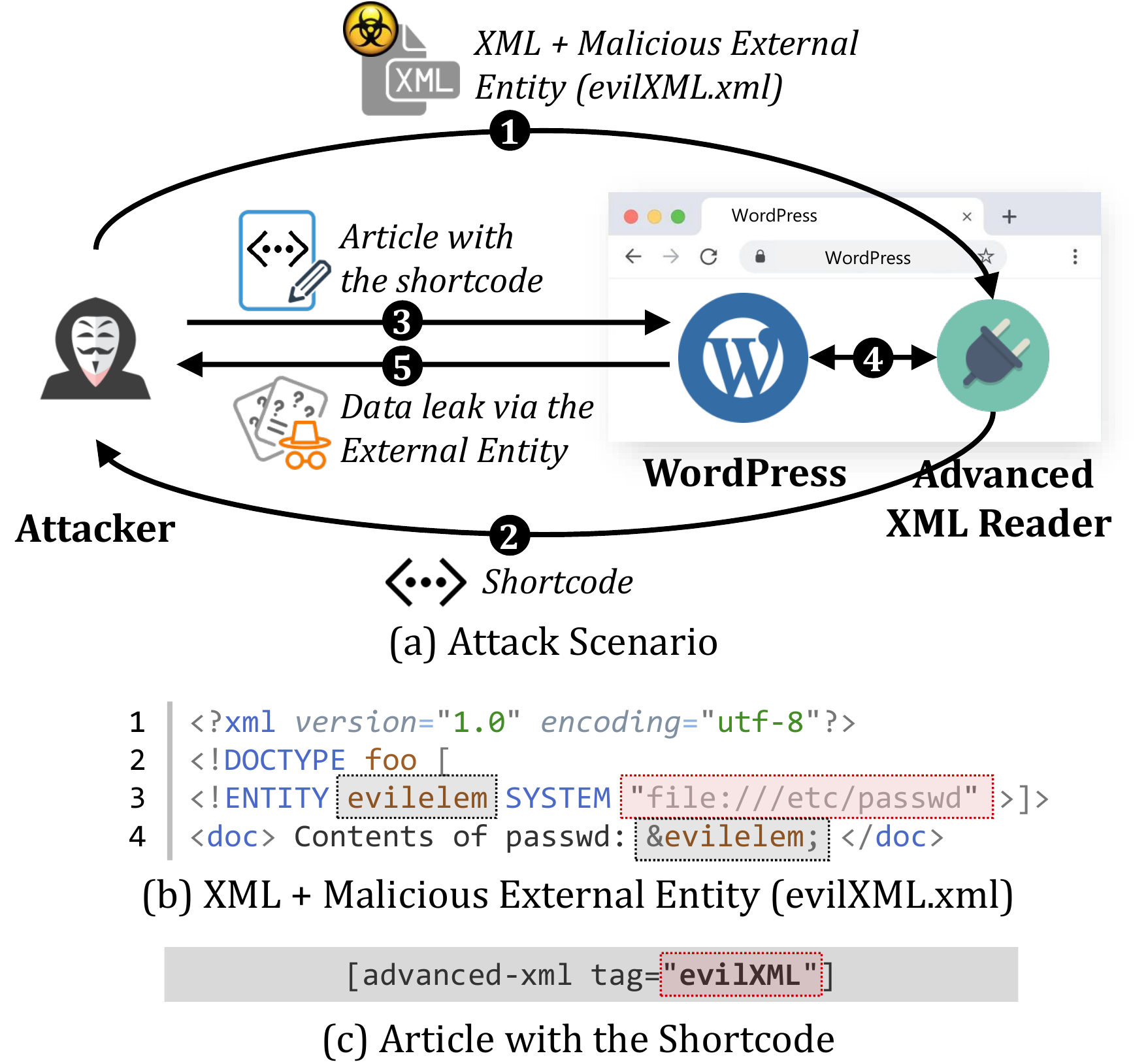}
    \vspace{-1em}
     \caption{XXE Injection Attack Scenario}
     \vspace{-1em}
     \label{fig:case_xee_attack}
     %\vspace{-1em}
\end{figure}

\autoref{fig:case_xee_attack}-(a) shows an attack scenario. The attacker first sends a malicious XML file with a malicious external entity (\blkcc{1}). The malicious XML file's content is shown in \autoref{fig:case_xee_attack}-(b). The XML file defines an \code{$<$!ENTITY SYSTEM$>$} tag with a file path \code{/etc/passwd}. The tag is used in line 4, which will be the content of the XML file when it is requested. 
The vulnerable plugin uploads the XML file and returns a shortcode (\blkcc{2}), which is essentially the name of the uploaded file to refer to the XML content in the future.
Now, the attacker posts an article with the shortcode (\blkcc{3}) as shown in \autoref{fig:case_xee_attack}-(c). Note that the tag value indicates the uploaded file's name. 
Once the post is uploaded, WordPress sends it to the plugin (Advanced XML Reader), which will parse and resolve the XML file referred to in the post (\blkcc{4}). During the processing, the plugin reads the password file and returns the content. When the posted article is requested, the password file's content will be delivered (\blkcc{5}).

\begin{figure}[ht]
    \centering
    \includegraphics[width=1.0\columnwidth]{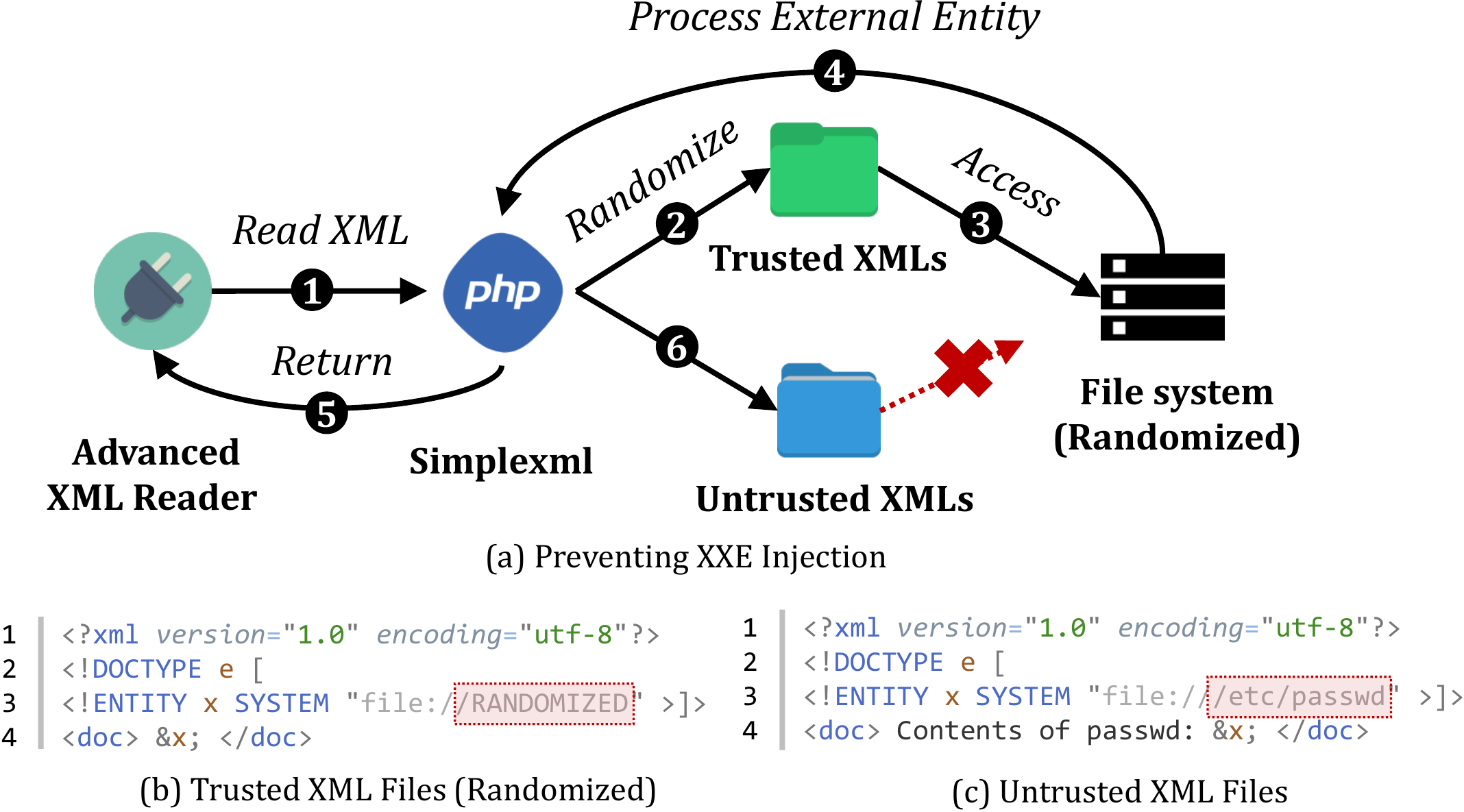}
    \vspace{-2em}
     \caption{Preventing an XXE Injection Attack}
     \vspace{-1em}
     \label{fig:case_xee_prevention}
         %\vspace{-0.5em}

\end{figure}

\autoref{fig:case_xee_prevention} shows how \sysname ensures benign operations while preventing the XXE injection attack described in \autoref{fig:case_xee_attack}.

\noindent
{\bf Benign Operation.}
When the plugin reads an XML file (\blkcc{1}), \sysname intercepts the file I/O and check whether it reads a trusted XML file or not. Note that \sysname maintains a list of trusted XML file paths. Typically, those are the XML files provided by an administrator, not the files that are uploaded by remote users.
If the file path of the XML file is in the list, \sysname randomizes the external entities' file contents (\blkcc{2}). Then, it tries to access the file system with the randomized file name. As the file paths of the file system are randomized by \sysname, it successfully reads the file and returns (\blkcc{3} and \blkcc{4}). Finally, the content is returned (\blkcc{5}).

\noindent
{\bf Preventing XXE Injection.}
When an attacker uploads a malicious XML file, it is not included in the trusted XML file list. 
When the plugin tries to read an XML file that is uploaded by a remote user (which is not trusted, \blkcc{6}), the XML file's entity will not be randomized. As a result, it will not be able to access the file system correctly, leading to a file open error.

\end{document}